\documentclass[twocolumn,showpacs,amsmath,amssymb,prd,epsf]{revtex4}

\usepackage{graphicx}
\usepackage{dcolumn}% Align table columns on decimal point
\usepackage{bm}% bold math
\usepackage{epsf}

\def\agt{\mathrel{\raise.3ex\hbox{$>$}\mkern-14mu\lower0.6ex\hbox{$\sim$}}}
\def\alt{\mathrel{\raise.3ex\hbox{$<$}\mkern-14mu\lower0.6ex\hbox{$\sim$}}}

\newcommand{\beq}{\begin{equation}}
\newcommand{\eeq}{\end{equation}}
\newcommand{\beqn}{\begin{eqnarray}}
\newcommand{\eeqn}{\end{eqnarray}}

\newcommand{\varep}{\varepsilon}

\begin{document}

\title{The mass ejection from the merger of binary neutron stars}

\author{Kenta Hotokezaka$^1$}
\author{Kenta Kiuchi$^2$}
\author{Koutarou Kyutoku$^3$}
\author{Hirotada Okawa$^4$}
\author{Yu-ichiro Sekiguchi$^2$}
\author{Masaru Shibata$^2$}
\author{Keisuke Taniguchi$^5$}

\affiliation{
$^1$Department of Physics, Kyoto University, Kyoto 606-8502, Japan \\
$^2$Yukawa Institute of Theoretical Physics, Kyoto University,
Kyoto 606-8502, Japan \\
$^3$Theory Center, Institute of Particles and Nuclear Studies, KEK,
Tsukuba, Ibaraki 305-0801, Japan \\
$^4$CENTRA, Departamento de F\'{i}sica,
Instituto Superior T\'ecnico, Universidade T\'ecnica de Lisboa - UTL,Av. Rovisco Pais 1, 1049 Lisboa, Portugal\\
$^5$Graduate School of Arts and Sciences, The University of Tokyo,
Tokyo 153-8902, Japan
}

\begin{abstract}
Numerical-relativity simulations for the merger of binary neutron
stars are performed for a variety of equations of state (EOSs) and for
a plausible range of the neutron-star mass, focusing primarily on the
properties of the material ejected from the system. We find that a
fraction of the material is ejected as a mildly relativistic and
mildly anisotropic outflow with the typical and maximum velocities
$\sim 0.15$ -- $0.25c$ and $\sim 0.5$ -- $0.8c$ (where $c$ is the
speed of light), respectively, and that the total ejected rest mass is
in a wide range $10^{-4}$ -- $10^{-2}M_{\odot}$, which depends
strongly on the EOS, the total mass, and the mass ratio. The total
kinetic energy ejected is also in a wide range between $10^{49}$ and
$10^{51}~{\rm ergs}$. The numerical results suggest that for a binary
of canonical total mass $2.7M_{\odot}$, the outflow could generate an
electromagnetic signal observable by the planned telescopes through
the production of heavy-element unstable nuclei via the
$r$-process~\cite{LP1998,Shri2005,Metzger2010} or through the
formation of blast waves during the interaction with the interstellar
matter~\cite{Nakar2011}, if the EOS and mass of the binary are
favorable ones. 
\end{abstract}

\maketitle

\section{Introduction} \label{secI}

Coalescence of binary neutron stars is one of the most promising
sources for next-generation kilo-meter-size gravitational-wave
detectors such as advanced LIGO, advanced VIRGO, and KAGRA
(LCGT)~\cite{LIGOVIRGO}. These detectors will detect gravitational
waves in the next 5 -- 10~yrs.  Statistical studies have predicted
that the detection rate of gravitational waves emitted by binary
neutron stars for these detectors will be $\sim 1$ -- 100 per
year~\cite{Rate,RateLIGO}. The typical signal-to-noise ratio for most
of these events will be $\sim 10$ or less. Thus, it will be quite
helpful if electromagnetic or other signals observable are associated
with the gravitational-wave bursts and the gravitational-wave
detection is accompanied by the detections of other
signals. Short-hard gamma-ray bursts (SGRB) have been inferred to be
associated with the binary neutron star merger~\cite{GRB-BNS}.
However, the jet of SGRB would be highly collimated~\cite{Fong2012},
and hence, it will not be always possible to detect SGRB associated
with the binary neutron star mergers. Moreover, it is not guaranteed
that the telescopes for the observation of SGRB will be in operation
with the gravitational-wave detectors.  Exploring other possible
signals that could be detected is a very important subject in the
fields of gravitational-wave
physics/astronomy~\cite{Metzger2010,Liebling2010,RKLR2011,GBJ2011,Nakar2011,LR,MB2012}.

This paper presents our latest results of numerical simulations
performed in the framework of numerical relativity, focusing in
particular on the exploration of the material ejected from binary
neutron star mergers. In the past decade, numerical simulations for
the merger of binary neutron stars in full general relativity, which
is probably the unique approach of the rigorous theoretical study for
this subject, have been extensively performed since the first success
in 2000~\cite{SU00} (see,~e.g.,~\cite{Duez,FR2012} for a review of
this field).  However, most of the simulations have focused on the
studies of gravitational waveforms and the resulting product formed in
the central region. Few attention has been paid to the study for the
material ejected (but see~\cite{GBJ2011} for a study in an approximate
framework of general relativity, and see \cite{Rosswog1999, Freiburghaus1999,Rosswog2000,Rosswog2012}
for an early effort in the context of Newtonian gravity).

The material ejected from binary neutron star mergers may generate
electromagnetic signals observable in the current and future-planned
telescopes. One possible signal could be generated by the radioactive
decay of unstable $r$-process nuclei, which are produced from the
neutron-rich material in the
ejecta~\cite{LP1998,Shri2005,Metzger2010,RKLR2011,GBJ2011,MB2012,Rosswog2012}. A
fraction of the unstable nuclei produced subsequently decay in a short
timescale and could heat up the ejecta, which emits a UV and visible
light that may be observable by current and future-planned optical
telescopes. In this case, the typical duration of a peak luminosity is
expected to be several hours to a day. Another possible signal could be
generated during the free expansion and the subsequent Sedov phase of
the ejecta which sweeps up the interstellar medium and forms blast
waves~\cite{Nakar2011}. In this process turning on, the shocked material
at the blast waves could generate magnetic fields and accelerate
particles that emit synchrotron radiation in the radio-wave band, for a
hypothetical amplification of the electromagnetic field and a
hypothetical electron injection. It is also pointed out that the binary
neutron star merger could drive ultra-relativistic outflows in every
direction and emit synchrotron radiation in x-ray-to-radio bands within
a second-to-day timescale~\cite{KIS2012}. All these studies illustrate
that exploring the process of the material ejection from binary neutron
star mergers in detail is an important subject.

For the detailed numerical study of the ejected material, we have to
be careful when following the motion of the materials in a low-density
outer region. Most of the numerical-relativity simulations of binary
neutron star mergers so far have been performed with a computational
domain that was not wide enough for this
purpose~\cite{Duez,FR2012}. We have to enlarge the computational
domain sufficiently widely to confirm that the outflowed material is
indeed ejected from the system (i.e., we have to confirm that the
material is indeed unbound by the system by following the motion of
the ejected material for a long time). Another subtle issue in the
hydrodynamics simulations is that we have to put an artificial
atmosphere when employing a conservative shock capturing scheme that
is a standard one in this field~\cite{Toni}. In our previous
simulations~\cite{STU,hotoke,SKSS}, we put an atmosphere with fairly
large density ($\sim 10^7~{\rm g/cm^3}$) that did not affect the
motion of neutron stars but did for the motion of the ejected material
of low density which might escape to a far region. For the study of
the mass ejection, we have to reduce the density of the atmosphere as
low as possible (which should be much lower than the density of the
ejected material), and in addition, we have to carefully confirm that
such an artificial atmosphere does not affect the properties of the
ejected material. In the simulation reported in this paper, we succeed
in the simulation reducing the atmosphere density to a low level
($\alt 10^5~{\rm g/cm^3}$) enough to obtain a scientifically
quantitative result.

The paper is organized as follows: In Sec. II, we summarize the
equations of state (EOSs) employed and models of binary neutron stars.
In Sec. III, we briefly summarize our formulation and numerics for
solving Einstein's equation and hydrodynamics equations as well as the
tools for diagnostics.  In Sec. IV, numerical results are presented,
focusing on the properties of the material ejected from the system.
Section V is devoted to a summary and discussion.  Throughout this
paper, we employ the geometrical units $c=1=G$ where $c$ and $G$ are
the speed of light and gravitational constant, respectively, although
we recover $c$ when we need to clarify the units.

\section{Equations of state and chosen models} \label{secII}

In this section, we summarize the model EOSs employed in this paper,
and initial condition of binary neutron stars chosen for numerical
simulations. As shown in Sec.~\ref{sec:res}, the properties of the
material ejected from binary neutron star mergers depend strongly on
these inputs.

\begin{table*}[t]
%\tbl{
\caption{ Parameters and key quantities for four piecewise polytropic
EOSs employed in this paper.  $P_2$ is shown in units of ${\rm
dyn/cm}^2$.  $M_{\rm max}$ is the maximum mass along the sequences of
spherical neutron stars (cf. Fig.~\ref{fig2}).  ($R_{1.35},
\rho_{1.35}$) and ($R_{1.5}, \rho_{1.5}$) are the circumferential
radius and the central density of $1.35M_{\odot}$ and $1.5M_{\odot}$
neutron stars, respectively.  We note that the values of the mass,
radius, and density listed are slightly different from those obtained
in the original tabulated EOSs (see the text for the reason).  MS1 is
referred to as this name in~\cite{rlof2009}, but in other references
(e.g., \cite{lattimerprakash}), it is referred to as MS0. We follow
\cite{rlof2009} in this paper.} 
%%%%%%%%%%%%%%
{\begin{tabular}{ccccccc} \hline
EOS &  $(\log(P_2), \Gamma_1, \Gamma_2, \Gamma_3)$ & $M_{\rm max}(M_{\odot})$ &
$R_{1.35}$(km) & $\rho_{1.35}({\rm g/cm^3})$ &
$R_{1.5}$(km) & $\rho_{1.5}({\rm g/cm^3})$
\\ \hline
APR4 & $(34.269, 2.830, 3.445, 3.348)$ & 2.20
& 11.1 & $8.9 \times 10^{14}$ & 11.1 & $9.6 \times 10^{14}$ \\
ALF2 & $(34.616, 4.070, 2.411, 1.890)$ & 1.99
& 12.4 & $6.4 \times 10^{14}$ & 12.4 & $7.2 \times 10^{14}$ \\
H4   & $(34.669, 2.909, 2.246, 2.144)$ & 2.03
& 13.6 & $5.5 \times 10^{14}$ & 13.5 & $6.3 \times 10^{14}$ \\
MS1  & $(34.858, 3.224, 3.033, 1.325)$ & 2.77
& 14.4 & $4.2 \times 10^{14}$ & 14.5 & $4.5 \times 10^{14}$ \\
%%SLy  & $(34.384, 3.005, 2.988, 2.851)$ & 2.049 & 11.74 \\
%%PS   & $(34.671, 2.216, 1.640, 2.365)$ & 1.755 & 15.47 \\
%%H3   & $(34.646, 2.787, 1.951, 1.901)$ & 1.788 & 13.84 \\
\hline
\end{tabular}
}
\label{table:EOS}
\end{table*}

\subsection{Equations of state} \label{sec:EOS}

The exact EOS for the high-density nuclear matter is still
unknown~\cite{lattimerprakash}. This implies that a numerical
simulation employing a single particular EOS, which might not be
correct, would not yield a scientific result. A study,
systematically employing a wide possible range of EOSs, is required
for binary neutron star mergers. Nevertheless, the latest discovery of
a high-mass neutron star PSR J1614-2230 with mass $1.97 \pm 0.04
M_{\odot}$~\cite{twosolar} significantly constrains the model EOS to
be chosen, because it suggests that the maximum mass for spherical
neutron stars for a given EOS has to be larger than $\sim 2M_{\odot}$.
This indicates that the EOS should be rather stiff, although there are
still many candidate EOSs. 

To model a variety of the candidate EOSs, specifically, we employ a
piecewise polytropic EOS proposed by Read et al.~\cite{rlof2009}.
This EOS is described assuming that neutron stars are cold (in a
zero-temperature state), i.e., the rest-mass density, $\rho$,
determines all other thermodynamical quantities. To systematically
model nuclear-theory-based EOSs at high density with a small number of
parameters, the pressure is written in a parameterized form as
\begin{equation}
P(\rho) = \kappa_i \rho^{\Gamma_i} ~~ {\rm for} ~~ \rho_i \le \rho <
 \rho_{i+1} ~~ (0 \le i \le n) ,
\end{equation}
where $n$ is the number of the pieces used to parameterize a
high-density EOS, $\rho_i$ is the rest-mass density at the boundary of
two neighboring $(i-1)$-th and $i$-th pieces, $\kappa_i$ is the
polytropic constant for the $i$-th piece, and $\Gamma_i$ is the
adiabatic index for the $i$-th piece. Here, $\rho_0=0$, $\rho_1$
denotes a nuclear density $\sim 10^{14}~{\rm g/cm^3}$ determined
below, and $\rho_{n+1} \to
\infty$.  Other parameters $(\rho_i , \kappa_i, \Gamma_i)$ are
determined by fitting with a nuclear-theory-based EOS.  Requiring the
continuity of the pressure at each $\rho_i$, $2n$ free parameters, say
$(\kappa_i,\Gamma_i)$, determine the EOS completely. The specific
internal energy, $\varepsilon$, and hence the specific enthalpy, $h$,
are determined by the first law of thermodynamics and the continuity
of each variable at boundary densities, $\rho_i$.

Read et al.~\cite{rlof2009} showed that a piecewise polytropic EOS
with three pieces above the nuclear density (i.e., $n=3$)
approximately reproduces most properties of the nuclear-theory-based
EOS at high density, and they derived the fitted parameters for a
large number of nuclear-theory-based EOSs.  In this paper, thus, we
employ this piecewise polytropic EOS, determining the free parameters
basically following~\cite{rmsucf2009,kst2010,kst2011} (in which a
piecewise polytrope with $n=1$ was used). First, the EOS below the
nuclear density $\rho_1$ is fixed by the following parameters
\begin{eqnarray}
\Gamma_0 &=& 1.35692395 , \\
\kappa_0 / c^2 &=& 3.99873692 \times 10^{-8} \;
( {\rm g} / {\rm cm}^3)^{1 - \Gamma_0} .
\end{eqnarray}
The EOS for the nuclear matter was determined in~\cite{rlof2009} as
follows: $\rho_2$ was fixed to be $\rho_2=10^{14.7} {\rm g/cm}^3$, and
$P_2$ at $\rho=\rho_2$ was chosen as a free parameter. The reason is
that $P_2$ is closely related to the radius and deformability of
neutron stars~\cite{lattimerprakash2001}. Namely, $P_2$ primarily
determines the stiffness of an EOS. Second, $\rho_3$ was fixed to be
$\rho_3=10^{15.0} {\rm g/cm}^3$. With these choices, the set of free
parameters becomes ($P_2, \Gamma_1, \Gamma_2, \Gamma_3$). These four
parameters are determined by a fitting procedure (see~\cite{rlof2009}
for the fitting procedure).

With the given values of $\Gamma_1$ and $P_2$, $\kappa_1$ and $\rho_1$
are subsequently determined by
\begin{eqnarray}
\kappa_1 &=& P_2 \rho_{2}^{-\Gamma_1} , \\
\rho_1 &=& ( \kappa_0 / \kappa_1 )^{1 / ( \Gamma_1 - \Gamma_0 )} .
\end{eqnarray}
By the same method, $\kappa_2$ and $\kappa_3$ are determined from
\begin{eqnarray}
\kappa_2 \rho_2^{\Gamma_2}= \kappa_1 \rho_2^{\Gamma_1},~~~~
\kappa_3 \rho_3^{\Gamma_3}= \kappa_2 \rho_3^{\Gamma_2}.
\end{eqnarray}

Table~\ref{table:EOS} lists the EOSs and their parameters which we
employ in this study.  We choose four types of the representative
EOSs.  APR4 was derived by a variational method with modern nuclear
potentials~\cite{APR4} for the hypothetical components composed of
neutrons, protons, electrons, and muons; MS1 was derived by a mean-field
theory for the hypothetical components composed of neutrons, protons,
electrons, and muons, as well~\cite{MS1}; H4 was derived by a
relativistic mean-field theory including effects of
hyperons~\cite{H4}; ALF2 is a hybrid EOS which describes a nuclear
matter for a low density and a quark matter for a high density with
the transition density is $3\rho_{\rm nuc}$ where $\rho_{\rm nuc}
\approx 2.8 \times 10^{14}~{\rm g/cm^3}$~\cite{ALF2}.
We note that the piecewise polytropic EOSs are slightly different from
the original tabulated ones, because of their simple fitting formula.
This results in a small error in the mass and radius of neutron stars.
However, as shown in~\cite{rlof2009}, the error is small (at most
several percent), and the semiquantitative properties of the original
EOSs are well captured by these simple EOSs.

\begin{figure}[t]
%\epsfxsize=3.2in
%\leavevmode
%\epsffile{fig1.eps}
\includegraphics[width=90mm,clip]{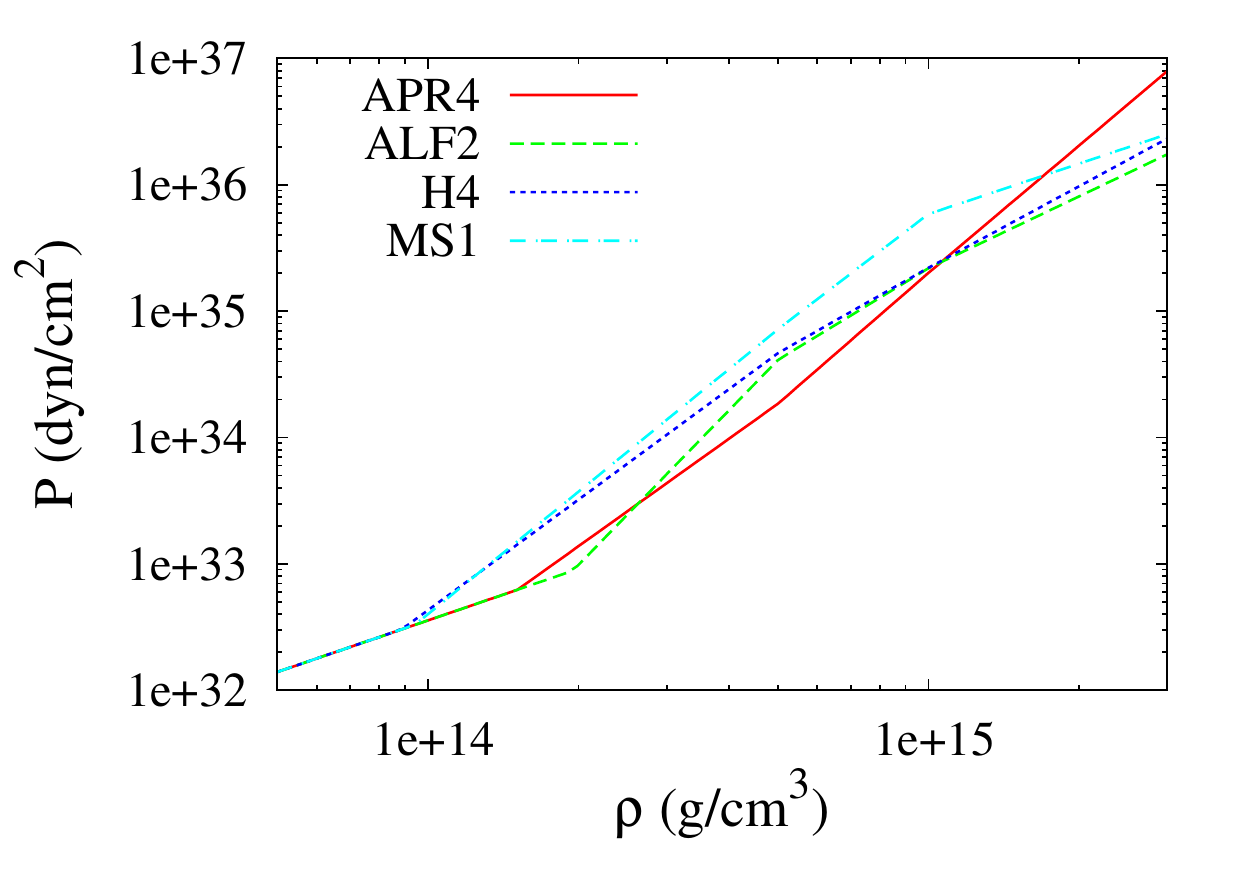}
\vspace{-7mm}
\caption{Pressure as a function of the rest-mass density for 
APR4, ALF2, H4, and MS1 EOSs (the solid, dashed, dotted, and
dash-dotted curves, respectively). }
\label{fig1}
\end{figure}

Figure~\ref{fig1} plots the pressure as a function of the rest-mass
density for four EOSs. APR4 has relatively small pressure for $\rho_1
\leq \rho \alt \rho_3$ while it has high pressure for $\rho \agt
\rho_3$. Thus, for $\rho < \rho_3$, which neutron stars of canonical
mass $1.3$ -- $1.4M_{\odot}$ have, this EOS is soft, and hence, the
value of $P_2$ is relatively small.  We note that for a small value of
$P_2$, $\Gamma_2$ and/or $\Gamma_3$ have to be large ($\sim 3$)
because the maximum mass of spherical neutron stars, $M_{\rm max}$ for
a given EOS has to be larger than $\sim 2M_{\odot}$. Thus, an EOS that
is soft at $\rho=\rho_2$ has to be in general stiff for $\rho \agt
\rho_3$. By contrast, H4 and MS1 have pressure higher than APR4 for
$\rho \alt \rho_3$, although the EOSs become softer for a high-density
region $\rho \agt \rho_3$. In particular, MS1 has extremely high
pressure (i.e., a higher value of $P_2$) among many other EOSs for
$\rho \alt \rho_3$, and thus, it is the stiffest EOS as far as the
canonical neutron stars are concerned.  ALF2 has small pressure for
$\rho \leq \rho_2$ as in the case of APR4, but for $\rho_2 \alt \rho
\leq \rho_3$, the pressure is higher than that for APR4. For $\rho
\geq \rho_2$ the pressure of ALF2 is as high as that for H4.  All the
properties mentioned above are reflected in the radius, $R_{1.35}$,
and central density, $\rho_{1.35}$, of spherical neutron stars with
the canonical mass $M=1.35M_{\odot}$ where $M$ is the gravitational
(Arnowitt-Deser-Misner; ADM) mass of the spherical neutron stars in
isolation: see Table~\ref{table:EOS}. The pressure at $\rho=\rho_2$
($P_2$) is correlated well with this radius and central density (see
below). 

Here, a word of caution is necessary for our APR4. The pressure in
this piecewise polytropic EOS is extremely (unphysically) high in the
high-density region with $\rho \agt 10^{16}~{\rm g/cm^3}$. This
results pathologically in the situation that the sound velocity
exceeds the speed of light for the high-density state. In reality,
such a high density is achieved only in the formation of a black hole
(i.e., inside the horizon), and such a pathology may not affect the
evolution of the system for the outside of the horizon. However, this
pathology could still break a numerical simulation after the formation
of a black hole. To avoid this happens, we artificially set the
maximum density as $10^{16}~{\rm g/cm^3}$ when employing this EOS.

\begin{figure*}[t]
\includegraphics[width=86mm,clip]{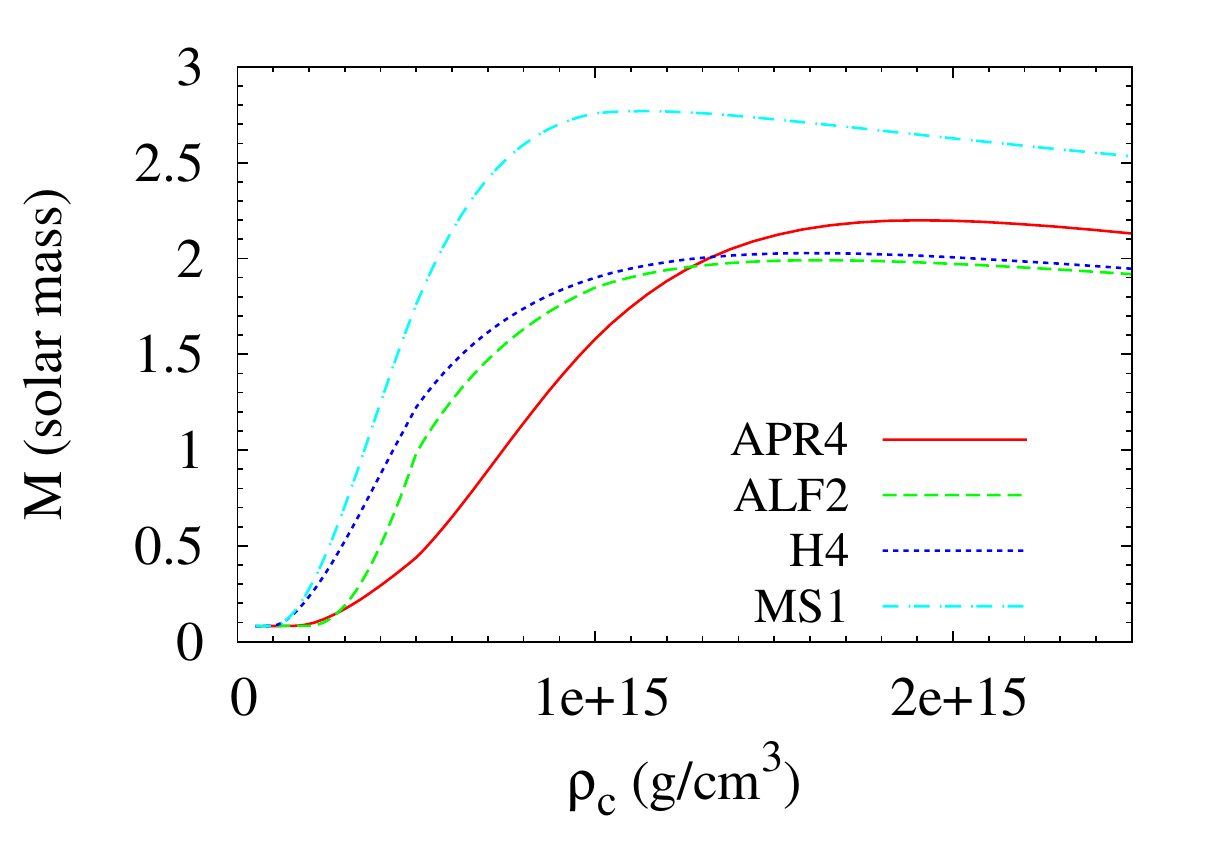}
\includegraphics[width=86mm,clip]{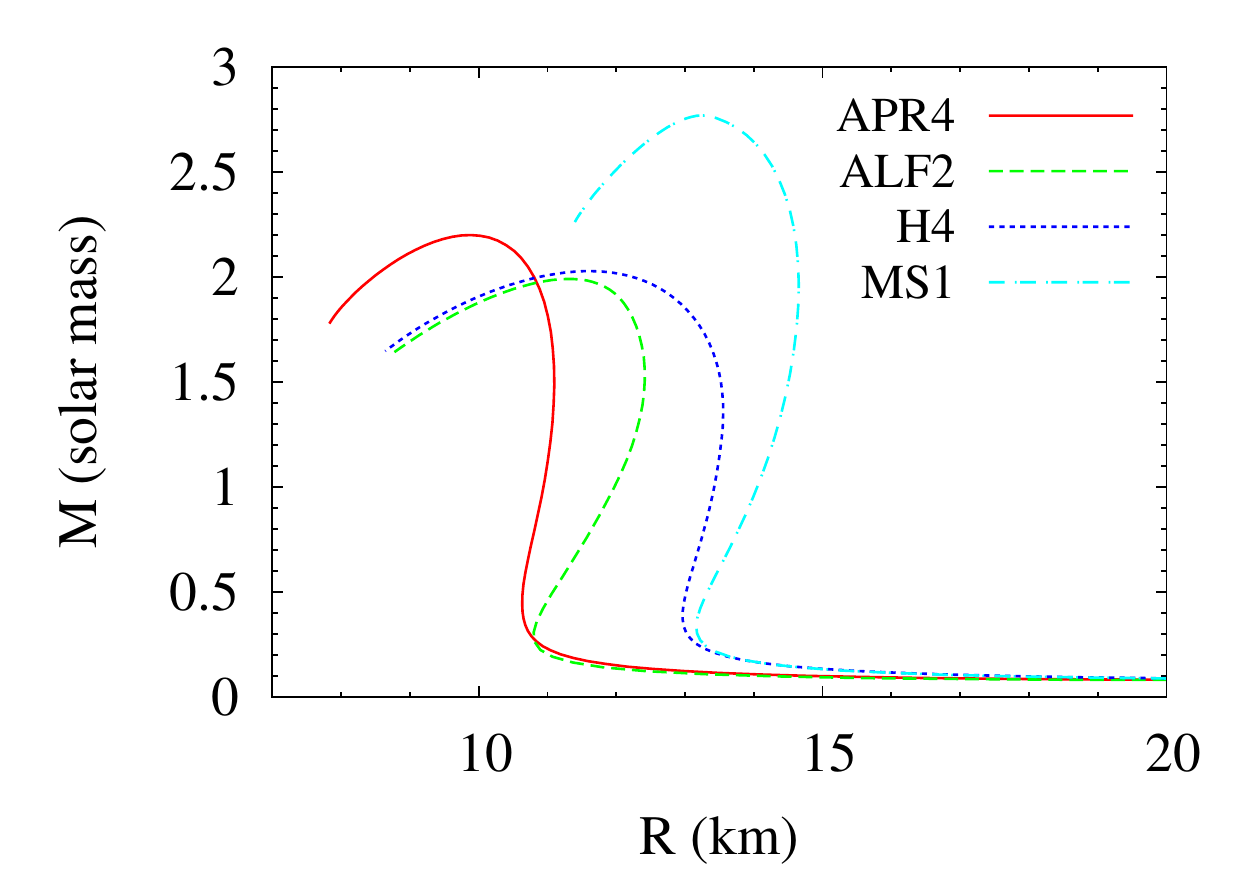}
\caption{Left: The gravitational mass as a function of the central density
$\rho_{\rm c}$ for spherical neutron stars in APR4, ALF2, H4, and MS1
EOSs (the solid, dashed, dotted, and dash-dotted curves).  Right: The
same as the left panel but for the gravitational mass as a function of
the circumferential radius.  }
\label{fig2}
\end{figure*}

Figure~\ref{fig2} plots the gravitational mass as a function of the
central density and as a function of the circumferential radius for
spherical neutron stars for four EOSs. All the EOSs chosen are stiff
enough that the maximum mass is larger than $1.97M_{\odot}$. Because
the pressure in a density region $\rho \alt 10^{15}~{\rm g/cm^3}$ is
relatively small (i.e., $P_2$ is small) for APR4 and ALF2, the radius
for these EOSs is relatively small as $\sim 11$~km and 12.5~km,
respectively, for the canonical mass of neutron stars 1.3 --
$1.4M_{\odot}$~\cite{Stairs}.  By contrast, for H4 and MS1 for which
$P_2$ is relatively large, the radius becomes a relatively large value
13.5 -- 14.5~km for the canonical mass. The radius has also the
correlation with the central density $\rho_{\rm c}$.  For APR4 and
ALF2 with $M=1.35M_{\odot}$, $\rho_{\rm c}\approx 8.9\times
10^{14}~{\rm g/cm^3}$ and $\rho_{\rm c}\approx 6.4\times 10^{14}~{\rm
g/cm^3}$.  For H4 and MS1 with $M=1.35M_{\odot}$, the central density
is rather low as $\rho_{\rm c}\approx 5.5\times 10^{14}~{\rm
g/cm^3}$ and $\rho_{\rm c}\approx 4.1\times 10^{14}~{\rm g/cm^3}$,
respectively. As we show in Sec.~\ref{sec:res}, the properties of the
material ejected from the merger of binary neutron stars depend
strongly on the radius of the neutron stars or $\rho_{\rm c}$.

\subsection{Initial conditions} \label{sec:ID}

\begin{table*}
\caption{List of the parameters of the initial condition for binaries
chosen in numerical simulations: Total mass, mass ratio, masses of
two components, initial value of angular velocity, and initial
frequency of gravitational waves ($f_0=\Omega_0/\pi$).  }
%%%%%%%%%%%%%%%%
{\begin{tabular}{c|cccccc} \hline
Model & ~$m (M_{\odot})$~ & $q$ & $m_1 (M_{\odot})$ & $m_2 (M_{\odot})$
& $m\Omega_0$ & $f_0$ (Hz)
\\ \hline \hline
APR4-130160 & 2.90 & 0.813 & 1.30 & 1.60 & 0.026 & 579 \\
APR4-140150 & 2.90 & 0.933 & 1.40 & 1.50 & 0.026 & 579 \\
APR4-145145 & 2.90 & 1.000 & 1.45 & 1.45 & 0.026 & 579 \\
APR4-130150 & 2.80 & 0.867 & 1.30 & 1.50 & 0.026 & 600 \\
%%APR4-135145 & 2.80 & 0.931 & 1.35 & 1.45 & 0.026 & 600 \\
APR4-140140 & 2.80 & 1.000 & 1.30 & 1.50 & 0.026 & 600 \\
APR4-120150 & 2.70 & 0.800 & 1.20 & 1.50 & 0.026 & 622 \\
APR4-125145 & 2.70 & 0.862 & 1.25 & 1.45 & 0.026 & 622 \\
APR4-130140 & 2.70 & 0.929 & 1.30 & 1.40 & 0.026 & 622 \\
APR4-135135 & 2.70 & 1.000 & 1.35 & 1.35 & 0.026 & 622 \\
APR4-120140 & 2.60 & 0.857 & 1.20 & 1.40 & 0.026 & 646 \\
APR4-125135 & 2.60 & 0.926 & 1.25 & 1.35 & 0.026 & 646 \\
APR4-130130 & 2.60 & 1.000 & 1.30 & 1.30 & 0.026 & 646 \\ \hline
ALF2-140140 & 2.80 & 1.000 & 1.40 & 1.40 & 0.026 & 600 \\
ALF2-120150 & 2.70 & 0.800 & 1.20 & 1.50 & 0.026 & 622 \\
ALF2-125145 & 2.70 & 0.862 & 1.25 & 1.25 & 0.026 & 622 \\
ALF2-130140 & 2.70 & 0.929 & 1.30 & 1.40 & 0.026 & 622 \\
ALF2-135135 & 2.70 & 1.000 & 1.35 & 1.35 & 0.026 & 622 \\
ALF2-130130 & 2.60 & 1.000 & 1.30 & 1.30 & 0.026 & 646 \\
\hline
H4-130150  & 2.80 & 0.867 & 1.30 & 1.50 & 0.025 & 577 \\
H4-140140  & 2.80 & 1.000 & 1.40 & 1.40 & 0.025 & 577 \\
H4-120150  & 2.70 & 0.800 & 1.20 & 1.50 & 0.025 & 598 \\
H4-125145  & 2.70 & 0.862 & 1.25 & 1.25 & 0.025 & 598 \\
H4-130140  & 2.70 & 0.929 & 1.30 & 1.40 & 0.025 & 598 \\
H4-135135  & 2.70 & 1.000 & 1.35 & 1.35 & 0.025 & 598 \\
H4-120140  & 2.60 & 1.000 & 1.30 & 1.30 & 0.025 & 621 \\
H4-125135  & 2.60 & 1.000 & 1.30 & 1.30 & 0.025 & 621 \\
H4-130130  & 2.60 & 1.000 & 1.30 & 1.30 & 0.025 & 621 \\ \hline
MS1-140140 & 2.80 & 1.000 & 1.40 & 1.40 & 0.025 & 577 \\
MS1-120150 & 2.70 & 0.800 & 1.20 & 1.50 & 0.025 & 598 \\
MS1-125145 & 2.70 & 0.862 & 1.25 & 1.25 & 0.025 & 598 \\
MS1-130140 & 2.70 & 0.929 & 1.30 & 1.40 & 0.025 & 598 \\
MS1-135135 & 2.70 & 1.000 & 1.35 & 1.35 & 0.025 & 598 \\
MS1-130130 & 2.60 & 1.000 & 1.30 & 1.30 & 0.025 & 621 \\
\hline
\hline
\end{tabular}
}
\label{table:ID}
\end{table*}

We employ binary neutron stars in quasiequilibria for the initial
condition of numerical simulations as in our series of
papers~\cite{STU,hotoke}. The quasiequilibrium state is computed in
the framework described in~\cite{TS2010} to which the reader may
refer. The computation of quasiequilibrium states is performed using
the spectral-method library LORENE~\cite{LORENE}.

Numerical simulations were performed, systematically choosing wide
ranges of the total mass and mass ratio of binary neutron stars.
Because the mass of each neutron star in the observed binary systems
is in a narrow range between $\sim 1.2$ -- $1.45M_{\odot}$
\cite{Stairs}, we basically choose the neutron-star mass $1.20$,
$1.25$, $1.30$, $1.35$, $1.40$, $1.45$, and $1.5M_{\odot}$. Also, the
mass ratio of the observed system $q:=m_1/m_2 (\leq 1)$ where $m_1$
and $m_2$ are lighter and heavier masses, respectively, is in a narrow
range $\sim 0.85$ -- 1. Thus, we choose $q$ as $0.8 \leq q \leq 1$.
Specifically, the simulations were performed for the initial data
listed in Table~\ref{table:ID}.

The initial data were prepared so that the binary has about 3 -- 4
quasicircular orbits before the onset of the merger. For four EOSs
chosen, this requirement is approximately satisfied with the choice of
the initial angular velocity $m\Omega_0=0.026$ for APR4 and ALF2 and
$m\Omega_0=0.025$ for H4 and MS1. Here, $m=m_1 + m_2$. For the
following, the model is referred to as the name ``EOS''-``$m_1$''``$m_2$'';
e.g., the model employing APR4, $m_1=1.2M_{\odot}$, and
$m_2=1.5M_{\odot}$ is referred to as model APR4-120150.

\section{Formulation and numerical methods} \label{sec3}

Numerical simulations were performed using an adaptive-mesh refinement
(AMR) code {\tt SACRA}~\cite{yst2008} (see also~\cite{bsy2010} for the
reliability of {\tt SACRA}). The formulation, the gauge conditions, and
the numerical scheme are basically the same as those described
in~\cite{yst2008}, except for the improvement in the treatment of the
hydrodynamics code for a far region. Thus, we here only briefly review 
them and describe the present setup of the computational domain for
the AMR algorithm and grid resolution. 

\subsection{Formulation and numerical methods} \label{subsec:sim_method}

{\tt SACRA} solves Einstein's evolution equations in the
Baumgarte-Shapiro-Shibata-Nakamura formalism with a moving-puncture
gauge \cite{BSSN}. It evolves a conformal factor $W := \gamma^{-1/6}$,
the conformal three-metric $\tilde{\gamma}_{ij} := \gamma^{-1/3}
\gamma_{ij}$, the trace of the extrinsic curvature $K$, a
conformally-weighted trace-free part of the extrinsic curvature
$\tilde{A}_{ij} := \gamma^{-1/3} ( K_{ij} - K \gamma_{ij}/3 )$, and an
auxiliary variable $\tilde{\Gamma}^i := - \partial_j
\tilde{\gamma}^{ij}$. Introducing an additional auxiliary variable
$B^i$ and a parameter $\eta_s$, which we typically set to be $\approx
0.8/m$ in units of $c = G = M_\odot =1$, we employ a moving-puncture
gauge in the form \cite{bghhst2008}
\begin{eqnarray}
 ( \partial_t - \beta^j \partial_j ) \alpha &=& - 2 \alpha K , \\
 ( \partial_t - \beta^j \partial_j ) \beta^i &=& (3/4) B^i , \\
 ( \partial_t - \beta^j \partial_j ) B^i &=& ( \partial_t - \beta^j
  \partial_j ) \tilde{\Gamma}^i - \eta_s B^i .
\end{eqnarray}
We evaluate the spatial derivative by a fourth-order central finite
difference except for the advection terms, which are evaluated by a
fourth-order lopsided upwind finite differencing scheme, and employ a
fourth-order Runge-Kutta method for the time integration.

To solve hydrodynamics equations, we evolve $\rho_* := \rho
\alpha u^t W^{-3}$, $\hat{u}_i := h u_i$, and $e_* := h \alpha
u^t - P / ( \rho \alpha u^t )$. Here, $u^\mu$ denotes the four
velocity of the fluid. The advection terms are handled with a
high-resolution central scheme by Kurganov and Tadmor~\cite{KT} with a
third-order piecewise parabolic interpolation for the cell
reconstruction. We note that the total rest mass of the system is
calculated by
\beqn
M_*=\int \rho_* d^3x.
\eeqn

For the EOS employed in the numerical simulation, we decompose the
pressure and specific internal energy into cold and thermal parts as
\begin{equation}
 P = P_{\rm cold} + P_{\rm th} \; , \; \varepsilon = \varepsilon_{\rm
  cold} + \varepsilon_{\rm th} .
\end{equation}
We calculate the cold parts of both variables using the piecewise
polytropic EOS (see section~\ref{sec:EOS}) from the primitive variable
$\rho$, and then the thermal part of the specific internal energy is
defined from $\varepsilon$ as $\varepsilon_{\rm th} = \varepsilon -
\varepsilon_{\rm cold}(\rho)$. Because $\varepsilon_{\rm th}$ vanishes in
the absence of shock heating, $\varepsilon_{\rm th}$ is regarded as
the finite temperature part determined by the shock heating in the
present context.  In this paper, we adopt a $\Gamma$-law ideal gas EOS
for the thermal part as
\begin{equation}
 P_{\rm th} = ( \Gamma_{\rm th} - 1 ) \rho \varepsilon_{\rm th}.
\end{equation}
Following the conclusion of a detailed study in~\cite{BJO2010},
$\Gamma_{\rm th}$ is chosen in the range 1.6 -- 2.0 with the canonical
value 1.8. For several models, we performed simulations varying the
value of $\Gamma_{\rm th}$, and explored the effects of the shock
heating; as shown in Sec.~\ref{sec:res}, numerical results depend
fairly strongly on the value of $\Gamma_{\rm th}$ (although the
dependence on $\Gamma_{\rm th}$ is not as strong as the dependence on
the EOS, $P_{\rm cold}$).

Because the vacuum is not allowed in any conservative hydrodynamics
scheme (e.g., to derive the velocity by dividing the momentum density
by the density), we put an artificial atmosphere of small density
outside the neutron stars. The atmosphere has to be as tenuous as
possible because a dense atmosphere may significantly affect the
motion of the material ejected from binary neutron stars.
Specifically, we set the density of the atmosphere in the following
simple rule
\begin{eqnarray}
 \rho_{\rm at}=\left\{
\begin{array}{ll}
f_{\rm at}~\rho_{\rm max}                     &~~(r\leq r_{\rm uni}),\\
f_{\rm at}~\rho_{\rm max}(r/r_{\rm uni})^{-n} &~~(r\geq r_{\rm uni}),\\
 \end{array}
 \right.
\end{eqnarray}
where $\rho_{\rm max}$ denotes the maximum rest-mass density of the
neutron stars at the initial state $\alt 10^{15}~{\rm g/cm^3}$ (see
Table~\ref{table:EOS}). We typically set $f_{\rm at}=10^{-10}$, $n=3$,
and $r_{\rm uni}=16L_{\rm min}$ where $2L_{\rm min}$ denotes the side
length of the finest computational domain in the AMR algorithm (see
Sec.~\ref{subsec:sim_grids} and Table~\ref{table:grid}).  For MS1, a
computational region is wider and we set $f_{\rm at}=10^{-11}$ to
reduce the atmosphere mass.  In these settings, the total rest mass of
the atmosphere is always $\sim 10^{-6} M_\odot$ or less.  In test
simulations, we also adopted $n=2$ and $f_{\rm at}=10^{-10}$ --
$10^{-12}$, and found that the numerical results on the ejected
material such as its mass and its total energy depend only weakly on
the values of $n$ and $f_{\rm at}$ (e.g., the ejected mass increases
by $\sim 10\%$ if we change $n$ from 3 to 2 (denser one) for some
models of APR4 and H4).  Hence, we could safely conclude that the
tenuous atmosphere chosen in this work does not significantly affect
the properties of the ejected material.

We extracted $l=|m|=2$ modes of gravitational waves, $h_+$ and
$h_\times$ , by calculating the outgoing part of the complex Weyl
scalar $\Psi_4$ at finite coordinate radii $r \approx 200M_{\odot}$ --
$400M_\odot$ and by integrating $\Psi_4$ twice in time as
in~\cite{kst2011}, to which the reader may refer (see
also~\cite{RP2010}). We also analyzed the evolution of
gravitational-wave frequency, which is determined by extracting the
phase of $\Psi_4$, arg$(\Psi_4)$, and by taking the time derivative as
$2\pi f:=d({\rm arg}(\Psi_4))/dt$. To find the characteristic
frequency of gravitational waves, we also define the average value of
$f$ by
%%%%%%%%%%%%%
\beqn
f_{\rm ave}:={\displaystyle \int f |h| dt \over \displaystyle \int |h|
dt},\label{eq:fave1}
\eeqn
where we used $|h|=(h_+^2+h_\times^2)^{1/2}$ as the weight
factor. Then, we define the physical dispersion of $f$ by
\beqn
\sigma_f^2:={\displaystyle \int (f-f_{\rm ave})^2 |h| dt
\over \displaystyle \int |h| dt}. \label{eq:fave2}
\eeqn
In the following, $f_{\rm ave}$ and $\sigma_f$ are calculated 
for gravitational waves emitted by the remnant massive neutron stars. 

\subsection{Analysis of the ejected material} \label{subsec3.2}

\begin{table*}
 \caption{The grid structure for the simulation in our AMR
 algorithm. $\Delta x$ is the grid spacing in the finest-resolution
 domain with $L$ being the location of the outer boundaries along each
 axis and $L_{\rm min}=N\Delta x$. $R_{\rm diam}/\Delta x$ denotes the
 numbers of grid assigned inside the semi-major diameter of the
 lighter and heavier neutron stars in the finest level.  $\lambda_0$
 is the gravitational wavelength for the initial configuration.  The
 last column shows the values of $\Gamma_{\rm th}$ employed.  }
%%%%%%%
{\begin{tabular}{c|cccccc} \hline
Model & $\Delta x$(km) & $R_{\rm diam}/\Delta x$
& $L$ (km) & $L_{\rm min}$ (km) &
$\lambda_0$ (km) & $\Gamma_{\rm th}$ \\ \hline \hline
APR4-130160 & 0.172 & 102, ~96 & 2636 & 10.3 & 518 & 1.8\\
APR4-140150 & 0.167 & 102, 101 & 2572 & 10.0 & 518 & 1.8\\
APR4-145145 & 0.166 & 102, 102 & 2550 & 10.0 & 518 & 1.8\\
APR4-130150 & 0.172 & 102, ~98 & 2636 & 10.3 & 500 & 1.8\\
APR4-140140 & 0.167 & 102, 102 & 2572 & 10.0 & 500 & 1.8\\
APR4-120150 & 0.172 & 103, ~98 & 2644 & 10.3 & 482 & 1.6, 1.8, 2.0\\
APR4-125145 & 0.174 & 102, 100 & 2665 & 10.4 & 482 & 1.8 \\
APR4-130140 & 0.170 & 103, 101 & 2609 & 10.2 & 482 & 1.8 \\
APR4-135135 & 0.169 & 102, 102 & 2601 & 10.2 & 482 & 1.6, 1.8, 2.0 \\
APR4-120140 & 0.174 & 102, ~99 & 2679 & 10.5 & 464 & 1.8 \\
APR4-125135 & 0.174 & 102, 100 & 2665 & 10.4 & 464 & 1.8 \\
APR4-130130 & 0.171 & 102, 102 & 2629 & 10.3 & 464 & 1.8 \\ \hline
ALF2-140140 & 0.195 & 102, 102 & 3001 & 11.7 & 500 & 1.8 \\
ALF2-120150 & 0.200 & 102, ~98 & 3065 & 12.0 & 482 & 1.8 \\
ALF2-125145 & 0.199 & 102, 100 & 3054 & 11.9 & 482 & 1.8 \\
ALF2-130140 & 0.198 & 102, 101 & 3044 & 11.9 & 482 & 1.8 \\
ALF2-135135 & 0.195 & 103, 103 & 3001 & 11.7 & 482 & 1.8 \\
ALF2-130130 & 0.199 & 102, 102 & 3054 & 11.9 & 464 & 1.8 \\
\hline
H4-130150  & 0.222 & 102, ~98 & 3429 & 13.4 & 480 & 1.8 \\
H4-140140  & 0.219 & 102, 102 & 3358 & 13.1 & 480 & 1.8 \\
H4-120150  & 0.228 & 102, ~96 & 3501 & 13.7 & 463 & 1.6, 1.8, 2.0\\
H4-125145  & 0.226 & 102, ~98 & 3465 & 13.5 & 463 & 1.8 \\
H4-130140  & 0.223 & 102, 100 & 3430 & 13.4 & 463 & 1.8 \\
H4-135135  & 0 221 & 102, 102 & 3393 & 13.3 & 463 & 1.6, 1.8, 2.0 \\
H4-120140  & 0.230 & 101, ~98 & 3537 & 13.8 & 446 & 1.8 \\
H4-125135  & 0.227 & 102, 100 & 3494 & 13.6 & 446 & 1.8 \\
H4-130130  & 0.223 & 103, 103 & 3430 & 13.4 & 446 & 1.8 \\ \hline
MS1-140140 & 0.237 & 103, 103 & 3644 & 14.2 & 480 & 1.8 \\
MS1-120150 & 0.249 & 101, ~97 & 3823 & 14.9 & 463 & 1.8 \\
MS1-125145 & 0.244 & 102, ~99 & 3751 & 14.7 & 463 & 1.8 \\
MS1-130140 & 0.244 & 101, 100 & 3751 & 14.7 & 463 & 1.8 \\
MS1-135135 & 0.242 & 102, 102 & 3715 & 14.5 & 463 & 1.8 \\
MS1-130130 & 0.244 & 102, 102 & 3751 & 14.7 & 446 & 1.8 \\
\hline
\end{tabular}
}
\label{table:grid}
\end{table*}

In this section, we describe the method for analyzing the material
ejected from the merger of binary neutron stars.  Here, the ejected
material is composed of a fluid element which is unbound by the
gravitational potential of binary neutron stars and an object formed
after the merger.  Thus, first of all, we have to determine which
fluid elements are unbound. To assess this point for all the fluid
elements, we calculate $u_\mu t^\mu=u_t$ at each grid point.  Here,
$t^\mu$ is a timelike vector $(1,0,0,0)$ which is a Killing vector at
spatial infinity.  If $|u_t| > 1$, we consider that the fluid element
there is unbound.

Then we calculate the total rest mass, total energy (excluding
gravitational potential energy), and total internal energy of the
fluid element of $|u_t| > 1$ by
\beqn
M_{*{\rm esc}} &=& \int_{|u_t| > 1} \rho_* d^3x,\\
E_{{\rm tot,esc}} &=&
\int_{|u_t| > 1} T_{\mu\nu}n^{\mu}n^{\nu} \sqrt{\gamma}d^3x \nonumber \\
&=& \int_{|u_t| > 1} \rho_* e_* d^3x,\\
U_{\rm esc}&=&\int_{|u_t| > 1} \rho_* \varep d^3x,
\eeqn
where $T_{\mu\nu}$ is the stress-energy tensor, 
\beqn
T_{\mu\nu}=\rho h u_{\mu}u_{\nu} + P g_{\mu\nu},
\eeqn
and $n^{\mu}$ is the unit timelike hypersurface normal.  We note that
the total energy is not uniquely defined by $E_{{\rm tot,esc}}$ for
dynamical spacetimes, and thus, the total energy defined here should
be considered as an approximate measure for it. We here choose this
expression for simplicity. We then define kinetic energy approximately
by
\beqn
T_{*\rm esc}:=E_{\rm tot,esc}-M_{*{\rm esc}}-U_{\rm esc}.
\eeqn
We found irrespective of models that $T_{*\rm esc}$ is much (by
about 1 -- 2 orders of magnitude) larger than $U_{\rm esc}$. 
%% and will focus only on $T_{*\rm esc}$ in the following.

To approximately analyze the configuration of the ejected material,
we also calculate the moments of inertia defined by
\beqn
I_{ii,{\rm esc}} &=& \int_{|u_t| > 1} \rho_* (x^i)^2d^3x,~~~~
({\rm no~sum~for}~i), 
\eeqn
and then, define
\beqn
\bar X =\sqrt{{I_{xx,{\rm esc}} \over M_{*{\rm esc}}}},~~
\bar Y =\sqrt{{I_{yy,{\rm esc}} \over M_{*{\rm esc}}}},~~
\bar Z =\sqrt{{I_{zz,{\rm esc}} \over M_{*{\rm esc}}}},\nonumber \\
\eeqn
and $\bar R=\sqrt{\bar X^2 + \bar Y^2}$. From $d\bar R/dt$ and $d\bar
Z/dt$, we can determine the typical (average) velocity of the ejected
material, which is denoted by $\bar V^R_{\rm esc}$ and $\bar V^Z_{\rm
esc}$ in the following.

We consider a model that the configuration of the ejected material is
approximated by an axisymmetric anisotropic shell of uniform density as
\beqn
\rho=\left\{
\begin{array}{ll}
\rho_{\rm esc} &~~
\pi/2 - \theta_0 \leq \theta \leq \pi/2 + \theta_0 \\
&~~~~~ {\rm and}~R_- \leq r \leq R_+,\\
0 &~~ {\rm otherwise},
\end{array}
\right.
\eeqn
where $\rho_{\rm esc}$, $R_{\pm}$, and $\theta_0$ are
time-varying parameters. In this case,
\beqn
M_{*{\rm esc}}&=&{4\pi \over 3}\rho_{\rm esc}(R_+^3 - R_-^3) \sin\theta_0, \\
\bar R^2&=&{1 \over 5} {R_+^5 - R_-^5 \over R_+^3 - R_-^3}(3-\sin^2\theta_0),\\
\bar Z^2&=&{1 \over 5} {R_+^5 - R_-^5 \over R_+^3 - R_-^3}\sin^2\theta_0.
\eeqn
Thus for an axial ratio,
\beqn
\eta_R = {\bar Z \over \bar R},
\eeqn
$\sin\theta_0$ is calculated as
\beqn
\sin^2\theta_0 = {3 \eta_R^2 \over 1+ \eta_R^2}.
\eeqn
Hence, from the axial ratio calculated for a numerical result of
the ejected material, we can approximately define the extent in the
$\theta$ direction; e.g., for $\eta_R=0.4$ and 0.5,
$\theta_0 \approx 40^\circ$ and $51^\circ$, respectively.

\subsection{Setup of AMR grids} \label{subsec:sim_grids}

An AMR algorithm implemented in {\tt SACRA} can prepare a
fine-resolution domain in the vicinity of compact objects as well as a
sufficiently wide domain that covers a local wave zone.  In the
present study, we prepare additional domains wider than those used in
our previous studies \cite{kst2010,kst2011,hotoke}, to follow the
motion of the material ejected during the merger of binary neutron
stars for a sufficiently long time (longer than 10~ms).

The chosen AMR grids consist of a number of computational domains,
each of which has the uniform, vertex-centered Cartesian grids with
$(2N+1, 2N+1, N+1)$ grid points for $(x,y,z)$ with the equatorial
plane symmetry at $z=0$. Since we chose that the grid spacing for
three directions is identical, the shape of each AMR domain is a half
cube.  We chose $N=60$ for the best resolved runs in this work, and
all the results shown in the following were obtained with this
resolution. We also performed simulations with $N=40$ and 50 (or 48)
for several chosen models to check the convergence of the results (see
Appendix A).

%%The rest mass and kinetic energy obtained for the ejected
%%material converge within $\sim 10$ -- $20\%$ fluctuation for the
%%unequal-mass case for $N \geq 50$. For the equal-mass case, the
%%convergence is poor, but the magnitude of the fluctuation is still in
%%an acceptable level for the present purpose.

We classify the domains of the AMR algorithm into two categories: one
is a coarser domain, which covers a wide region including both neutron
stars with its origin fixed at the approximate center of mass
throughout the simulation. The other is a finer domain, two sets of
which comove with two neutron stars and cover the region in their
vicinity. We denote the side length of the largest domain, number of
the coarser domains, and number of the finer domains by $2L$, $l_c$,
and $2 l_f$, respectively. In this work, $l_c=5$ and $l_f=4$ (in
total, 13).  The grid spacing for each domain is $h_l = L / (2^l N)$,
where $l=0$ -- $l_{\rm max}(=l_c + l_f - 1)$ is the depth of each
domain.  In the following, we denote $L/2^{l_{\rm max}}$ by $L_{\rm
min}$ and $h_{l_{\rm max}}$ by $\Delta x$.

Table~\ref{table:grid} summarizes the parameters of the grid structure
for the simulations.  For all the simulations, $L$ is set to be $L/c
\agt 10$~ms. This implies that the material cannot escape from the
computational domain in $\sim 10$~ms after the onset of the merger,
even if it could move with the speed of light. In reality, the speed
of most of the ejected material is smaller than $\sim 0.5c$, and thus,
the material stays in the second coarsest level for more than 10~ms.
$L$ is also much larger than the gravitational wavelengths at the
initial instant $\lambda_0 := \pi /\Omega_0$. This implies that a
spurious effect caused by outer boundaries when extracting
gravitational waves is excluded in the present work more efficiently
than in the previous works.  The semi-major diameter of each neutron
star is covered approximately by 100 grid points for $N=60$.

\section{Numerical results} \label{sec:res}

Table~\ref{table:result} summarizes the remnant formed, the rest mass
and kinetic energy of the ejected material measured at 10~ms after the
onset of the merger $t=t_{\rm merge}$, and the characteristic (average)
frequency of gravitational waves emitted by the hypermassive neutron
star (HMNS) for $N=60$ \footnote{For MS1, for which the maximum mass
of spherical neutron stars is quite large (see Table~\ref{table:EOS}),
the remnant neutron stars are not hypermassive nor supramassive for $m
\leq 2.8M_{\odot}$ (see Refs.~\cite{BSS00} and \cite{CST92} for the
definition of the hypermassive and supramassive neutron stars,
respectively). We should call the remnant neutron star for this EOS
normal massive neutron star (MNS). However, in this paper, we do not
distinguish MNS from HMNS for simplicity.}.  Here, $t_{\rm merge}$ is
chosen to be the time at which the amount of the rest mass of the
ejected material steeply increases.  In the following two subsections,
we summarize the results for the formation of HMNSs and black holes
separately.

\subsection{Properties of the merger and mass ejection: HMNS case}
\label{sec4.1}

Binary neutron stars in quasicircular orbits evolve due to the 
gravitational-wave emission. Their orbital separation decreases
gradually, and eventually, the merger sets in. Previous studies (e.g.,
\cite{hotoke}) clarified that soon after the onset of the merger, either
a long-lived HMNS or a black hole is formed.  For most of the
simulations in this paper performed with stiff EOSs and with the
canonical total mass $2.6$ -- $2.8M_{\odot}$, we found that a
long-lived HMNS is formed with its lifetime much longer than its
dynamical timescale $\sim 0.1$~ms and its rotation period $\sim 1$~ms;
the lifetime is longer than 10~ms for most of the models employed in
this paper. In this section, we pay attention to the case that such a
HMNS is formed.

%% THIS PARAGRAPH SHOULD BE LATER
%% Snapshots for APR1.2-1.5, 1.35-1.35, H4_1.2-1.5, H41.35-1.35

\begin{figure*}[p]
\begin{tabular}{c}
\includegraphics[bb=0 0 960 360,width=180mm]{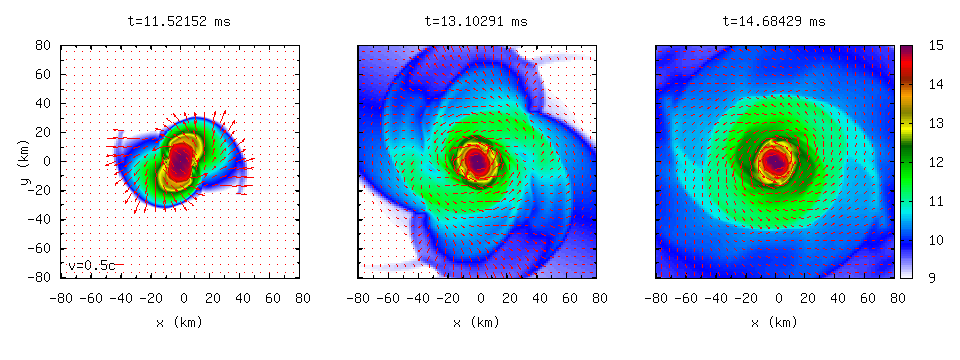} \\
\includegraphics[bb=0 -35 960 325,width=180mm]{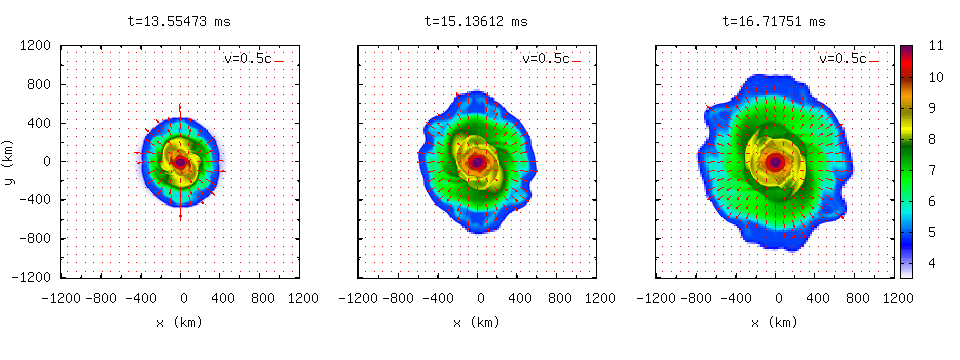}\\
\includegraphics[bb=0 -50 960 142,width=180mm]{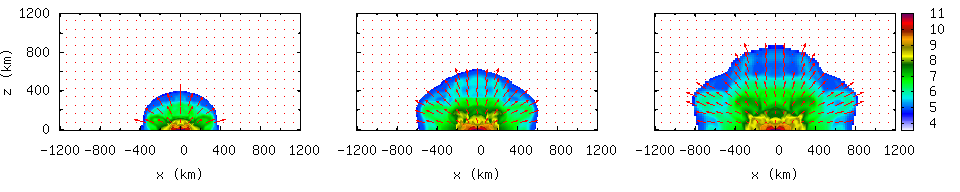} \\
\includegraphics[bb=0 -50 960 142,width=180mm]{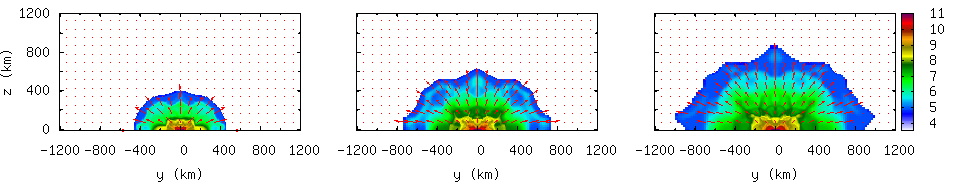}
\end{tabular}
\caption{Snapshots of the density profile for the merger of 
binary neutron stars for an equal-mass model APR4-135135. The first
row shows the density profiles in the equatorial plane and in the
central region, and second -- fourth ones show the density profile for
a wide region in the $x$-$y$, $x$-$z$, and $y$-$z$ planes. $t_{\rm
merge} \approx 11.3$~ms for this model.} 
\label{fig3A}
\end{figure*}

\begin{figure*}[p]
\begin{tabular}{c}
\includegraphics[bb=0 0 960 360,width=180mm]{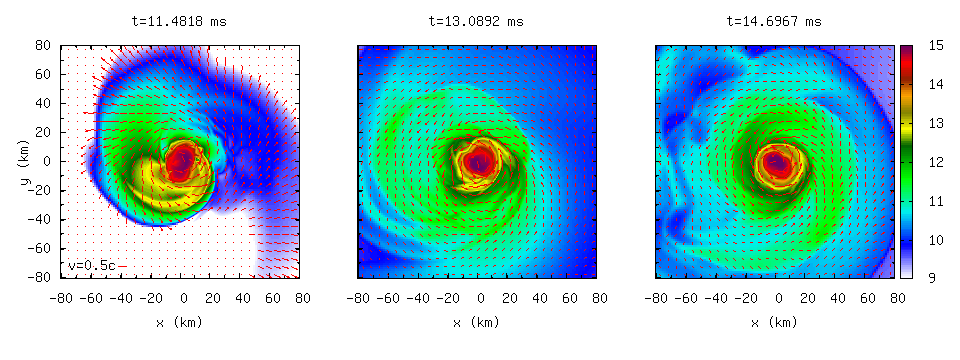} \\
\includegraphics[bb=0 -35 960 325,width=180mm]{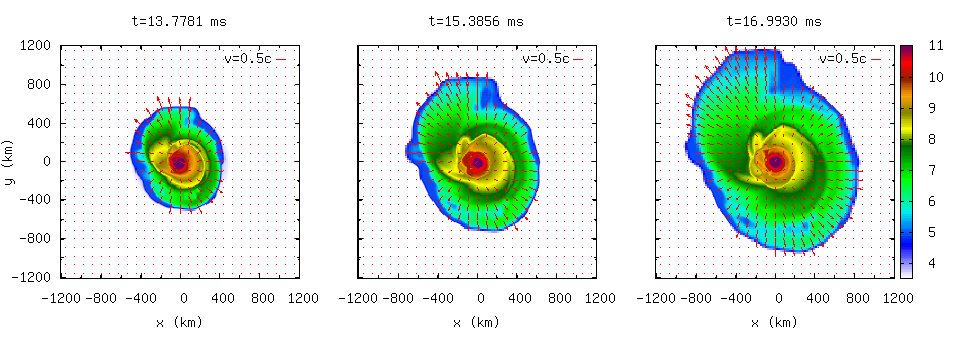}\\
\includegraphics[bb=0 -50 960 142,width=180mm]{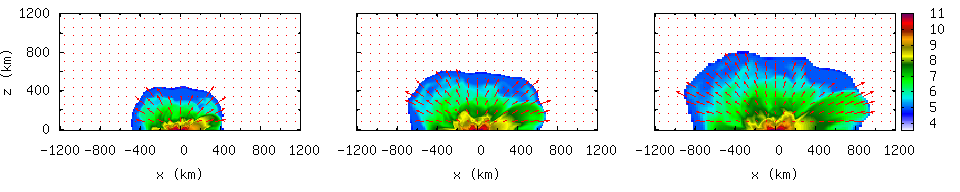} \\
\includegraphics[bb=0 -50 960 142,width=180mm]{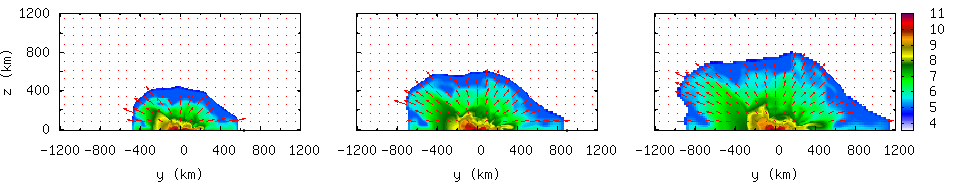}
\end{tabular}
\caption{The same as Fig.~\ref{fig3A}, but for unequal-mass model APR4-120150.
 $t_{\rm merge} \approx
10.3$~ms for this model.}
\label{fig3B}
\end{figure*}

\begin{figure*}[p]
\begin{tabular}{c}
\includegraphics[bb=0 0 960 360,width=180mm]{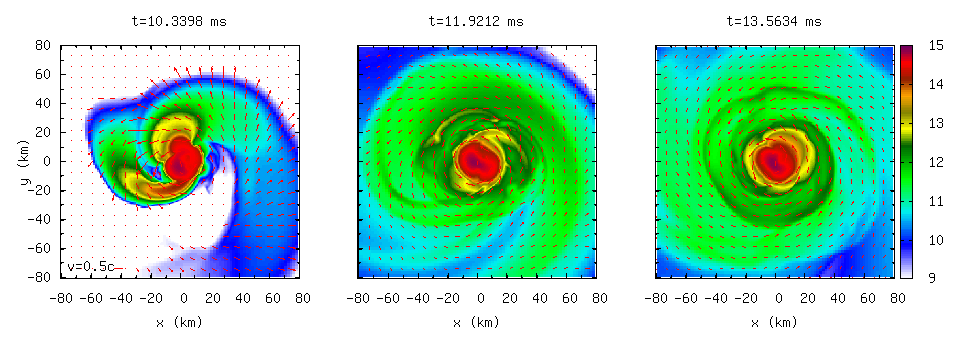} \\
\includegraphics[bb=0 -35 960 325,width=180mm]{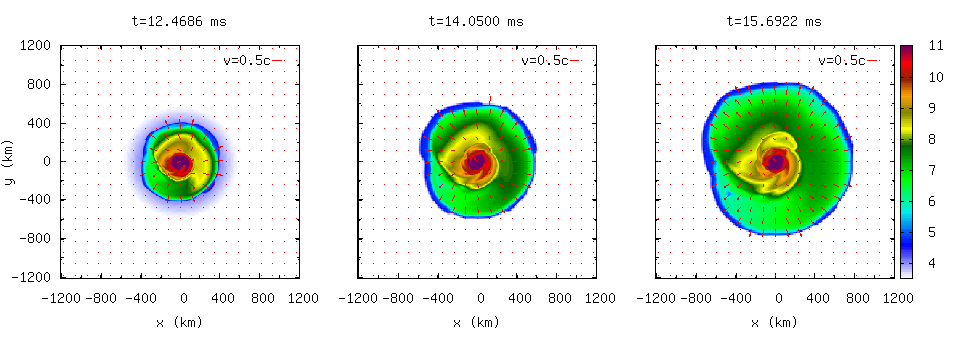}\\
\includegraphics[bb=0 -50 960 142,width=180mm]{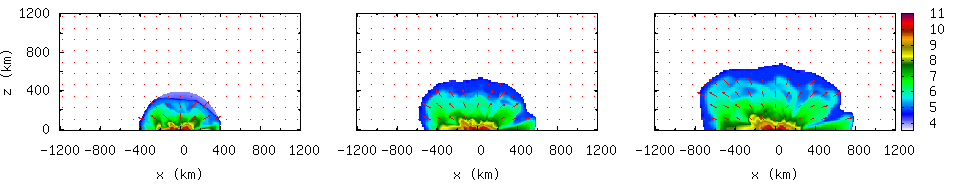} \\
\includegraphics[bb=0 -50 960 142,width=180mm]{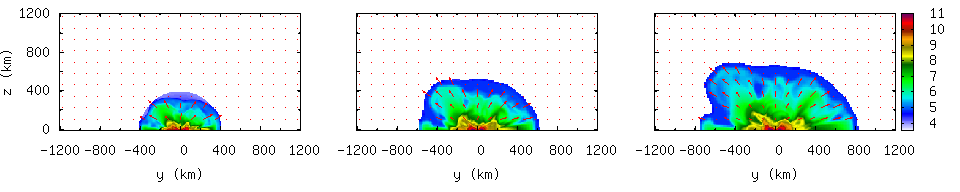}
\end{tabular}
\caption{The same as Fig.~\ref{fig3B} but for models H4-120150. 
 $t_{\rm merge} \approx
8.8$~ms for this model.}
\label{fig3C}
\end{figure*}

\begin{figure*}[t]
\begin{tabular}{cc}
\includegraphics[width=85mm,clip]{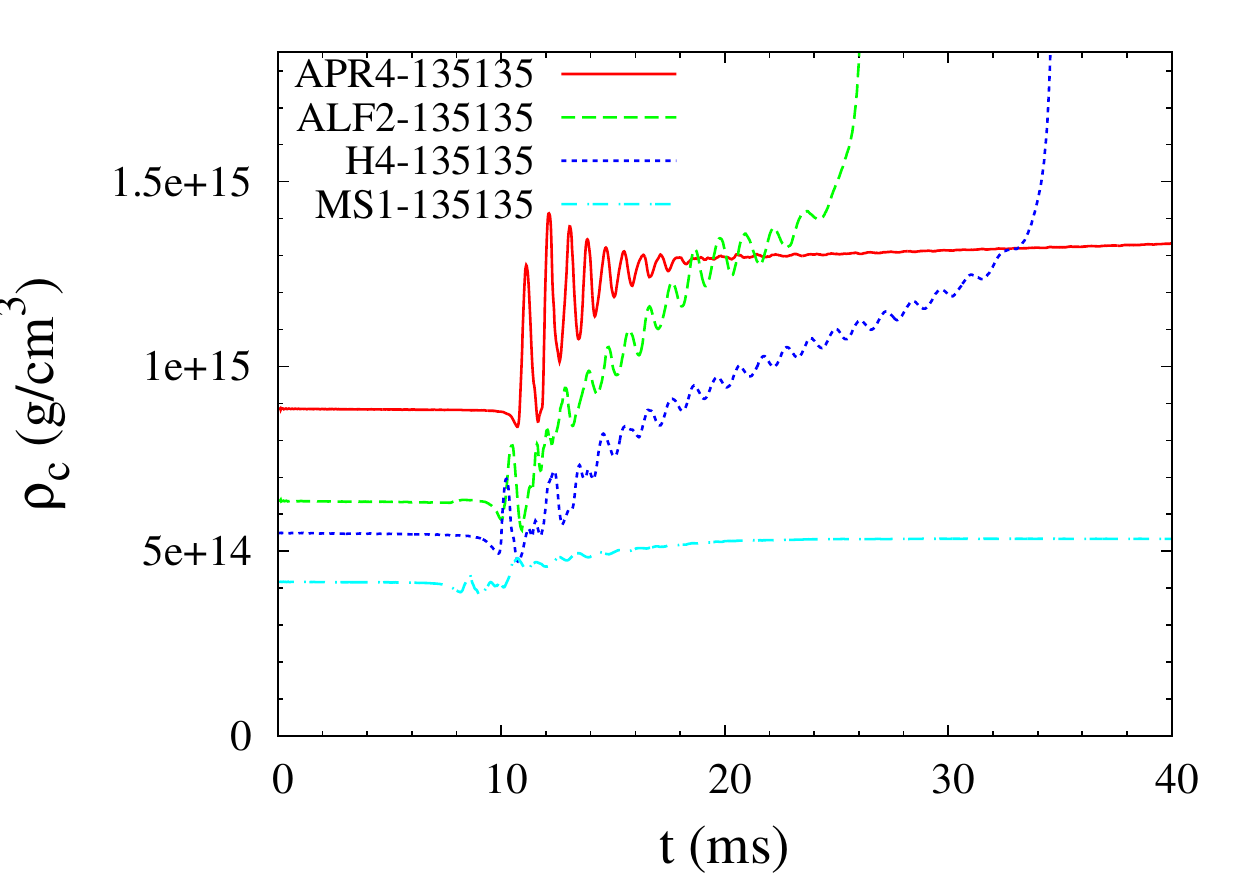}
\includegraphics[width=85mm,clip]{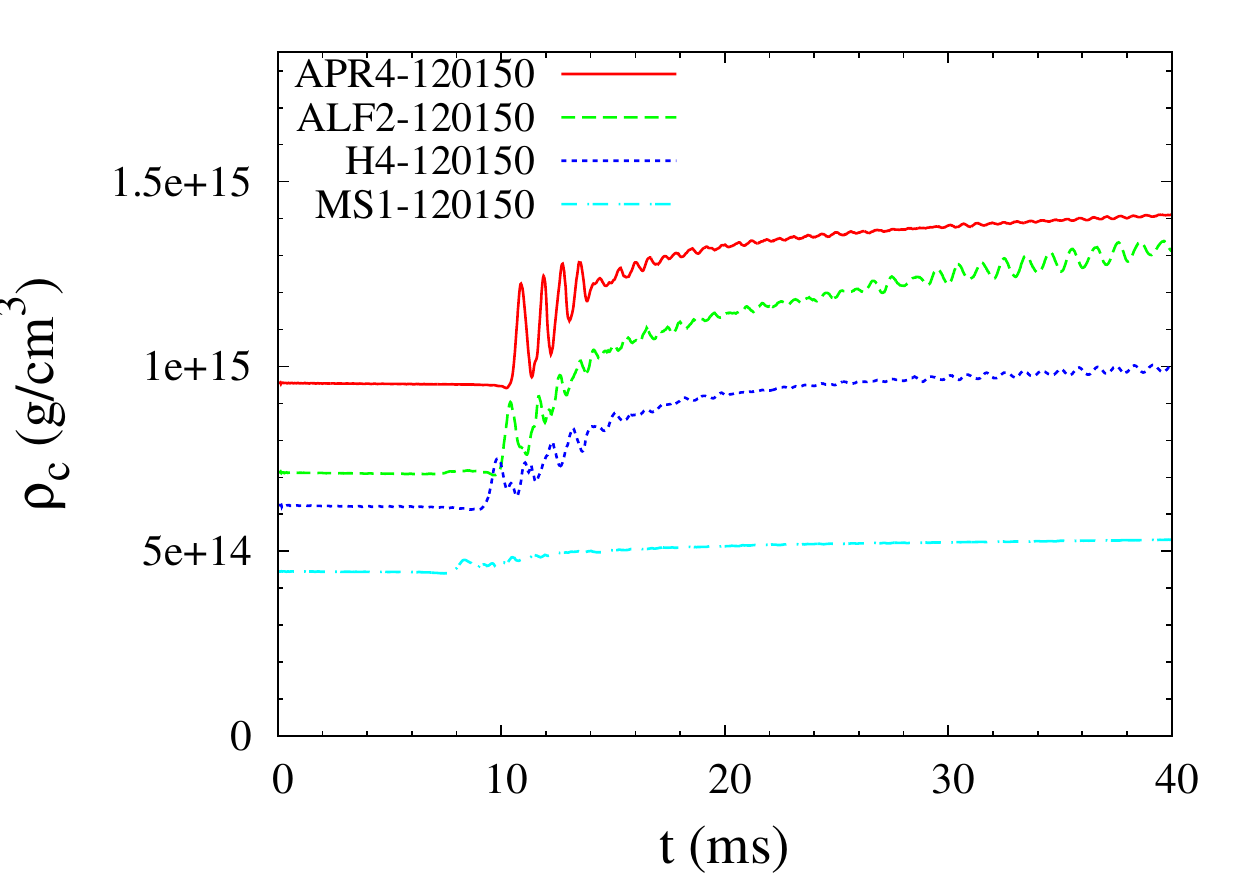}
\end{tabular}
\caption{The central density as a function of time for models with
$m_1=m_2=1.35M_{\odot}$ (left), and $m_1=1.2M_{\odot}$ and
$m_2=1.5M_{\odot}$ (right). Before the merger of unequal mass binaries,
the central density of heavier neutron stars are plotted. $\Gamma_{\rm
th}=1.8$ is employed for the results presented here. }
\label{figrho}
\end{figure*}

Figures~\ref{fig3A} -- \ref{fig3C} display snapshots of the density
profiles in the merger for models APR4-135135, APR4-120150, and
H4-120150, respectively.  Figure~\ref{figrho} also displays the central
density as a function of time for the models with
$m_1=m_2=1.35M_{\odot}$ (left), and $m_1=1.2M_{\odot}$ and
$m_2=1.5M_{\odot}$ (right). These figures show that a compact and
nonaxisymmetric object (proto HMNS) is formed in the central region soon
after the onset of the merger. The shape and compactness of the HMNS
depend strongly on the EOS and mass ratio; e.g., the presence of the
asymmetric spiral arms found in the top panels of Figs.~\ref{fig3B} and
\ref{fig3C} is the feature only for the asymmetric binaries; the
amplitude of the quasiradial oscillation is larger for the equal-mass
binaries; a high-amplitude quasiradial oscillation is a unique property
found only for models with APR4 (see Fig.~\ref{figrho}). However, it is
universal that the HMNSs are rapidly rotating and nonaxisymmetric,
irrespective of the EOS, total mass ($m \leq 2.8M_{\odot}$), and mass
ratio, as found in previous studies \cite{SU00,STU,hotoke}.  This rapid
rotation together with the nonaxisymmetric configuration not only
results in the emission of strong gravitational waves but also is the key for
an efficient mechanism of angular momentum transport from the HMNS to
the surrounding material because the HMNS exerts the torque.

\begin{figure*}[t]
\begin{tabular}{c}
\includegraphics[bb=0 0 960 360,width=180mm]{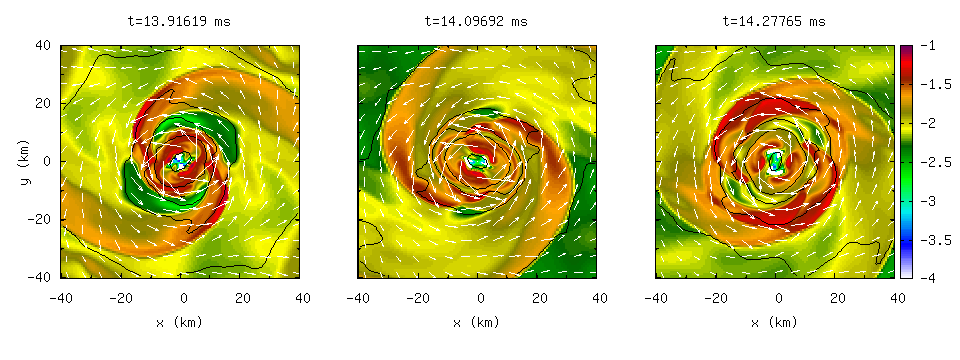} \\
\includegraphics[bb=0 0 960 215,width=180mm]{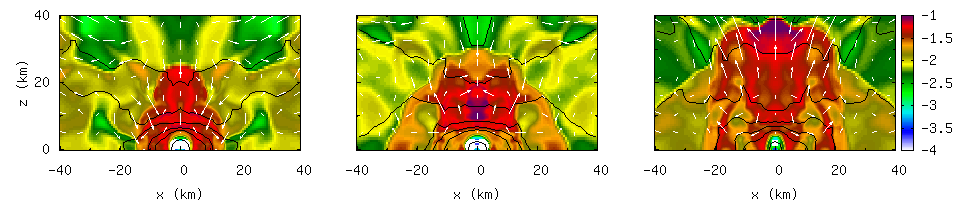}
\end{tabular}
\caption{Snapshots of the thermal part of the specific internal energy
 ($\varepsilon_{\rm th}$) profile in the vicinity of HMNSs on the
 equatorial (top) and $x$-$z$ (bottom) planes for an equal-mass model
 APR4-135135. The rest-mass density contours are overplotted for every
 decade from $10^{15}~{\rm g/cm^3}$.}
\label{figeth}
\end{figure*}

\begin{figure*}[t]
\begin{tabular}{c}
\includegraphics[bb=0 0 960 360,width=180mm]{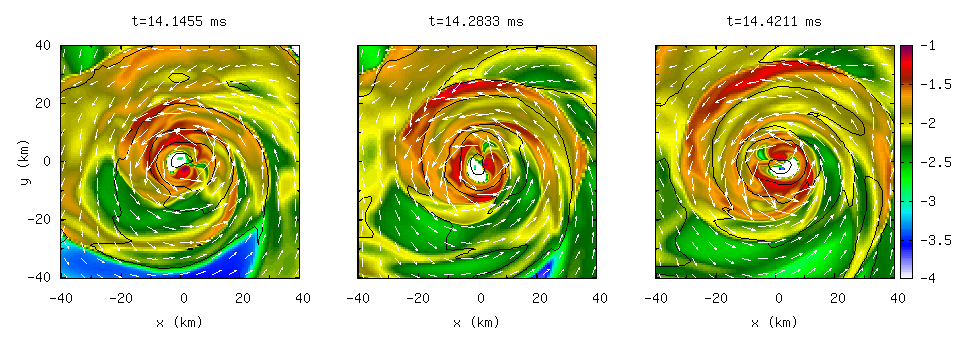} \\
\includegraphics[bb=0 0 960 215,width=180mm]{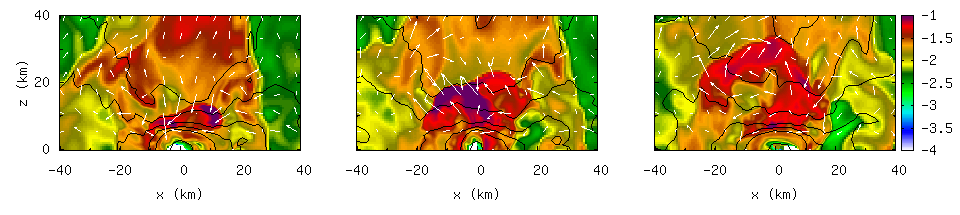}
\end{tabular}
\caption{The same as Fig.~\ref{figeth2}, but for an unequal-mass model
 APR4-120150.}
\label{figeth2}
\end{figure*}

Figures~\ref{fig3A} -- \ref{fig3C} indicate that there are two
important processes for the mass ejection. The first one is the
heating by shocks formed at the onset of the merger between the inner
surfaces of two neutron stars. Figures~\ref{figeth} and \ref{figeth2}
display snapshots of the thermal part of the specific internal energy,
$\varepsilon_{\rm th}$, in the vicinity of HMNSs for APR4-135135 and
APR4-120150, respectively. These figures show clearly that hot
materials with $\varepsilon_{\rm th} \alt 0.1$ ($\alt$~100MeV) are indeed ejected from the
HMNSs, in particular, to bidirectional regions on the equatorial plane
and to the polar region. This suggests that the shock heating works
efficiently to eject materials from the HMNSs. This occurs in an
outstanding manner in particular for the equal-mass (and only slightly
asymmetric) binaries. The heated-up material is pushed outwards by the
thermal pressure generated by the shock approximately in the plane
parallel to the (rotating) shock surface. Subsequently, it expands
outwards with rotation, and eventually forms hot spiral arms around the
HMNS.
This component subsequently gains angular momentum (and hence kinetic
energy) due to the torque exerted by the HMNS of a nonaxisymmetric
configuration, and a fraction of the material eventually gains the
kinetic energy that is large enough for it to escape from the
system. This effect plays a primary role for the early mass ejection
that occurs in the first a few ms after the onset of the merger.

A stronger shock appears to play basically a positive role for
increasing the amount of the ejected material, because the amount of the
heated-up material can be more, and as a result, the materials in the
spiral arm and ejected fraction increase. A stronger shock is formed for
softer EOSs or for binaries composed of more compact neutron stars
(e.g., APR4 in the context of canonical-mass neutron stars). The reason
is that neutron stars for such an EOS can achieve a more compact state
(cf. Fig.~\ref{figrho}) and at the merger, the collision velocity of two
neutron stars is larger (the minimum separation between two stars is
smaller).  This point will be in more detail described in
Sec.~\ref{sec:EOSdep}.  A strong shock could be also formed for binaries
with the total mass close to the critical value for the collapse to a
black hole even for stiff EOSs, because a highly compressed state is
realized by the strong gravity.

The shocks are also formed continuously in the outer part of the HMNS
during its evolution through the interaction with spiral arms formed
in its envelope due to a torque exerted by the HMNS (see below).  This
effect plays an important role in a relatively longer-term mass
ejection with the duration $\sim 10$ -- 20~ms.

The secondly important process for the mass ejection is a hydrodynamic
interaction induced by the HMNS of a nonaxisymmetric configuration
that exerts the torque to the surrounding material and transports the
angular momentum outwards. Since it is rapidly rotating, the HMNS
works as an efficient torque supplier.  Our simulations show that this
process is important in particular in the early phase of the merger:
For the nearly equal-mass binaries, a fraction of the material that
spreads outwards by the shock formed at the merger subsequently gains
angular momentum from the HMNS and eventually obtains kinetic energy
large enough to escape from the system; for sufficiently asymmetric
binaries (for small values of $q$), a less-massive neutron star is
tidally elongated during the early phase of the merger, a fraction of
its material forms spiral arms, and it subsequently gains angular
momentum from the HMNS enough to escape from the system.  In the early
mass ejection caused by the torque exerted by the HMNS, the material
is primarily ejected in the direction near the equatorial plane, and
the typical velocity of the escaping material in this early stage is
quite high $\sim 0.5$ -- $0.8c$ (follow the locations of the head of
the ejected materials in Figs.~\ref{fig3A} -- \ref{fig3C}). The
maximum velocity is larger for the EOS that yields smaller-radius
neutron stars; for APR4, it is $\sim 0.8c$ and for MS1, it is $\sim
0.5c$. This also depends on the mass ratio for models with a large
neutron-star radius (for models of H4 and MS1).

In the later phase, the mass ejection appears to occur by the
combination of the shock heating and by the torque exerted by the
HMNS. As mentioned already, the continuous shock heating occurs in the
envelope of the HMNS in the presence of spiral arms. Due to this, a
fraction of the material gains large kinetic energy. In addition, the
material in the outer region gains angular momentum by the torque
exerted by the HMNS.  These two effects give a fraction of the
material the escape velocity. By this process, the material is
gradually ejected from the system in a quasispherical manner; the
anisotropy of the configuration of the ejected material is not as
large as that of the material ejected in the early stage. This
indicates that the shock heating plays a relatively important role.
The average velocity of the escaping material in this process is
sub-relativistic $\sim 0.15$ -- $0.25c$ (see
Table~\ref{table:result}).

In the mass ejection process, these two nonlinearly coupled effects
(shock heating and torque exerted by the HMNS) play a substantial
role.  As a result, the amount of the ejected material depends on the
EOS, the total mass of the system, and the mass ratio in a nonlinear
manner.  Thus, a small change (associated, e.g., with the grid
resolution, the initial orbital separation, configuration of the
atmosphere, and presence or absence of the $\pi$ symmetry for
equal-mass binaries) results in the change in the rest mass and
kinetic energy of the ejected material; this fluctuation is in general
small, $\sim 10$ -- 20\%, for unequal-mass binaries for which the
torque plays a primary role (see Appendix A). For the equal-mass case,
the convergence is poor because a strong shock often occurs at the
merger and plays a primary role in the mass ejection. The possible
reason for this poor convergence is that shocks are always computed by
the first-order accuracy in the spatial grid resolution, and hence,
the accuracy is low and in addition, the ejected mass is a tiny part
of the entire system. A random error for the entire system computed
with a low accuracy significantly (and randomly) affects a tiny part
(i.e., the ejected material), resulting in the poor convergence.  (We
note that for global quantities, the convergence is usually good.)
For some models (such as ALF2-135135 and MS1-135135), the ejected mass
increases steeply with the grid resolution, and for such cases, the
results in this paper might give the lower bound. 

In the following subsections, we describe the properties of the
ejected material in more detail. 

%%%%%%%%%%%%%%%%%%%%%%%%%%%%%%%%%%%%%%%%%%%%%%%%%%%%%%%%%%%%%%%%%%%%

\begin{table*}
\caption{Summary of numerical results. The remnant, the total rest
mass, $M_{*{\rm esc}}$, the kinetic energy, $T_{*{\rm esc}}$, the $R$
and $Z$ components of the average velocity of escaping material, $\bar
V^R_{\rm esc}$ and $\bar V^Z_{\rm esc}$, of the ejected material, and
characteristic frequencies of gravitational waves emitted by HMNSs for
5 and 10 ms time integration after the formation of the HMNSs.  The
total rest mass, kinetic energy, and average velocity are measured at
$\approx 10$~ms after the onset of the merger.  The dispersion of
$f_{\rm ave}$ shown here is $\sigma_f$. BH denotes black hole. The
remnant is judged at $\approx 30$ ms after the onset of the merger.
All the results shown are those in the run with $N=60$ and our
standard setting of atmosphere. The rest mass and kinetic energy of
the ejected material have the uncertainty of order $10$\%. The
approximate lifetime of HMNSs for APR4-130150, APR4-140140,
ALF2-140140, ALF2-130140, ALF2-135135, H4-130150, H4-140140, H4-135135
($\Gamma_{\rm th}=1.6)$), and H4-135135 ($\Gamma_{\rm th}=1.8)$) is
$\sim 30$, 30, 5, 10, 15, 20, 10, 15, 25~ms for $N=60$, respectively.}
%%%%%%%
{\begin{tabular}{cc|ccccccc} \hline
Model & $\Gamma_{\rm th}$ & Remnant & $M_{*{\rm esc}}(10^{-3}M_{\odot})$
& $T_{*\rm esc}(10^{50}{\rm ergs})$
& $\bar V^R_{\rm esc}/c$ & $\bar V^Z_{\rm esc}/c$ 
& $f_{\rm ave, 5 ms}$~(kHz) & $f_{\rm ave, 10 ms}$~(kHz)
\\ \hline \hline
APR4-130160 &1.8& BH    & 2.0 & 1.5 & 0.24 & 0.08 & --- & --- \\
APR4-140150 &1.8& BH    & 0.6 & 0.9 & 0.35 & 0.12 & --- & --- \\
APR4-145145 &1.8& BH    & 0.1 & $<0.1$ & 0.29 & 0.13 & --- & --- \\
APR4-130150 &1.8& HMNS$\rightarrow$BH  & 12  & 8.5 & 0.23 & 0.12 
& $3.48 \pm 0.47$ & $3.46 \pm 0.37$ \\
APR4-140140 &1.8& HMNS$\rightarrow$BH  & 14  & 10 & 0.22 & 0.15 
& $3.53 \pm 0.52$ & $3.52 \pm 0.48$ \\
APR4-120150 &1.6& HMNS  & 9 & 5 & 0.20 & 0.10 
& $3.47 \pm 0.30$ & $3.44 \pm 0.27$ \\
APR4-120150 &1.8& HMNS  & 8 & 5.5 & 0.23 & 0.11 
& $3.44 \pm 0.30$ & $3.41 \pm 0.24$ \\
APR4-120150 &2.0& HMNS  & 7.5 & 5.5 & 0.24 & 0.12 
& $3.32 \pm 0.32$ & $3.27 \pm 0.26$ \\
APR4-125145 &1.8& HMNS  & 7 & 4.5 & 0.22 & 0.11 
& $3.36 \pm 0.31$ & $3.31 \pm 0.25$ \\
APR4-130140 &1.8& HMNS  & 8 & 5 & 0.19 & 0.12 
& $3.30 \pm 0.29$ & $3.27 \pm 0.28$ \\
APR4-135135 &1.6& HMNS  & 11  & 6 & 0.19 & 0.13 
& $3.46 \pm 0.42$ & $3.45 \pm 0.37$\\
APR4-135135 &1.8& HMNS  & 7 & 4 & 0.19 & 0.12 
& $3.31 \pm 0.35$ & $3.31 \pm 0.32$ \\
APR4-135135 &2.0& HMNS  & 5 & 3 & 0.19 & 0.13 
& $3.35 \pm 0.39$ & $3.33 \pm 0.33$ \\
APR4-120140 &1.8& HMNS  & 3 & 2 & 0.21 & 0.12 
& $3.15 \pm 0.21$ & $3.13 \pm 0.19$ \\
APR4-125135 &1.8& HMNS  & 5 & 3 & 0.18 & 0.10 
& $3.22 \pm 0.25$ & $3.19 \pm 0.24$ \\
APR4-130130 &1.8& HMNS  & 2 & 1 & 0.19 & 0.10 
& $3.22 \pm 0.28$ & $3.19 \pm 0.26$ \\ \hline
%%%%%%%%%%%%%%%%%%%%%%%%%%%%
ALF2-140140 & 1.8& HMNS$\rightarrow$BH   & 2.5 & 1.5 & 0.21 & 0.13
& $2.93 \pm 0.42$ & --- \\
ALF2-120150 & 1.8& HMNS                  & 5.5 & 3   & 0.21 & 0.10 
& $2.70 \pm 0.19$ & $2.71 \pm 0.16$ \\
ALF2-125145 & 1.8& HMNS                  & 3 & 1.5 & 0.20 & 0.10 
& $2.66 \pm 0.14$ & $2.66 \pm 0.13$ \\
ALF2-130140 & 1.8& HMNS $\rightarrow$ BH & 1.5 & 0.8 & 0.16 & 0.11 
& $2.73 \pm 0.19$ & $2.75 \pm 0.17$ \\
ALF2-135135 & 1.8& HMNS $\rightarrow$ BH & 2.5 & 1.5 & 0.22 & 0.12 
& $2.75 \pm 0.18$ & $2.76 \pm 0.16$ \\
ALF2-130130 & 1.8& HMNS                  & 2 & 1.0 & 0.19 & 0.10 
& $2.58 \pm 0.18$ & $2.56 \pm 0.16$ \\
%%%%%%%%ALF2-130130 & 1.8& & & & & & &\\
\hline
H4-130150 & 1.8& HMNS$\rightarrow$BH  & 3 & 2 
& 0.19 & 0.10 & $2.44 \pm 0.17$ & $2.45 \pm 0.15$ \\
H4-140140 & 1.8& HMNS$\rightarrow$BH  & 0.3 & 0.2 
& 0.17 & 0.13 & $2.63 \pm 0.23$ & $2.77 \pm 0.41$ \\
H4-120150 & 1.6& HMNS  & 4.5 & 2 & 0.19 & 0.10 
& $2.28 \pm 0.16$ & $2.29 \pm 0.14$\\
H4-120150 & 1.8& HMNS  & 3.5 & 2 & 0.21 & 0.09 
& $2.30 \pm 0.18$ & $2.31 \pm 0.15$ \\
H4-120150 & 2.0& HMNS  & 4 & 2 & 0.21 & 0.09 
& $2.24 \pm 0.15$ & $2.23 \pm 0.14$ \\
H4-125145 & 1.8& HMNS  & 2 & 1.5 & 0.19 & 0.10 
& $2.41 \pm 0.15$ & $2.41 \pm 0.13$ \\
H4-130140 & 1.8& HMNS  & 0.7 & 0.4 & 0.18 & 0.10 
& $2.42 \pm 0.17$ & $2.42 \pm 0.15$ \\
H4-135135 & 1.6& HMNS$\rightarrow$BH  & 0.7 & 0.4 & 0.21 & 0.11 
& $2.49 \pm 0.19$ & $2.54 \pm 0.16$ \\
H4-135135 & 1.8& HMNS$\rightarrow$BH  & 0.5 & 0.2 & 0.19 & 0.11 
& $2.44 \pm 0.20$ & $2.48 \pm 0.16$ \\
H4-135135 & 2.0& HMNS  & 0.4 & 0.2 & 0.20 & 0.10 
& $2.39 \pm 0.21$ & $2.43 \pm 0.17$ \\
H4-120140 & 1.8& HMNS  & 2.5 & 1 & 0.19 & 0.10 
& $2.30 \pm 0.15$ & $2.30 \pm 0.14$ \\
H4-125135 & 1.8& HMNS  & 0.6 & 0.3 & 0.18 & 0.10 
& $2.29 \pm 0.17$ & $2.27 \pm 0.14$ \\
H4-130130 & 1.8& HMNS  & 0.3 & 0.1 & 0.16 & 0.10 
& $2.35 \pm 0.18$ & $2.38 \pm 0.14$ \\ \hline
%%%%%%%%%%%%%%%%%%%%%%%%%%%%%%%%%%%%
MS1-140140 & 1.8& MNS  & 0.6 & 0.2 & 0.13 & 0.09 
& $2.09 \pm 0.14$ & $2.06 \pm 0.12$ \\
MS1-120150 & 1.8& MNS  & 3.5 & 1.5 & 0.19 & 0.10 
& $2.08 \pm 0.11$ & $2.09 \pm 0.09$ \\
MS1-125145 & 1.8& MNS  & 1.5 & 0.8 & 0.19 & 0.11 
& $2.02 \pm 0.14$ & $1.99 \pm 0.15$ \\
MS1-130140 & 1.8& MNS  & 0.6 & 0.2 & 0.17 & 0.09 
& $2.05 \pm 0.14$ & $2.02 \pm 0.13$ \\
MS1-135135 & 1.8& MNS  & 1.5 & 0.6 & 0.14 & 0.08 
& $1.98 \pm 0.18$ & $1.95 \pm 0.16$\\
MS1-130130 & 1.8& MNS  & 1.5 & 0.5 & 0.15 & 0.08 
& $1.93 \pm 0.19$ & $1.90 \pm 0.17$\\
\hline
\end{tabular}
}
\label{table:result}
\end{table*}

\subsubsection{Gravitational waves}\label{sec:GW}

\begin{figure*}[p]
\begin{tabular}{cc}
\includegraphics[width=80mm,clip]{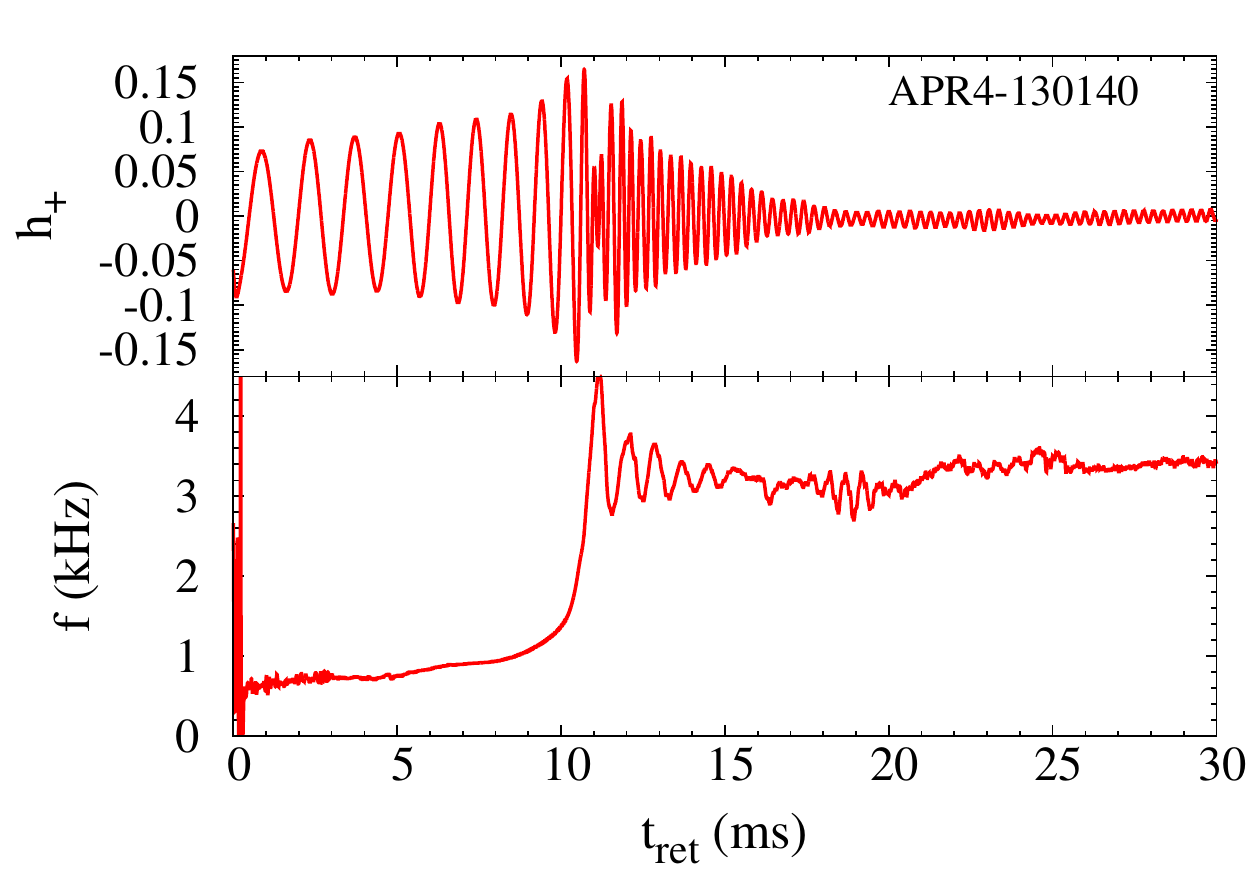}
\includegraphics[width=80mm,clip]{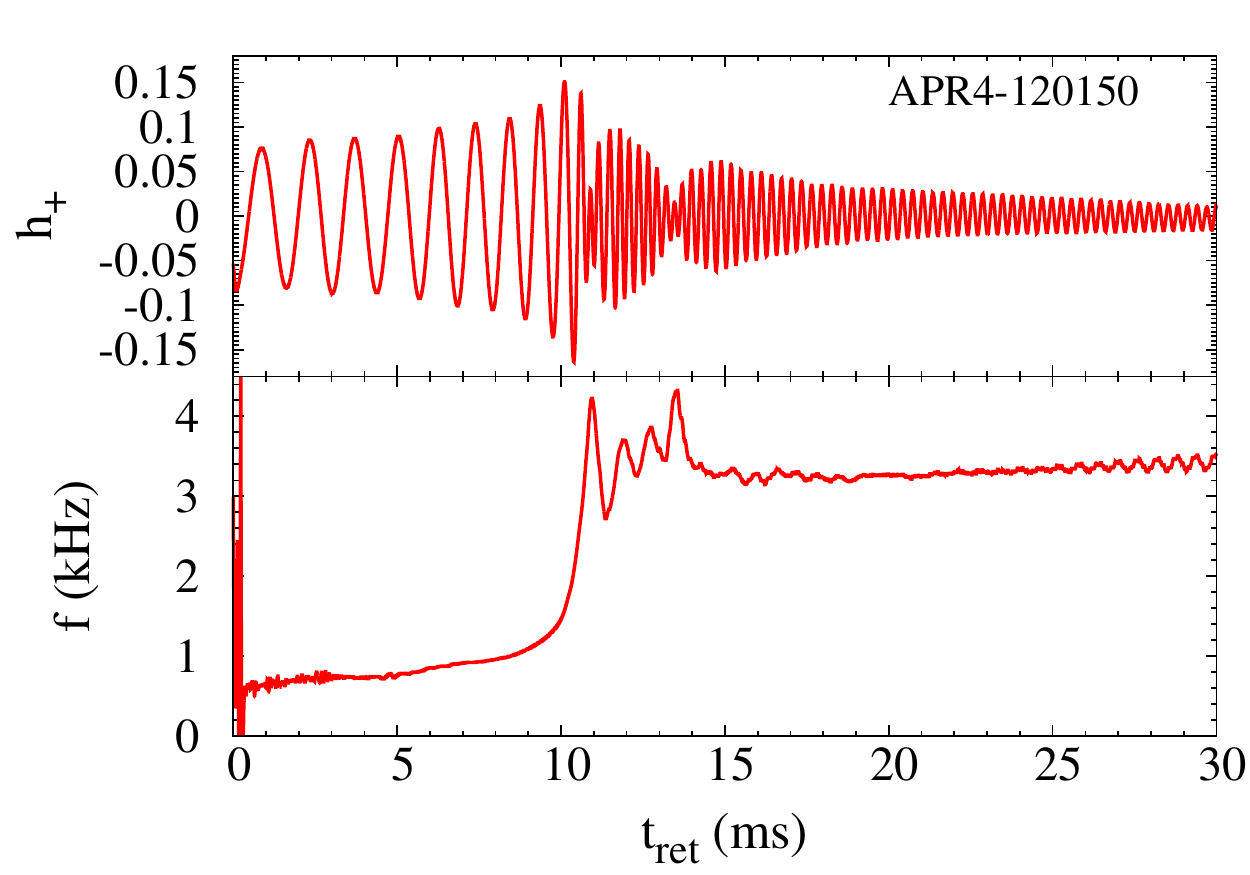}\\
\includegraphics[width=80mm,clip]{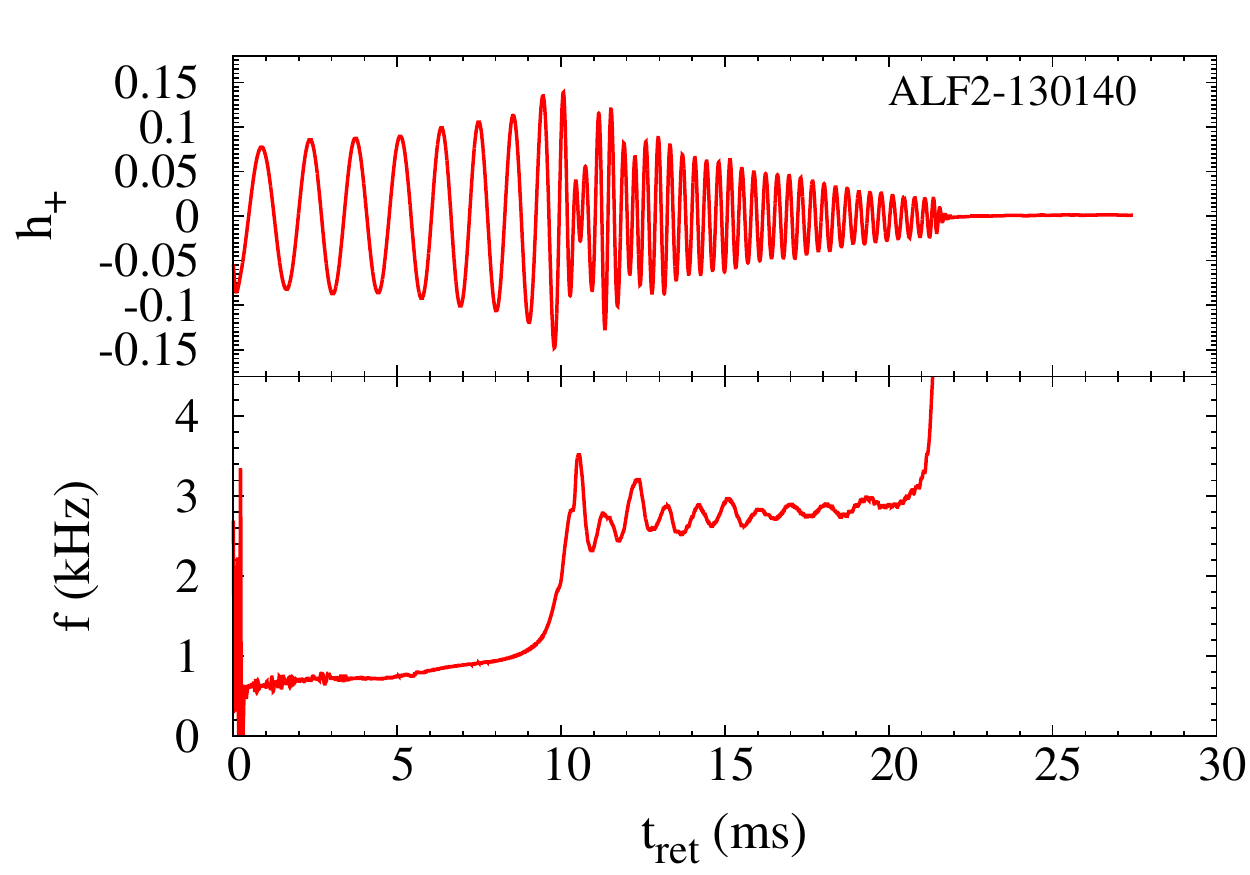}
\includegraphics[width=80mm,clip]{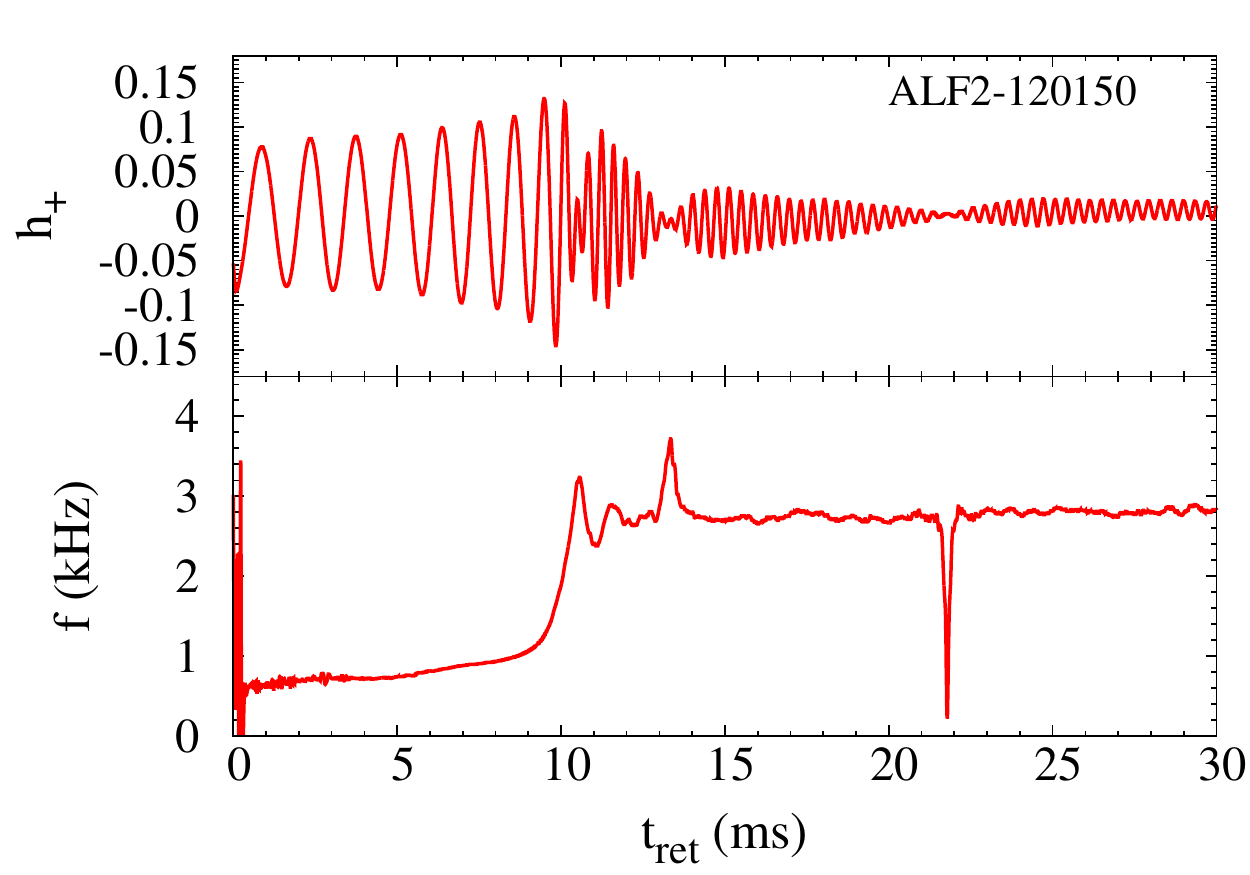}\\
\includegraphics[width=80mm,clip]{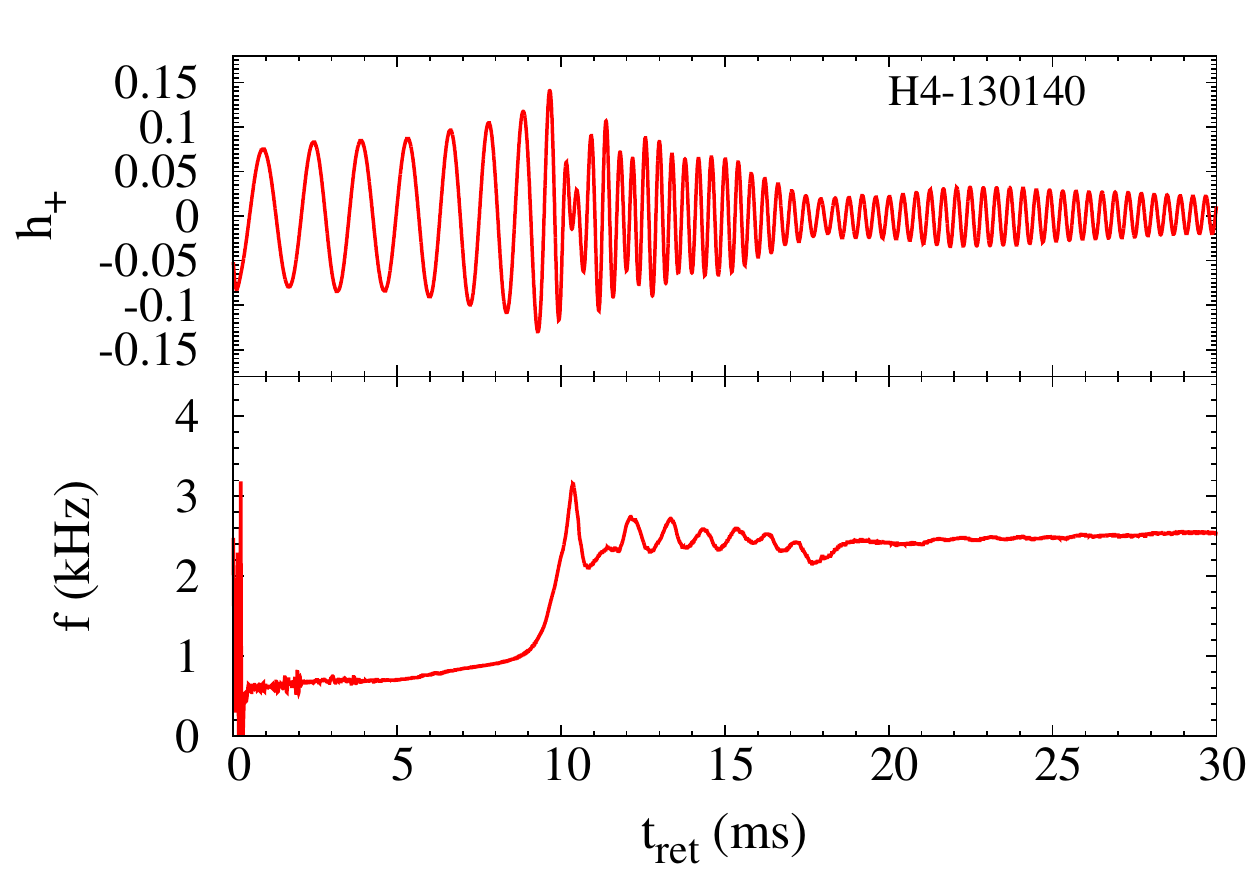}
\includegraphics[width=80mm,clip]{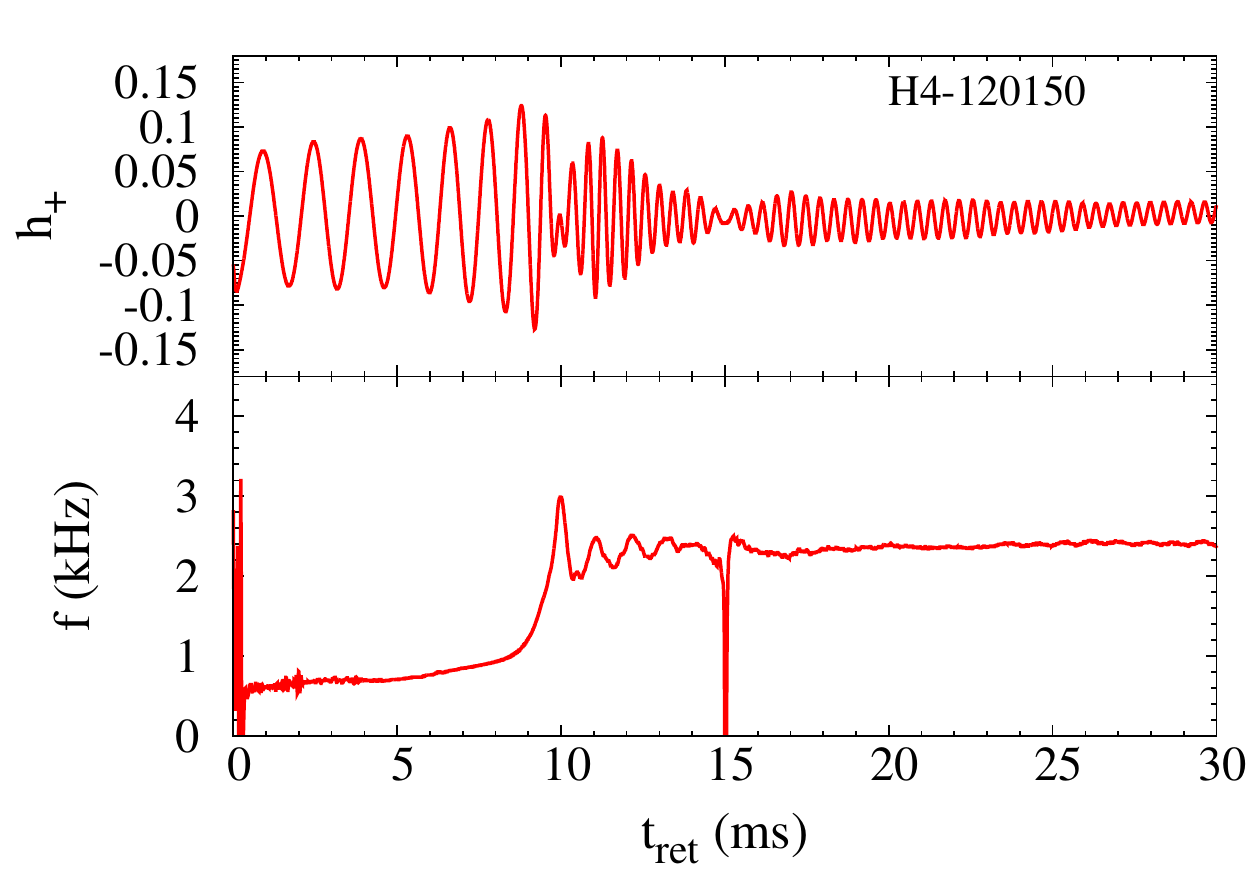}\\
\includegraphics[width=80mm,clip]{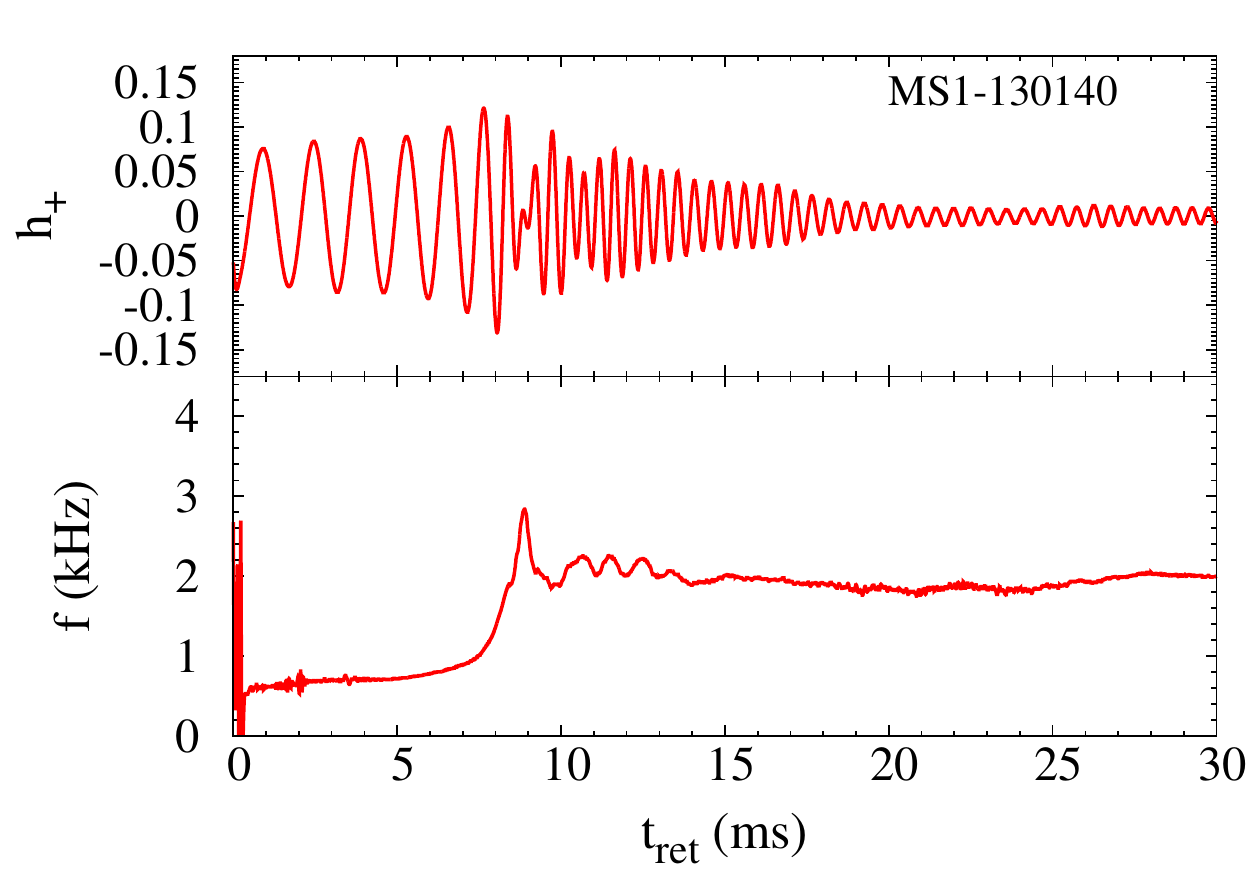}
\includegraphics[width=80mm,clip]{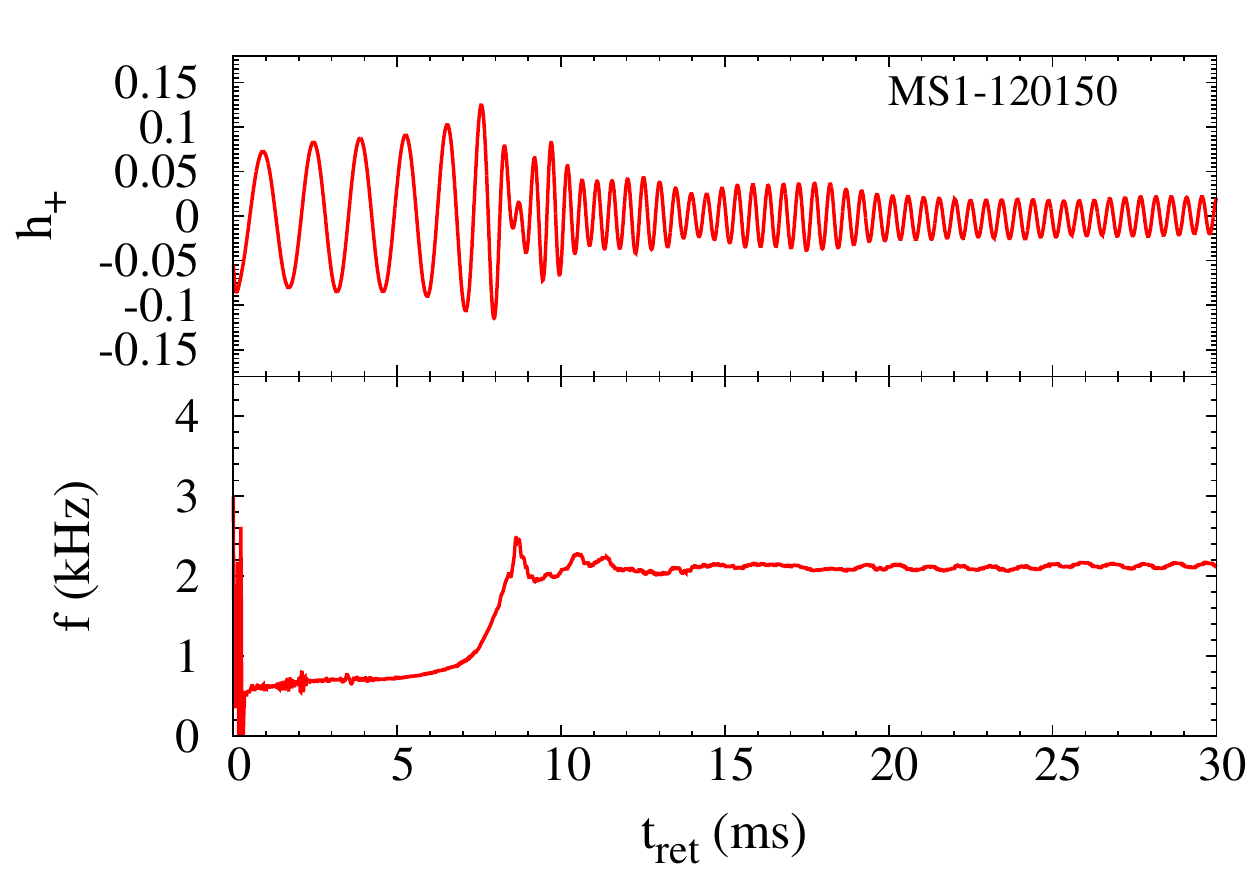}
\end{tabular}
\caption{Gravitational waves ($h_+ D/m$) and the frequency of
gravitational waves $f$ as functions of time for models APR4-130140
(top left), APR4-120150 (top right), ALF2-130140 (second top left),
ALF2-120150 (second top right), H4-130140 (third left), H4-120150
(third right), MS1-130140 (bottom left), and MS1-120150 (bottom
right).  For ALF2-130140, a black hole is formed at 11~ms after the
onset of the merger, and ringdown gravitational waves are emitted in
the final phase. For all the panels, the vertical axis shows the
non-dimensional amplitude, $h_+ D/m$, with $D$ being the distance to
the source.}
\label{figGW}
\end{figure*}

First of all, we summarize the properties of gravitational waves
emitted by the HMNS, because its gravitational-wave frequency, which
is determined by the spin of the HMNS, has a correlation with the
amount of the ejected material

As mentioned already, HMNSs exert the torque to its surrounding
material. The efficiency of the angular momentum transport is higher,
in general, for the faster rotating and more compact HMNS. Associated
with this property, the frequency of gravitational waves and the
efficiency of the angular momentum transport are expected to be
closely related.  The characteristic spin frequency for these deformed
HMNSs can be determined from gravitational waves emitted by them.
Figure~\ref{figGW} displays gravitational waves and their frequency as
functions of time for eight models with mass
$(m_1,m_2)=(1.3M_{\odot},1.4M_{\odot})$ and
$(m_1,m_2)=(1.2M_{\odot},1.5M_{\odot})$ and with four EOSs.  These
plots show that quasiperiodic gravitational waves are emitted by the
HMNSs for all the models. Namely, the gravitational-wave frequency
does not change significantly during the evolution of the
HMNSs. However, the frequency is not constant exactly and actually
varies with time. This is natural because (i) the HMNSs quasiradially
oscillate with time in their early stage of the evolution, and (ii)
the HMNSs lose the energy and angular momentum due to the
gravitational-wave emission and hydrodynamic angular momentum
transport process, and hence, their configuration evolves. These two
effects result in the variation in the characteristic spin velocity
and frequency of gravitational waves. The degree of the variation in
the frequency of gravitational waves is larger (a) for the HMNS with
the EOS that yields a compact neutron star (we often call such an EOS
soft EOS in this paper), and (b) for the HMNS for which the mass is
close to the critical value to the collapse to a black hole; see,
e.g., the gravitational-wave frequency for model ALF2-130140.  The
case (a) is due to the fact that at the merger, the central density
significantly increases in the soft EOSs, resulting in subsequent
high-amplitude oscillations.  The case (b) is due to the fact that for
such a HMNS, a small change in the spin velocity results in a large
change in the central density. 

\begin{figure*}[t]
\begin{tabular}{cc}
\includegraphics[width=88mm,clip]{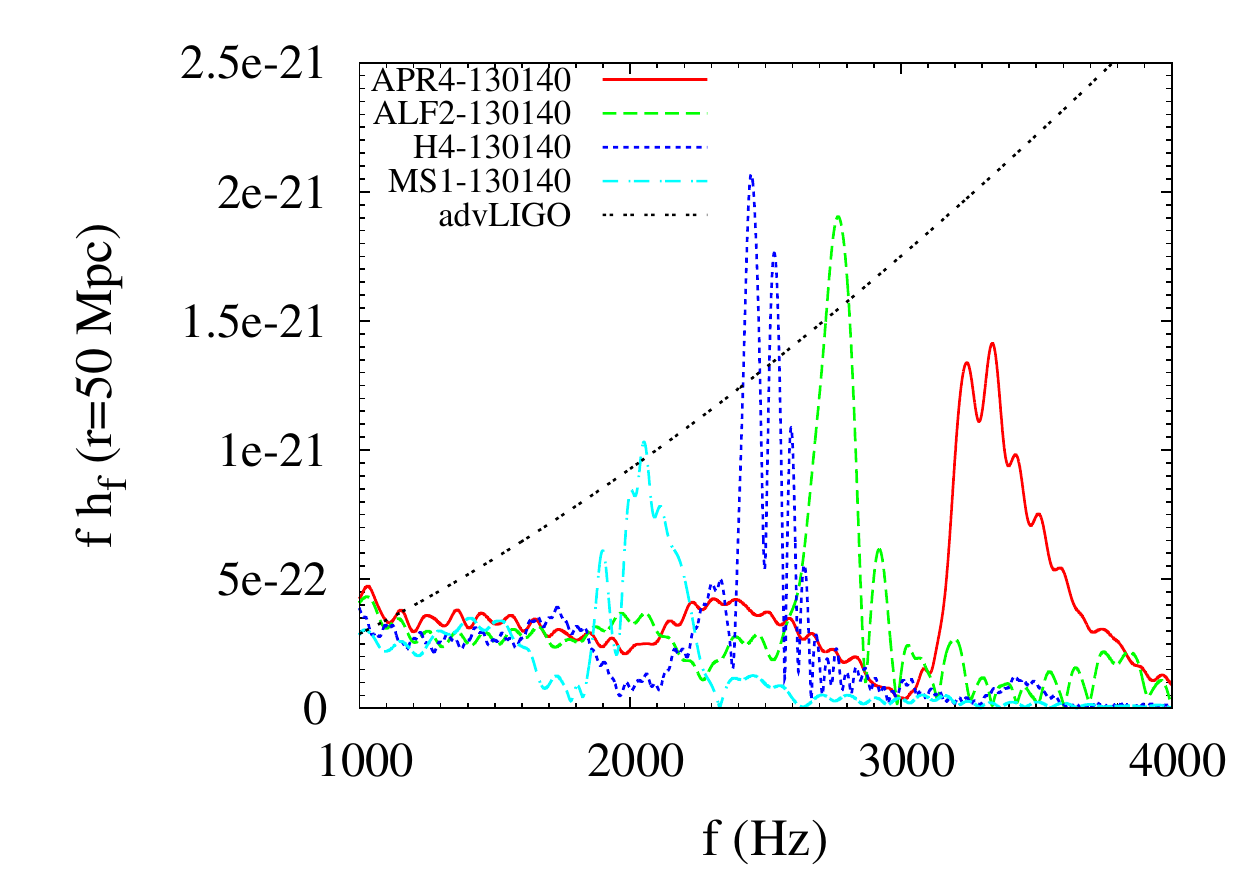}
\includegraphics[width=88mm,clip]{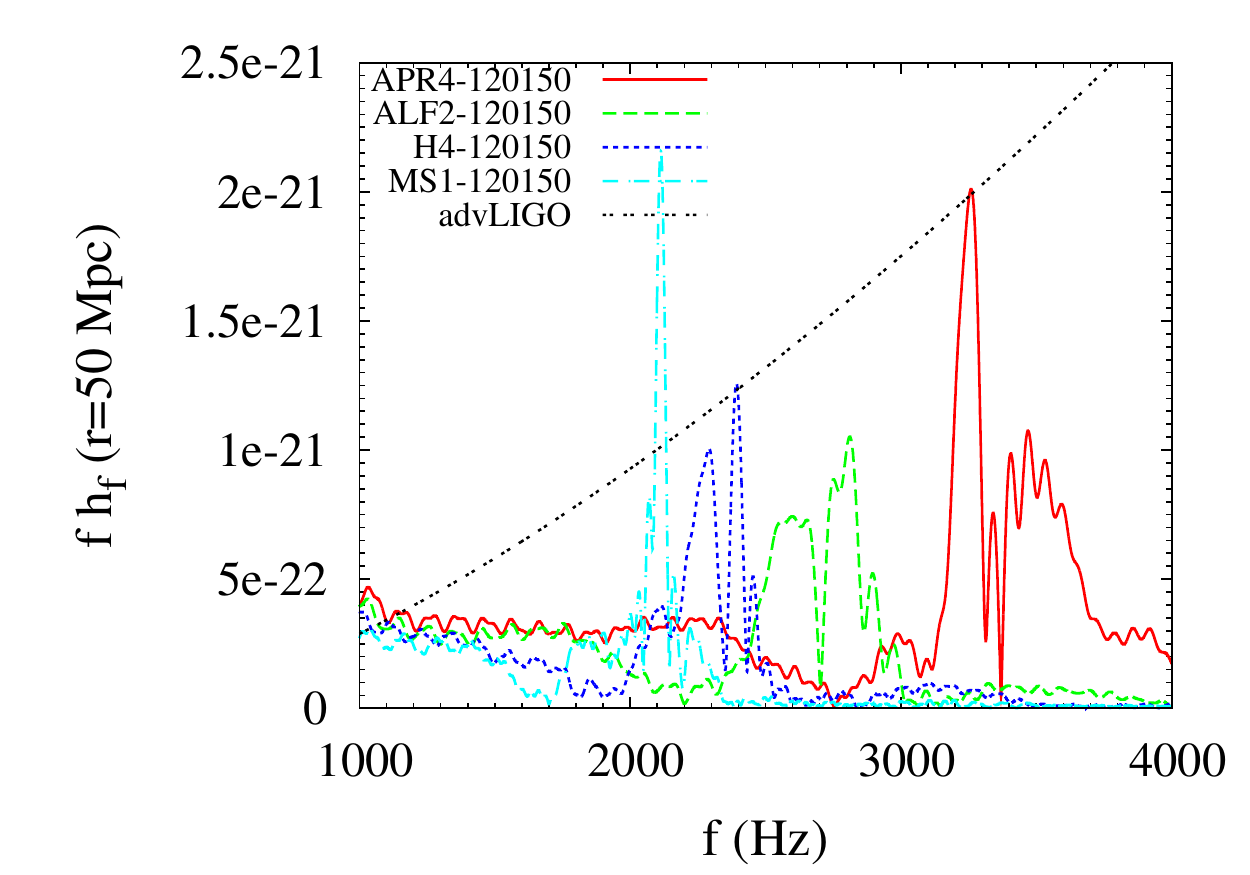}
\end{tabular}
\caption{Fourier spectra of gravitational waves for the results shown
in Fig.~\ref{figGW}.  The amplitude is shown for the hypothetical
event at a distance of 50~Mpc along the direction perpendicular to the
orbital plane (the most optimistic direction). The black dot-dot curve
is the noise spectrum of the advanced LIGO with an optimistic
configuration for the detection of high-frequency gravitational
waves (see https://dcc.ligo.org/cgi-bin/DocDB/ShowDocument?docid=2974).}
\label{figGWF}
\end{figure*}

%%% TIME INTEGRATION

Figure~\ref{figGWF} plots the Fourier spectra for gravitational waves
shown in Fig.~\ref{figGW}. This shows that there are peaks for a
high-frequency band $2~{\rm kHz} \alt f \alt 4~{\rm kHz}$ irrespective
of models. For a ``soft'' EOS that yields a compact neutron star for
the canonical mass, the peak frequency is higher (e.g., for the
spectra of APR4, the peak frequency is the highest among the four
EOSs), and a certain correlation exists between the peak frequency and
stellar radius~\cite{BJ2012}. The peak frequency is approximately
associated with the typical frequency of quasiperiodic oscillation of
gravitational waves found in Fig.~\ref{figGW}. However, as already
mentioned, the (nonaxisymmetric) oscillation frequencies of the HMNSs
vary during the evolution due to a quasiradial oscillation and the
back reaction due to the gravitational-wave emission and angular
momentum transport process, and hence, the peak frequencies change
with time, resulting in the broadening of the peak or appearance of
the multi peaks.  Therefore, it is not a very good idea to determine
the characteristic frequency from the peak of the Fourier spectrum.
Rather, the Fourier spectrum might provide an inaccurate message when
we determine the characteristic oscillation frequency.  Thus, we
determine the average frequency from the results of the frequency
shown in Fig.~\ref{figGW} in terms of Eq.~(\ref{eq:fave1}) with the
dispersion determined by Eq.~(\ref{eq:fave2}). Here, the time
integration is performed for 5 and 10 ms after the formation of the
HMNSs.  The last two columns of Table~\ref{table:result} list the
average frequency and the dispersion determined for 5 and 10 ms
integration.  Note that the typical nonaxisymmetric oscillation
frequency of the HMNSs is half as large as the values listed in
Table~\ref{table:result} because the listed ones are the
gravitational-wave frequencies.

The value of the oscillation frequency for a given mass of the HMNS
depends primarily on its radius, i.e., a stiffness of the EOS. For the
EOS that yields small-radius neutron stars (``soft'' EOS), the
oscillation frequency and peak frequency of gravitational waves are
higher, because the spin angular velocity of the HMNS is close to
the Kepler velocity, and thus, the oscillation and peak frequencies
are qualitatively proportional to $(M_{\rm HMNS}/R_{\rm
HMNS}^3)^{1/2}$ where $M_{\rm HMNS}$ and $R_{\rm HMNS}$ denote the
typical mass and radius of a HMNS. The oscillation frequency depends
also weakly on the value of $\Gamma_{\rm th}$: For the smaller value
of it, the frequency is slightly higher for many cases, because the
effect of shock heating is weaker, and the HMNS becomes more compact.

For a larger spin of the nonaxisymmetric HMNS, the material
surrounding the HMNS can receive a torque with a higher efficiency.
This suggests that for the merger of a binary neutron star composed of
smaller-radius neutron stars, the amount of the ejected material could
be larger. As shown in Sec.~\ref{sec:EOSdep}, this is indeed the case
(in particular for unequal-mass models), as long as the models in this
paper are concerned. 

Table~\ref{table:result} as well as Fig.~\ref{figGWF} also show that
the magnitude of the dispersion, $\sigma_f$, is not negligible. For
APR4 for which the neutron-star radius is rather small and the
amplitude of a quasiradial oscillation induced at the formation of the
HMNSs is rather large, the magnitude of the dispersion is 0.2 --
0.5~kHz; for $m=2.7M_{\odot}$, the typical value is 0.3 -- 0.4~kHz.
For other EOSs, the dispersion is relatively small. However, it is
still 0.1 -- 0.2~kHz. Thus, we conclude that the characteristic
frequency of gravitational waves from HMNSs varies with time in
general.

\subsubsection{Average velocity of the ejected material}

\begin{figure}[t]
\includegraphics[width=83mm,clip]{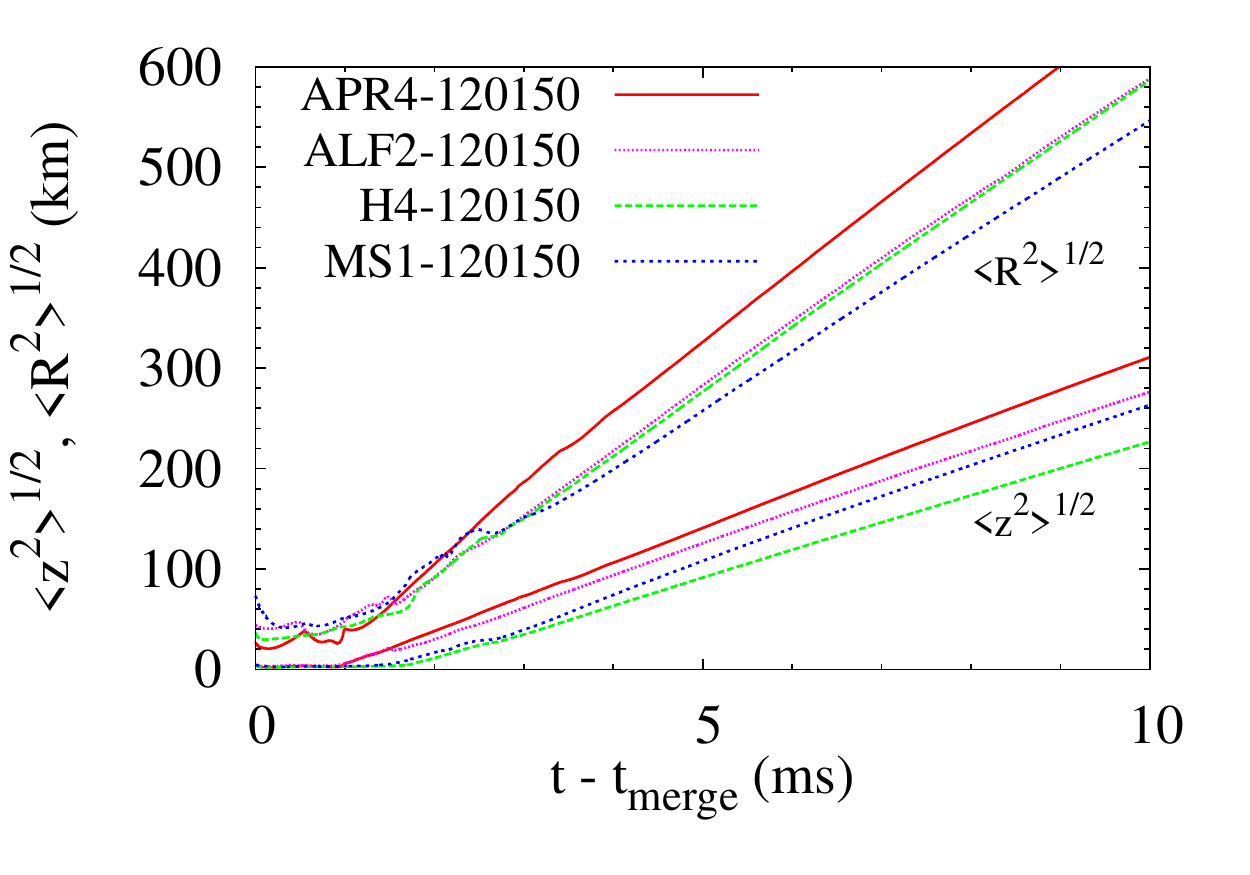}
\vspace{-3mm}
\caption{$\bar R$($=\langle R^2 \rangle^{1/2}$) and
$\bar Z$($=\langle z^2 \rangle^{1/2}$) as functions of time for
APR4-120150, ALF2-120150, H4-120150, and MS1-120150.  }
\label{figxyz}
\end{figure}

Figure~\ref{figxyz} plots $\bar R$ and $\bar Z$ as functions of time
for APR4-120150, ALF2-120150, H4-120150, and MS1-120150. Note that
$d{\bar R}/dt$ and $d{\bar Z}/dt$ may be considered as an average
velocity of the ejected material in the cylindrical and vertical
directions, respectively, and that a similar result is found for other
choices of mass. This shows that the material ejected expands with an
approximately constant sub-relativistic velocity $\sim 0.15$ --
$0.25c$ for $t-t_{\rm merger} \agt 2$~ms in the cylindrical direction
and the velocity in the vertical direction is 0.4 -- 0.5 times as
large as that in the cylindrical direction. This suggests that the
vertical thickness angle of the ejected material, $\theta_0$, is $\sim
40$ -- $50^\circ$. Namely, the ejected material expands in a
moderately anisotropic manner.  Note that the velocity in the
cylindrical direction is primarily caused by the torque exerted by the
HMNS, while the velocity in the vertical direction is primarily caused
by the shock heating. This implies that both effects play an important
role.

The velocity in the later phase, $t-t_{\rm merge} \agt 3$~ms, is
larger for APR4 than that for other EOSs employed in this paper. This
is due to the fact that with APR4, a more compact state is realized in
the HMNS, and hence, (i) a strong shock associated with the
compression by a strong gravity and a subsequent large-amplitude
oscillation (cf. Fig.~\ref{figrho}) occurs, resulting in an efficient
mass ejection, and (ii) the HMNS strongly exerts the torque to its
surrounding material.  For APR4, a relatively dense atmosphere
surrounding the HMNS is formed not only in the vicinity of the
equatorial plane but also in the vertical direction (compare
Figs.~\ref{fig3B} and~\ref{fig3C}). This also reflects the fact that a
strong shock heating occurs with this EOS (see Figs.~\ref{figeth} and
\ref{figeth2}). 

\subsubsection{Dependence on EOS} \label{sec:EOSdep}

\begin{figure*}[t]
\begin{tabular}{cc}
\includegraphics[width=82mm,clip]{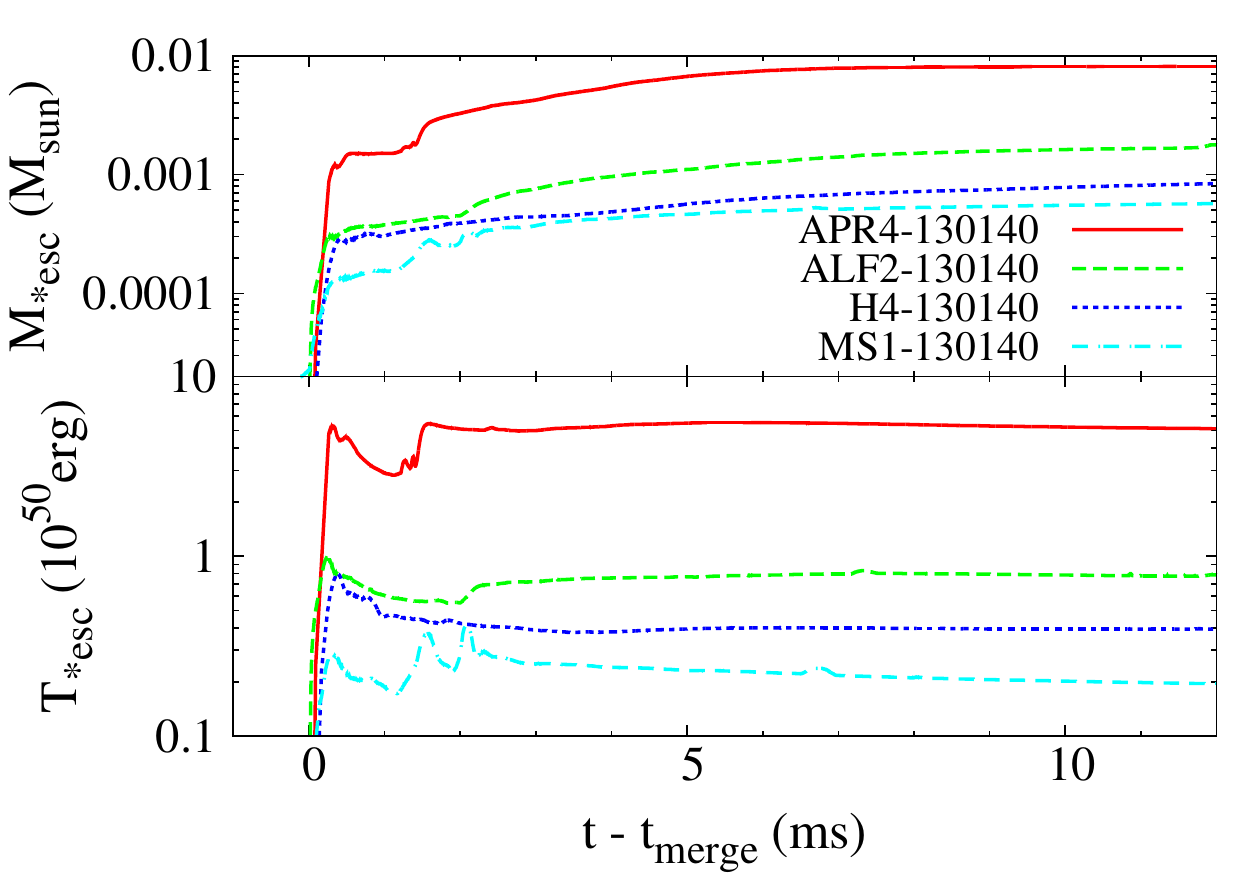}~~~
\includegraphics[width=82mm,clip]{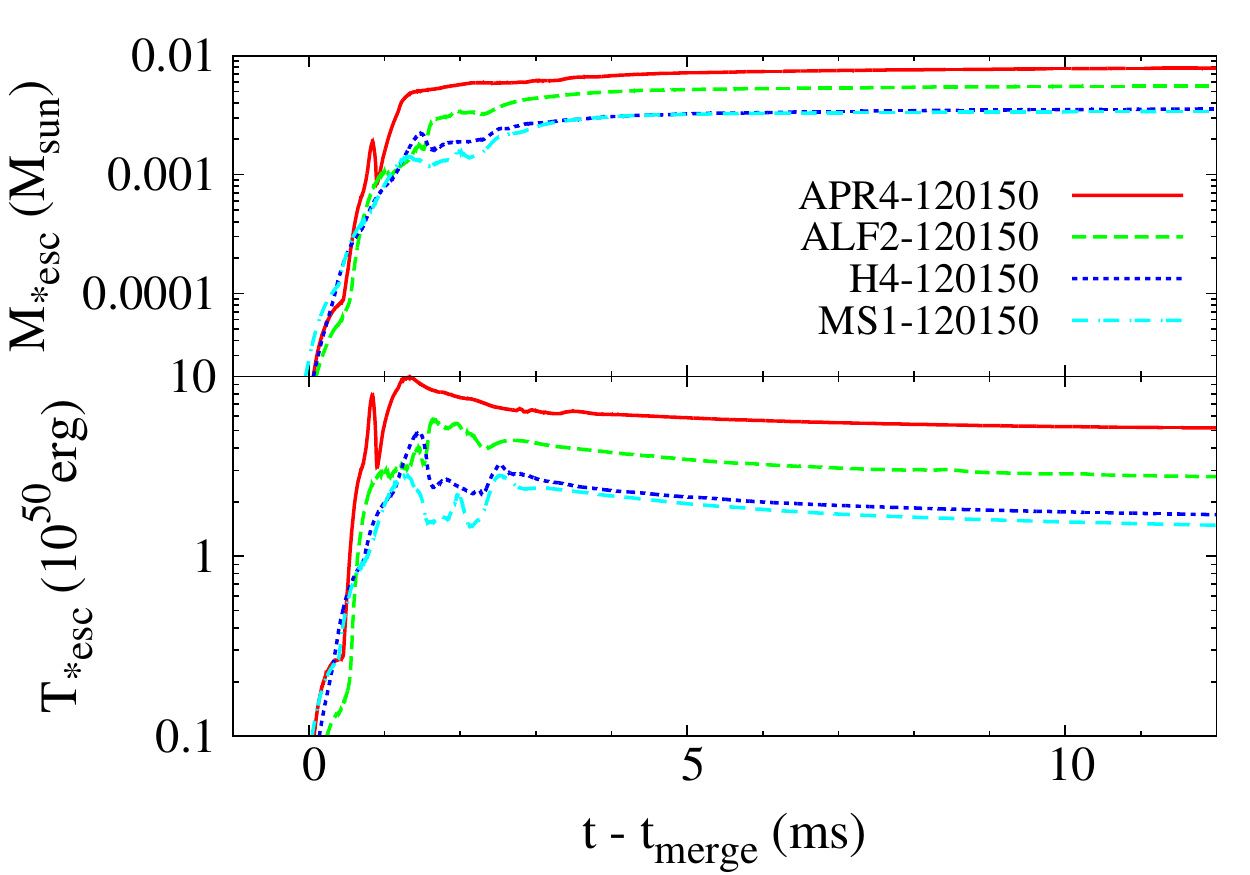}
\end{tabular}
\caption{$M_{*{\rm esc}}$ and $T_{*\rm esc}$ as functions of
$t-t_{\rm merge}$ (left) for models APR4-130140, ALF2-130140,
H4-130140, MS1-130140, and (right) for models APR4-120150,
ALF2-120150, H4-120150, MS1-120150.  }
\label{fig:escEOS}
\end{figure*}

Figure~\ref{fig:escEOS} plots the total rest mass and kinetic energy
of the material ejected from the HMNSs as functions of $t-t_{\rm
merge}$ for several models; for the left and right panels, the masses
of two neutron stars are $(1.3M_{\odot}, 1.4M_{\odot})$ and
$(1.2M_{\odot}, 1.5M_{\odot})$, respectively, with the total mass
$2.7M_{\odot}$, while four EOSs are chosen.  This shows that the rest
mass and kinetic energy of the ejected material depend strongly on the
EOS.  The primary reason is that the compactness of the HMNS depends
strongly on the EOS. For APR4 and ALF2, neutron stars of canonical
masses 1.2 -- $1.5M_{\odot}$ have a relatively small radius
(cf. Table~\ref{table:EOS}). This implies that the merger sets in at a
compact orbit, and the formed HMNS is more compact than that formed in
stiffer EOSs that yield large-radius neutron stars.

A high compactness of a HMNS affects the properties of the material
ejected from it in the following two ways. First, the HMNS is more
rapidly rotating, and hence, it exerts the torque, caused by its
nonaxisymmetric configuration and rapid rotation, to the material in
the outer region more efficiently than a less compact HMNS. As a
result of this effect, a fraction of the material that gains the
kinetic energy large enough to escape from the system is increased. In
addition, during the formation of such a compact HMNS, a quasiradial
oscillation with a high amplitude is often induced (see
Fig.~\ref{figrho}).  This is in particular the case for APR4 in which
the EOS becomes stiff for a high-density region although it is rather
soft for the density of canonical-mass neutron stars. This quasiradial
oscillation helps the material surrounding the HMNS to obtain kinetic
energy through shock heating (see section~\ref{sec:GW}).

A possibly important fact to be noted is that the material, which
eventually escapes from the system, initially stays in the vicinity of
the HMNS. Namely, this material stays in a deep gravitational
potential well initially, and thus, it is trapped.  For a more compact
HMNS, this potential should be deeper, and hence, the material there
needs to obtain more energy to escape from the HMNS. At the same time,
however, such a material can gain a stronger torque and thermal energy
for a longer timescale, because it is trapped for a longer duration,
and as a result, the material could get more kinetic energy if the
HMNS is more compact.

Namely, there are two competing effects, and it is not trivial at all
which effects are more important.  If the trapping effect due to the
deep potential well plays a more important role, we should find the
evidence that less material is ejected from more compact HMNS.
However, Fig.~\ref{fig:escEOS} shows that the ejected rest mass is
smaller for EOS with larger neutron star radii (less compact neutron
star).  This indicates that the trapping mechanism is less important
than the effects of the quasiradial oscillation and the torque exerted
from the HMNS, as long as the comparison among four model EOSs is
concerned (but see Sec.~\ref{sec4.1.5} for an evidence that this may
not be always the case).

%%%  FIGURE OF f_peak vs M_*esc

\begin{figure}[t]
\includegraphics[width=85mm,clip]{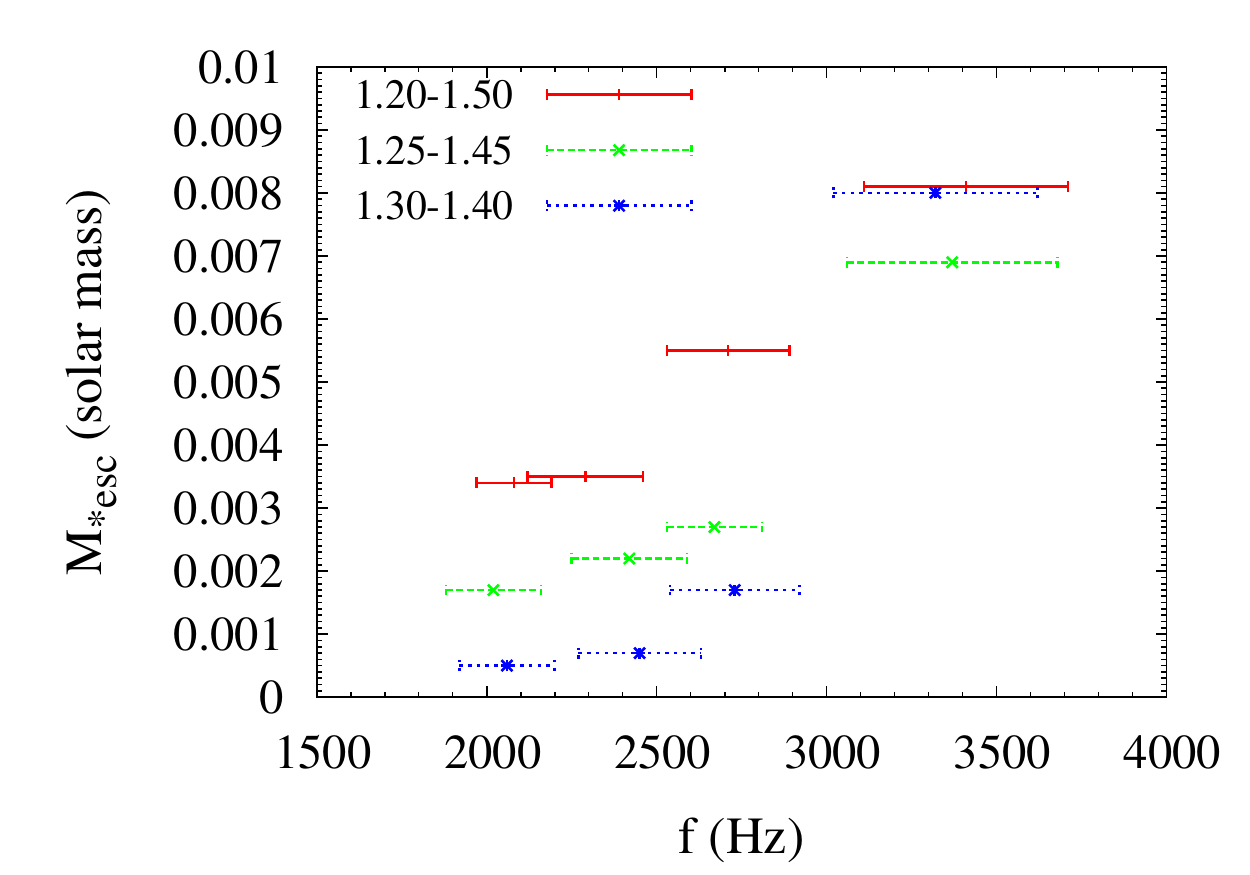}
\caption{$M_{*{\rm esc}}$ as a function of the characteristic 
gravitational-wave frequency emitted by the HMNS.  Here, the typical
frequency is determined by the EOS; from the highest to the 
lowest, APR4, ALF2, H4, and MS1. }
\label{figFM}
\end{figure}

The compactness of HMNSs, and hence, the EOS of neutron stars, are
well reflected in the frequency of gravitational waves emitted by the
HMNS, as already described in Sec.~\ref{sec:GW}. For a given total
mass and mass ratio of the binary system, the frequencies are higher
for binaries composed of more compact neutron stars (``softer'' EOS),
because the formed HMNS is more compact and hence the rotational
angular velocity approximately proportional to $(M_{\rm HMNS}/R_{\rm
HMNS}^3)^{1/2}$ is larger. Since we found that the rest mass and
kinetic energy of the ejected material are larger for the EOS that
yields more compact HMNSs, these quantities and the frequency of
gravitational waves should have a correlation.

Figure~\ref{figFM} plots the rest mass of the ejected material as a
function of the characteristic frequency of gravitational waves
emitted by the HMNS, $f_{\rm ave}$, for several models. Here, $f_{\rm
ave}$ is determined by the 5~ms integration using
Eq.~(\ref{eq:fave2}). For this plot, the results with $\Gamma_{\rm
th}=1.8$ are adopted. Note that the typical frequency for this plot is
determined primarily by the chosen EOS.  This figure shows that for a
given mass ratio $q$, these two quantities have a correlation; the
total rest mass of the ejected material increases with the
gravitational-wave frequency.

However, as already noted, there are counter examples (see
Sec.~\ref{sec4.1.5}). Namely, for some cases, the shallow potential
helps in enhancing the mass ejection.  For such models, the
correlation like that found in Fig.~\ref{figFM} does not hold. 

%%Such counter examples are found for the merger of equal-mass neutron
%%stars.
%%Thus, it should be noted that for a relatively low-frequency band
%%$\alt 2$~kHz, this correlation would not be very clear.

\subsubsection{Dependence on $\Gamma_{\rm th}$}

%% FIGURE
\begin{figure*}[t]
\begin{tabular}{cc}
\includegraphics[width=82mm,clip]{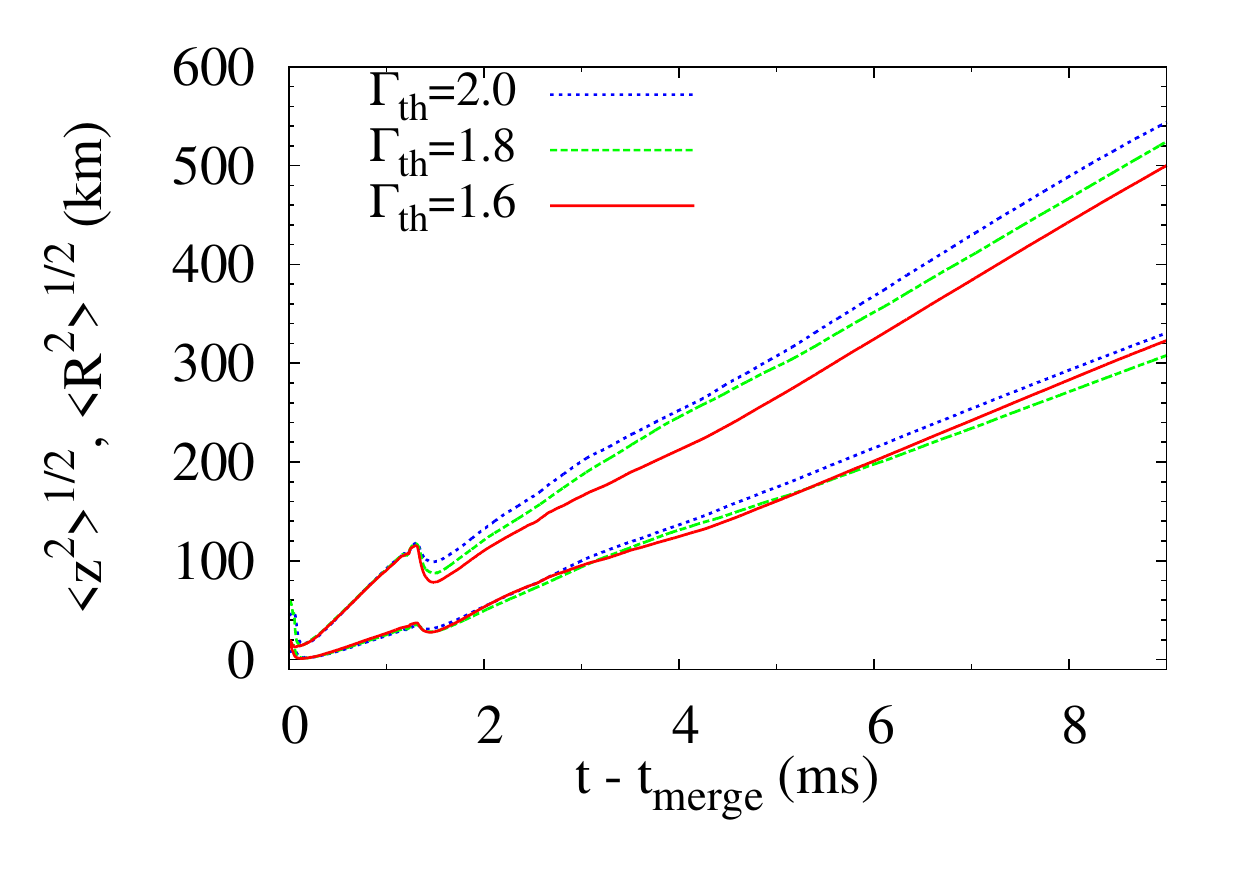}~~~
\includegraphics[width=82mm,clip]{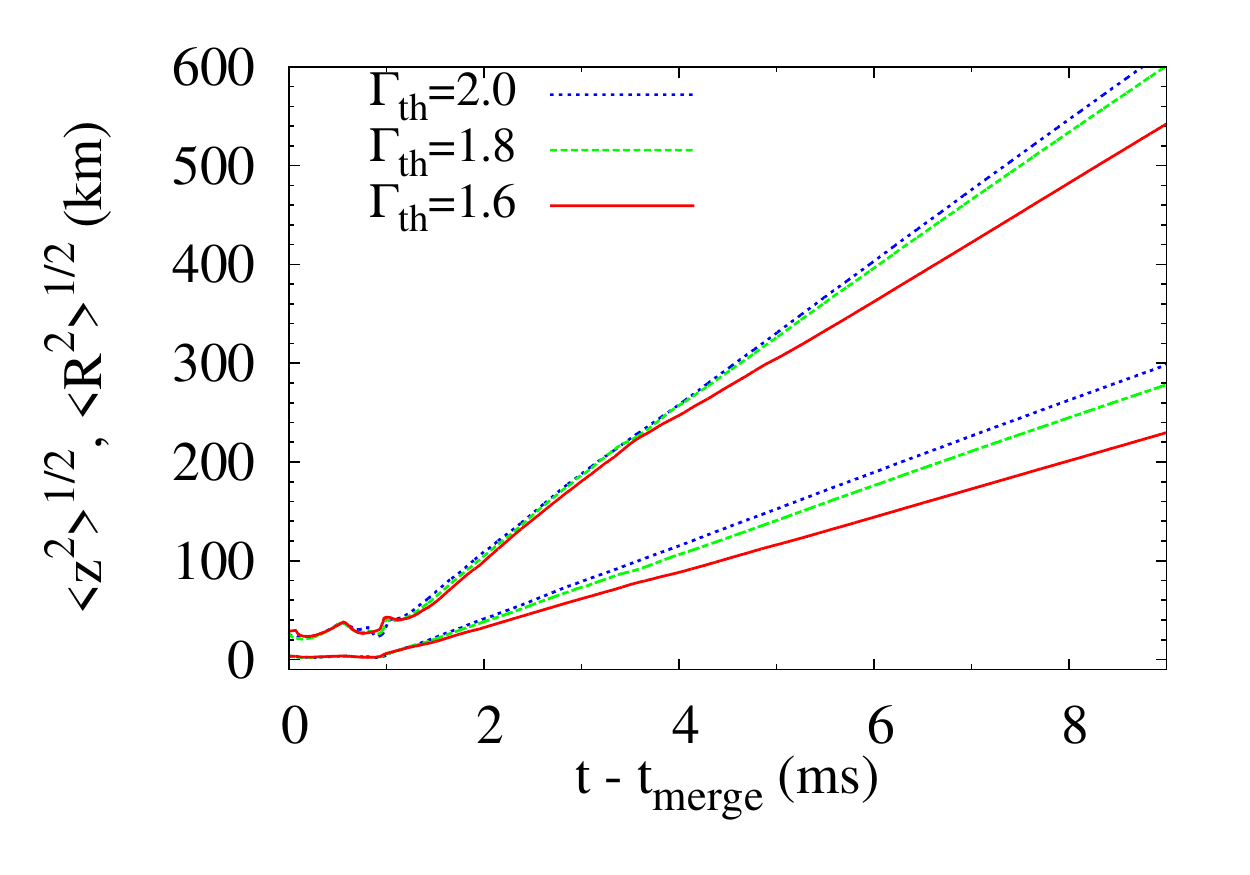}
\end{tabular}
\vspace{-3mm}
\caption{$\bar R$($=\langle R^2 \rangle^{1/2}$) (upper curves) and
$\bar Z$($=\langle z^2 \rangle^{1/2}$) (lower curves) as functions of
time for models APR4-135135 (left) and APR4-120150 (right) with
$\Gamma_{\rm th}=2.0$, 1.8, and 1.6.  }
\label{fig:xyzAPR}
\end{figure*}

%% FIGURE
\begin{figure*}[t]
\begin{tabular}{cc}
\includegraphics[width=82mm,clip]{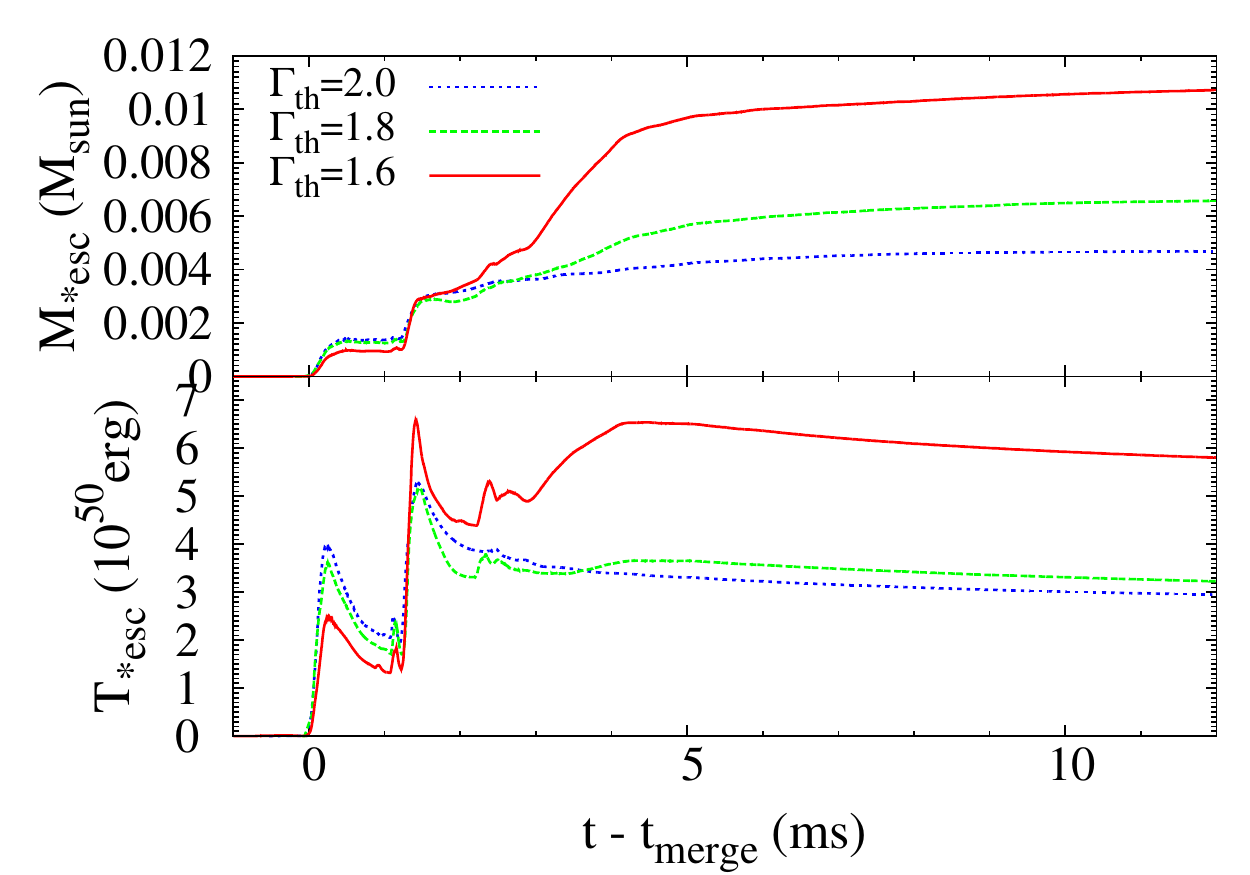}~~~
\includegraphics[width=82mm,clip]{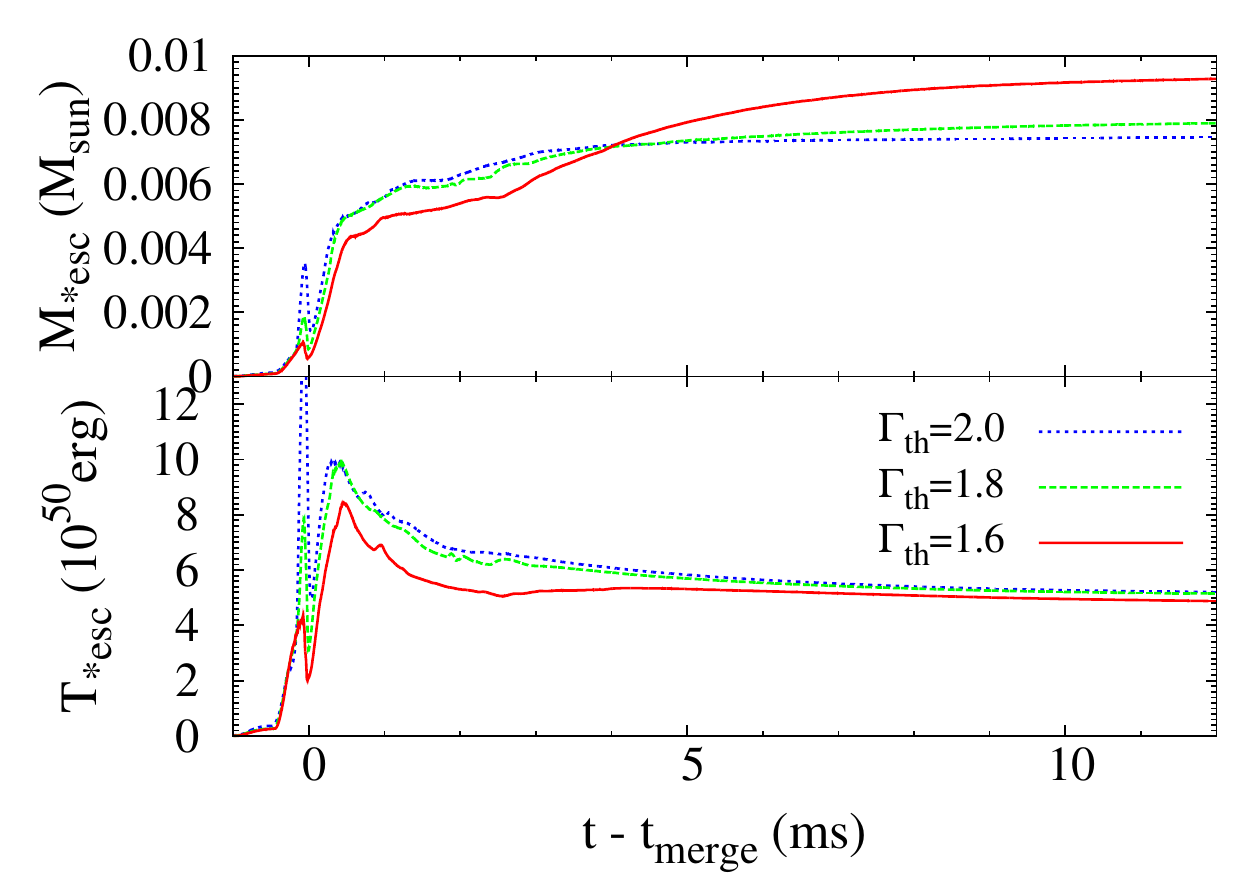}\\
\includegraphics[width=82mm,clip]{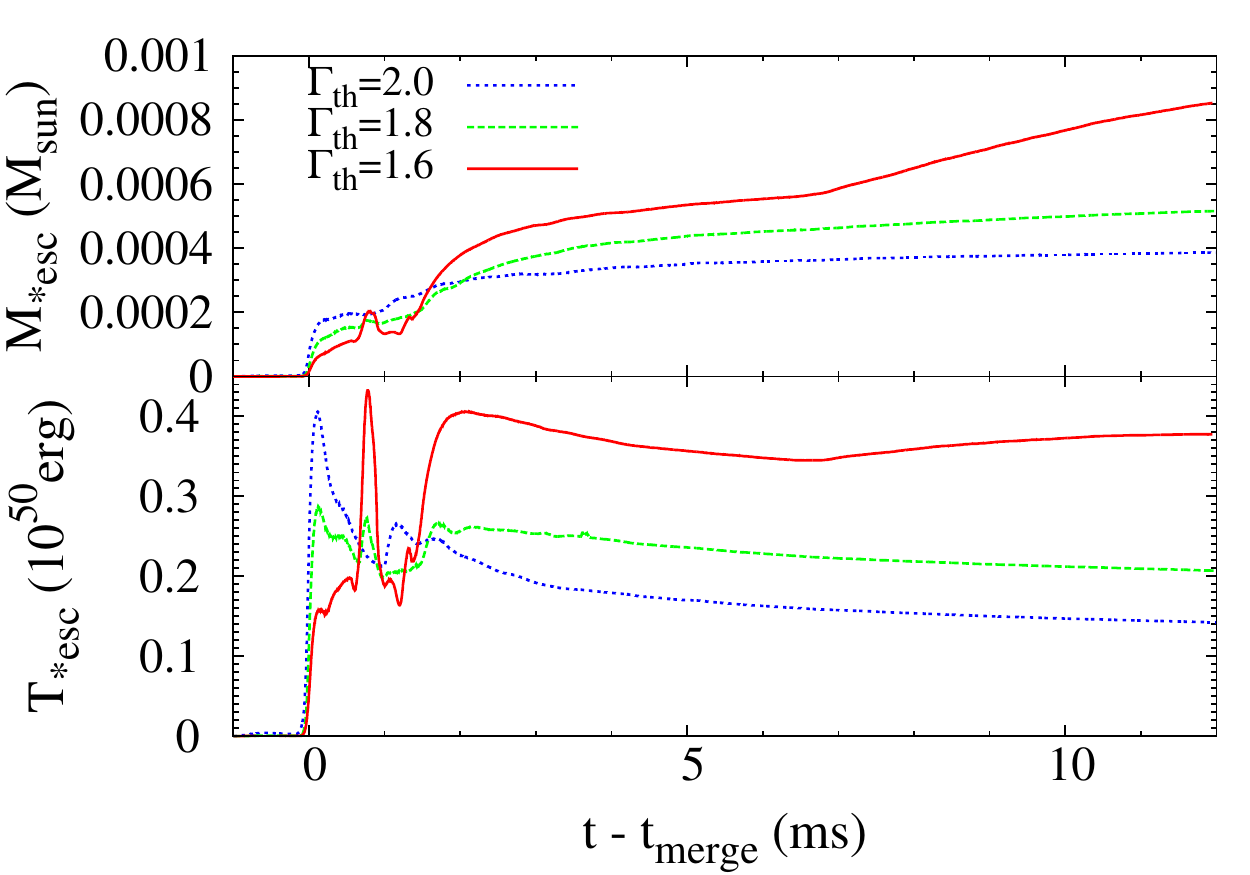}~~~
\includegraphics[width=82mm,clip]{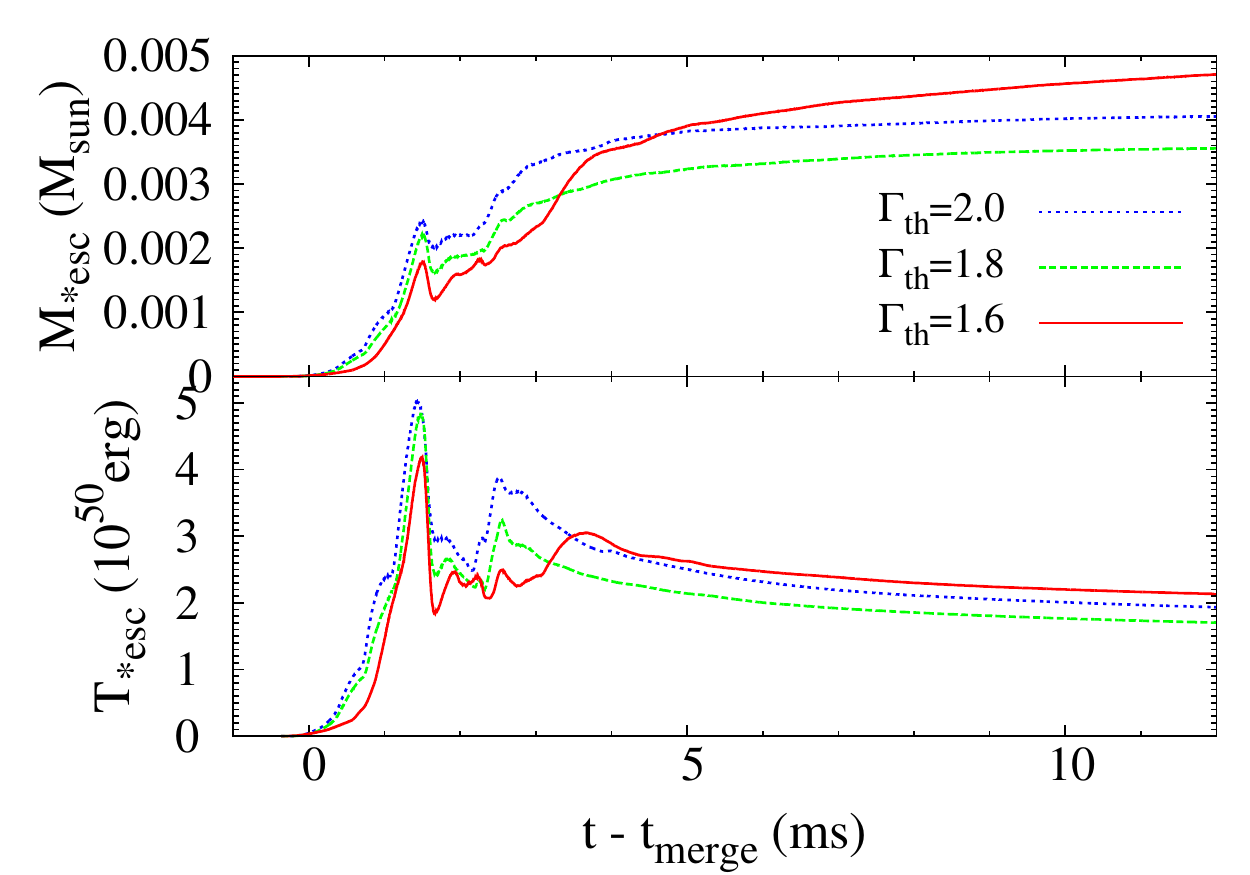}\\
\end{tabular}
\caption{$M_{*{\rm esc}}$ and $T_{*\rm esc}$ as functions of
$t-t_{\rm merge}$ (left) for models APR4-135135 (top left), APR4-120150
(top right), H4-135135 (bottom left), and H4-120150 (bottom right) 
with $\Gamma_{\rm th}=2.0$, 1.8, and 1.6.  }
\label{fig:APR1215}
\end{figure*}

\begin{figure*}[t]
\begin{tabular}{cc}
\includegraphics[width=82mm,clip]{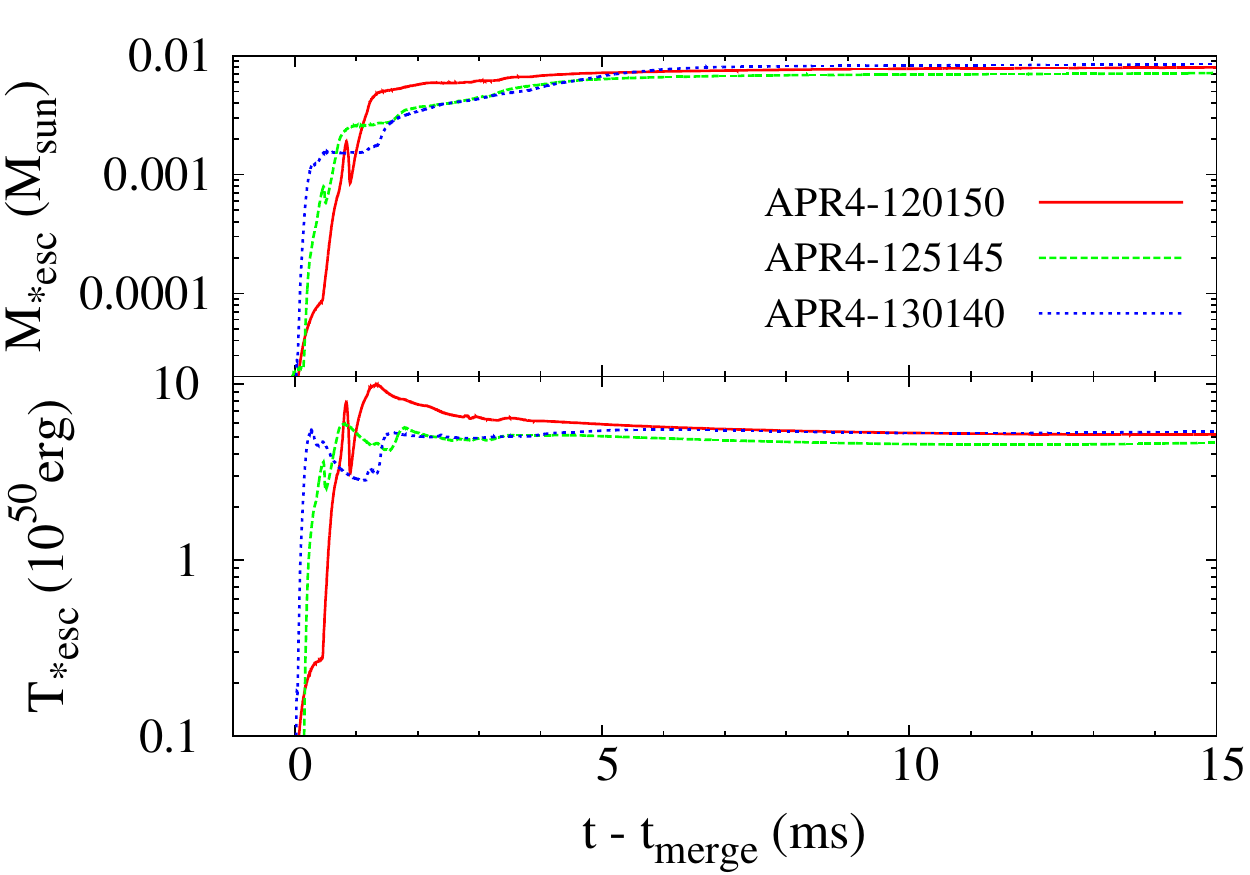}~~~
\includegraphics[width=82mm,clip]{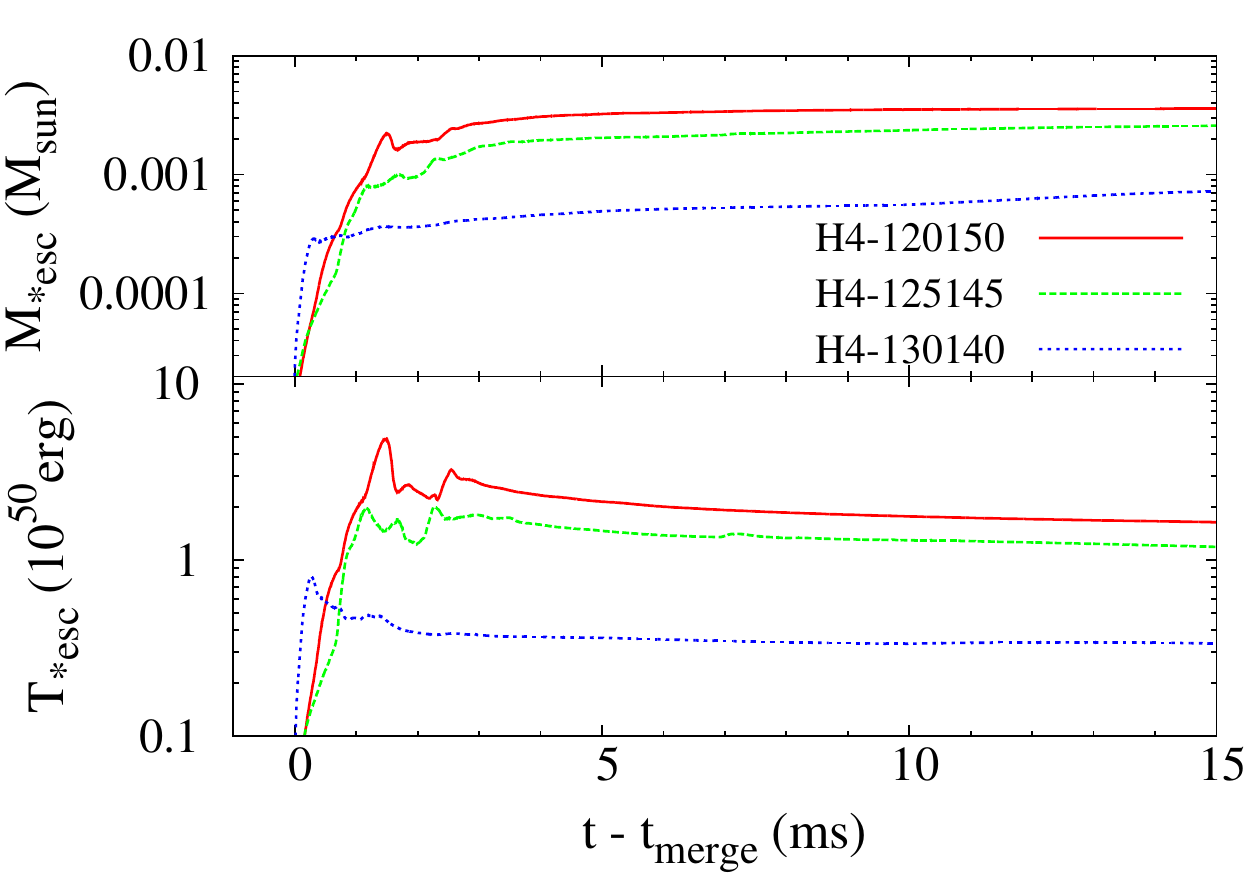}
\end{tabular}
\caption{$M_{*{\rm esc}}$ and $T_{*\rm esc}$ as functions of
$t-t_{\rm merge}$ (left) for models APR4-120150, APR4-125145,
APR4-130140, and (right) for models H4-120150, H4-125145, H4-130140.
}
\label{fig:escQ}
\end{figure*}

The total rest mass and kinetic energy for the ejected material
depend also on the value of $\Gamma_{\rm th}$.  The possible reason is
described as follows.

For larger values of $\Gamma_{\rm th}$, the effect of shock heating is
stronger. This implies that the thermal energy of the HMNS is
increased via the shock heating more efficiently, and thus, the
material located outside the HMNS that will eventually escape from the
system expands more efficiently at the merger and during the
subsequent shock heating. This effect could result in increasing the
ejected material.

On the other hand, the HMNS becomes less compact by more efficient
shock heating for the larger value of $\Gamma_{\rm th}$, and hence,
the amplitude of the quasiradial oscillation is smaller.  This
suggests that although the outward velocity of the material caused by
the shock heating is initially larger for the larger values of
$\Gamma_{\rm th}$, the subsequent gain of the kinetic energy via the
shock heating could be smaller. The less compact HMNS could be also
less favorable for exerting the torque to its surrounding material
because the rotational velocity is slower.  Therefore, the total rest
mass and kinetic energy of the material ejected from the system depend
on two competing nonlinear processes, as in a mechanism similar to
that mentioned in Sec.~\ref{sec:EOSdep}.  

%%Hence, the amount of the ejected material does not depend
%%monotonically on the value of $\Gamma_{\rm th}$. In the following, we
%%show this fact.

Figure~\ref{fig:xyzAPR} compares the evolution of $\bar R$ and $\bar
Z$ for $\Gamma_{\rm th}=1.6$, 1.8, and 2.0 for models APR4-135135 and
APR4-120150.  For APR4-135135, $\bar R$ is larger for the larger
values of $\Gamma_{\rm th}$ for $t-t_{\rm merge} \agt 1.5$~ms. This
agrees with the prediction that the shock heating effect is stronger
and the material expands in a wider region for the larger values of
$\Gamma_{\rm th}$. For $\bar Z$, the similar result is found for $3
\alt t-t_{\rm merge} \alt 5$~ms.  However, for $t-t_{\rm merge} \agt
5$~ms, $d\bar R/dt$ and $d\bar Z/dt$ have a similar magnitude
depending only weakly on the value of $\Gamma_{\rm th}$.  This is due
to the fact that the mass ejection is primarily driven by the torque
exerted by the HMNS.

For APR4-120150, soon after the onset of the merger, $\bar R$ and
$\bar Z$ are only slightly larger for the larger values of
$\Gamma_{\rm th}$. This is due to the fact that the mass ejection is
primarily driven by the tidal effect caused by the mass asymmetry
irrespective of the values of $\Gamma_{\rm th}$.  However, for
$t-t_{\rm merge} \agt 3$~ms, $d\bar R/dt$ and $d\bar Z/dt$ become
smaller for the smaller values of $\Gamma_{\rm th}$.  This is due to
the fact that for the smaller value of $\Gamma_{\rm th}$, in
particular, for $\Gamma_{\rm th}=1.6$, new materials with a smaller
velocity are gradually ejected for the later time.  This occurs due to
the fact that for the lower value of $\Gamma_{\rm th}$, a longer-term
mass ejection driven also by the torque exerted by the HMNS
occurs. Namely, for both equal-mass and unequal-mass cases, a longterm
mass ejection driven by the angular momentum transport from the HMNSs
play an important role for $\Gamma_{\rm th}=1.6$.

Figure~\ref{fig:APR1215} compares the evolution of the total rest mass
and kinetic energy for $\Gamma_{\rm th}=1.6$, 1.8, and 2.0 for models
APR4-135135, APR4-120150, H4-135135, and H4-120150. All the panels of
Fig.~\ref{fig:APR1215} clearly show that for the early time, $t
-t_{\rm merge} \alt 1.5$~ms for APR4-135135 and H4-135135, $\alt 5$~ms
for APR4-120150 and H4-120150, these two quantities are larger for the
larger value of $\Gamma_{\rm th}$. Namely the stronger shock heating
associated with the larger value of $\Gamma_{\rm th}$ plays an
important role.  However, after the early time, the rest mass tends to
be larger for the smaller value of $\Gamma_{\rm th}$.  In particular,
for $\Gamma_{\rm th}=1.6$, a rapid increase in the total rest mass is
found. Thus, a longterm mass ejection process driven by the torque
exerted by the HMNSs works for the smaller values of $\Gamma_{\rm th}$
(i.e., for more compact HMNSs), and this mechanism is remarkable for
$\Gamma_{\rm th}=1.6$. 

For APR4-120150 and H4-120150, the rest mass of the ejected material
is largest for $\Gamma_{\rm th}=1.6$. However, the kinetic energy
depends weakly on the value of $\Gamma_{\rm th}$. This implies that
although more materials are ejected, the gained kinetic energy is not
very large for $\Gamma_{\rm th}=1.6$, because the velocity of material
ejected later by the tidal torque is not very large.

The dependence of the rest mass and kinetic energy of the ejected
material on $\Gamma_{\rm th}$ is qualitatively similar for APR4 and
H4.  This indicates that the properties summarized in this subsection
would hold irrespective of the EOS.

\subsubsection{Dependence of the ejected material on the mass ratio
and total mass} \label{sec4.1.5}

The total rest mass and kinetic energy of the material ejected from
the HMNSs depend also on the mass ratio and total mass of binary
neutron stars.  The degree of the dependence depends on the EOS.
Figure~\ref{fig:escQ} plots $M_{*{\rm esc}}$ and $T_{*\rm esc}$ as
functions of $t-t_{\rm merge}$ for APR4 and H4 with three mass ratios
and with the total mass $2.7M_{\odot}$.  For the models with H4, the
total rest mass and kinetic energy of the ejected material depend
strongly on the mass ratio; e.g., the total rest mass and kinetic
energy for $q=0.8$ are by a factor of $\sim 5$ and 7 larger than those
for $q=0.929$ and $q=1$ with $m=2.7M_{\odot}$.  Essentially the same
results are found for the models with ALF2 and MS1 with $q<1$ (see
Table~\ref{table:result}).  By contrast, for the models with APR4, the
total rest mass and kinetic energy depend weakly on the mass ratio for
$m=2.7M_{\odot}$, and they are always larger than those with ALF2, H4
and MS1 for $m=2.6$ -- $2.8M_{\odot}$.  These facts indicate that (i)
for relatively stiff EOS such as ALF2, H4 and MS1, the asymmetry of
binary neutron stars enhances the efficiency of the angular momentum
transport via the tidal torque and increases the total amount of the
ejected material, and (ii) for a relatively soft EOS, APR4, which
yields a small-radius neutron star, the total amount of the ejected
material is always large irrespective of the mass ratio for the
canonical total mass $\sim 2.6$ -- $2.8M_{\odot}$. This is probably
because for APR4, the shock heating in the early evolution stage of
the HMNSs, in which they quasiradially oscillate with significant
amplitude, plays a primary role in the mass ejection irrespective of
the mass ratio; indeed, a large mass ejection is observed in the first
1 -- 2~ms after the onset of the merger.

The total rest mass and kinetic energy of the ejected material depend
also on the total mass of the system, and the degree of the dependence
depends also on the EOS: For APR4, these quantities are larger for
more massive system irrespective of the mass ratio (see
Table~\ref{table:result}). This property is consistent with the fact
that these quantities are larger for an EOS that yields compact
neutron stars. Namely, for the larger mass, the system can be in
general more compact for the binary neutron stars, and also the formed
HMNS can be more compact, more rapidly rotate, and quasiradially
oscillate with a larger amplitude.  Thus, the mass ejection is
enhanced through the angular momentum transport via the tidal torque
and the shock heating.

For H4, the similar results are obtained except for model H4-140140.
for which the rest mass and kinetic energy of the ejected material is
smaller than those for models H4-135135. The possible reason is that
for H4-140140 (for which a black hole is formed $\sim 10$~ms after the
onset of the merger), the HMNS formed is compact (i.e., it can trap
the material in its vicinity) and moreover, its shape (it is not a
sharp ellipsoid) could be unsuitable for efficiently exerting the
torque to the surrounding material.

For MS1 for which neutron stars and HMNSs are not very compact, the
quantities of the ejected material do not change very systematically.
As mentioned above, for $q<1$, the rest mass and kinetic energy increase
with the decrease of $q$ in the same manner as that for other
EOSs. However, for the equal-mass case ($q=1$) with $m=2.6$ and
$2.7M_{\odot}$, the ejected rest mass and kinetic energy are quite large
by contrast to that for $q=0.929$ and $m=2.7M_{\odot}$. This indicates
that for this system, the HMNS is not very compact and does not trap the
material strongly, and hence, angular momentum transport due to the
torque exerted by the HMNS and shock heating, which are not as efficient
as those in softer EOSs, are still large enough to overcome the trapping
effect.  For $m=2.7M_{\odot}$, in particular, the difference in the
results of $q=1$ and $q=0.929$ is quite large. The possible reason is
that (i) for the equal-mass case, the amplitude of the quasi-radial
oscillation is by a factor of $\sim 2$ larger than that for the
unequal-mass case, and thus, a larger amount of the materials are likely
to gain the escape velocity; (ii) the shape of the HMNS is quite
different between two models: For $q=1$, a sharp ellipsoid is formed,
and it appears to play a substantial role for a coherent angular
momentum transport from the HMNS to the material surrounding it. By
contrast, for $q=0.929$, the shape is not a clear ellipsoid but a
pear-shaped asymmetric object, and hence, the transport process does not
appear to proceed efficiently.

For MS1-140140, the quantities of the ejected material is much smaller
than those for MS1-135135 and as small as those for MS1-130140. The
possible reason is that the HMNS formed for MS1-140140 is slightly
more compact than that for MS1-135135, and the mass ejection is
suppressed by the trapping effect. This suggests that for these mass
ranges, a slight change in the compactness significantly affects the
efficiency of the mass ejection.

\subsection{Properties of the merger and mass ejection:
black hole formation case} \label{sec4.4}

\begin{figure*}[p]
\includegraphics[bb=0 0 960 360,width=180mm]{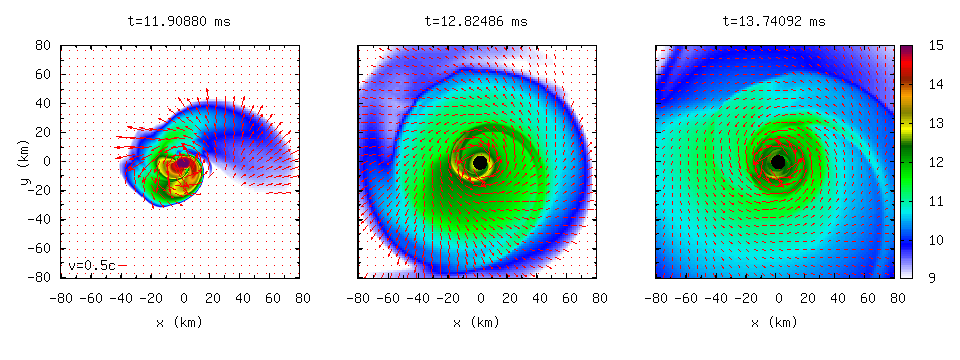} \\
\includegraphics[bb=0 -35 960 325,width=180mm]{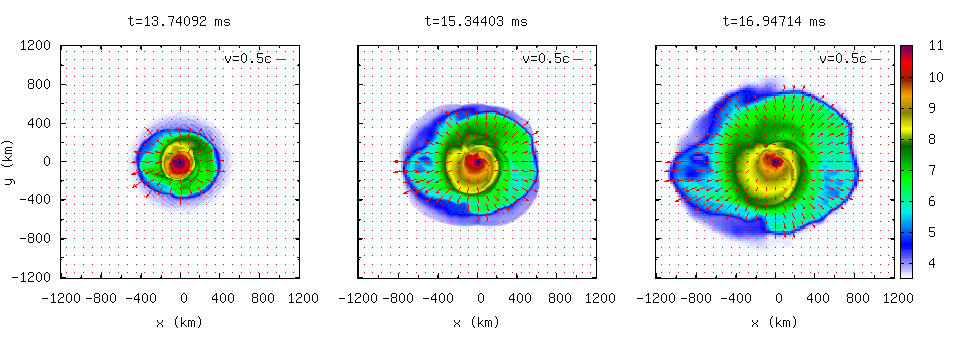}\\
\includegraphics[bb=0 -50 960 142,width=180mm]{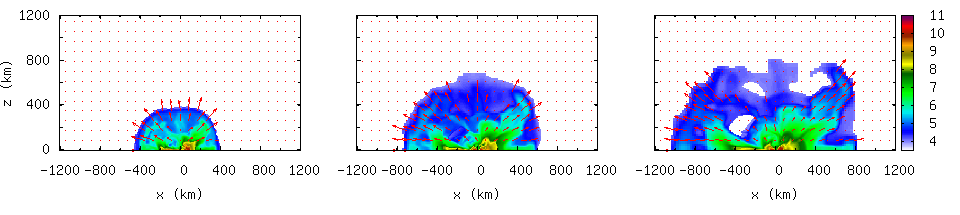} \\
\includegraphics[bb=0 -50 960 142,width=180mm]{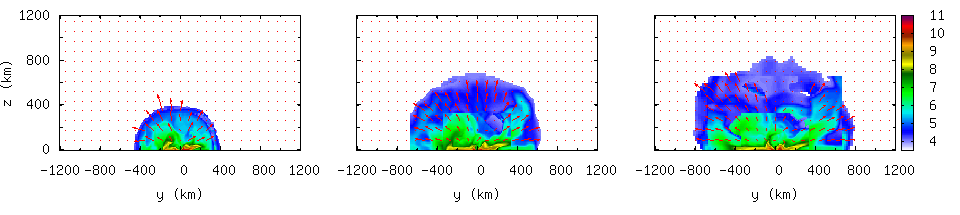}
\caption{The same as Fig.~\ref{fig3A} but for high-mass and 
unequal-mass model APR4-130160.  The filled black circles in the
middle and right panels of the top row denote black holes.}
\label{fig14}
\end{figure*}

We briefly summarize the properties of the ejected material for the
case that a black hole is promptly formed after the onset of the
merger. In this study, the prompt formation of a black hole occurs
only for APR4 with the total mass $2.9M_{\odot}$.

\begin{figure}[t]
\includegraphics[width=82mm,clip]{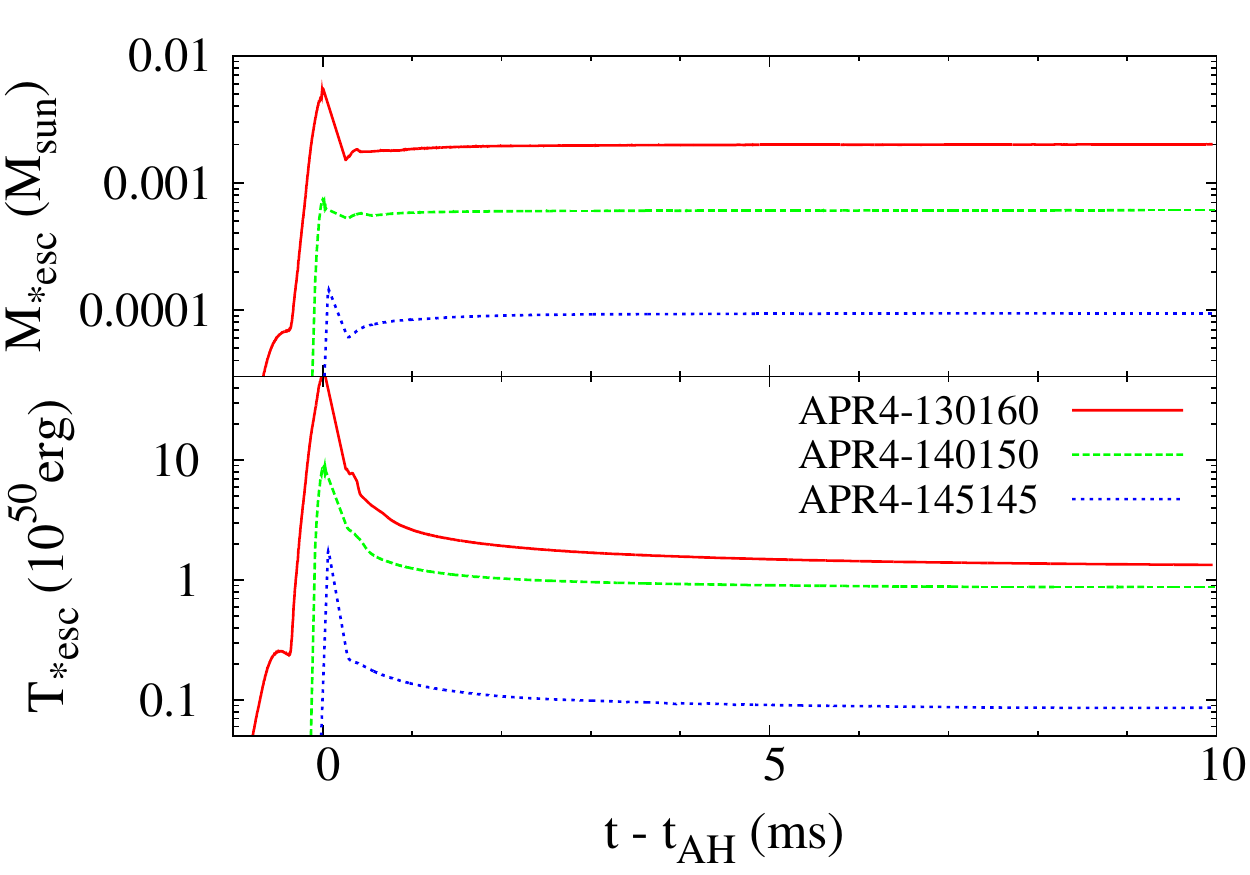}
\caption{$M_{*{\rm esc}}$ and $T_{*\rm esc}$ as functions of
$t-t_{\rm AH}$ for models APR4-130160, APR4-140150,
APR4-145145. Here, $t_{\rm AH}$ denotes the time at which 
an apparent horizon is formed. 
}
\label{fig15}
\end{figure}

For these models, the mass ejection primarily proceeds at the instance
of the merger, i.e., during a short duration before the formation of a
black hole. Because a black hole is promptly formed, a region
shock-heated at the collision of two neutron stars is soon swallowed
by the black hole, and thus, the shock heating does not play a primary
role in the mass ejection. A significant mass ejection occurs for the
case that the mass asymmetry is present, and the mass ejection is
induced primarily by a tidal torque. In the presence of mass
asymmetry, the less-massive neutron star is tidally elongated during
the merger, and a fraction of the tidally elongated neutron-star
material gains a sufficient torque from the merged object just before
the formation of a black hole and gets the escape velocity. For models
APR4-140150 and APR4-130160, this gain of the angular momentum is
large enough to eject materials of rest mass $\sim 6 \times
10^{-4}M_{\odot}$ and $2 \times 10^{-3}M_{\odot}$, respectively (see
Fig.~\ref{fig15}). In these cases, disks are also formed, and their
rest mass (for material bounded by the black hole) is $0.03M_{\odot}$
and $0.002M_{\odot}$, respectively.  The values for the mass ejection
depend only very weakly on the grid resolution with the fluctuation
within 10 -- 20\% level (see Appendix A). The reason is that strong
shocks do not play an important role in the mass ejection mechanism. 

The average velocity of the ejected material for these cases is $\sim
0.3c$ and larger than that in the case of the HMNS formation. The
reason is that the mass ejection is caused primarily by the tidal
interaction at the onset of the merger, and for this case, the induced
velocity is larger than that by subsequent longterm shock
heating. Because the tidal interaction plays a primary role, the
material is ejected primarily in the direction of the equatorial
plane. The motion to the $z$ direction is also induced by shock
heating that occurs when spiral arms surrounding the black hole
collide each other. However, this is a secondary effect. Hence, for
the case that a black hole is promptly formed from an asymmetric
binary, the value of $\theta_0$ is 30 -- $35^\circ$ which is smaller
than those for the case of the HMNS formation for which $\theta_0=40$
-- $50^\circ$.

For the equal-mass binary, the total rest mass of the ejected material
is quite small $\sim 10^{-4}M_{\odot}$ (see Fig.~\ref{fig15}), because
of the absence of the asymmetry and of the lack of the time during
which the material located in the outer region gains the torque from
the merged object (note that most of the fluid elements of binary
neutron stars just before the onset of the merger do not have the
specific angular momentum large enough to escape from the black hole
formed~\cite{SU00}). In this case, the disk mass surrounding the black
hole is also quite small, $\sim 10^{-4}M_{\odot}$. This is consistent
with our previous finding \cite{STU}.

\begin{figure*}[t]
\begin{tabular}{cc}
\includegraphics[width=80mm,clip]{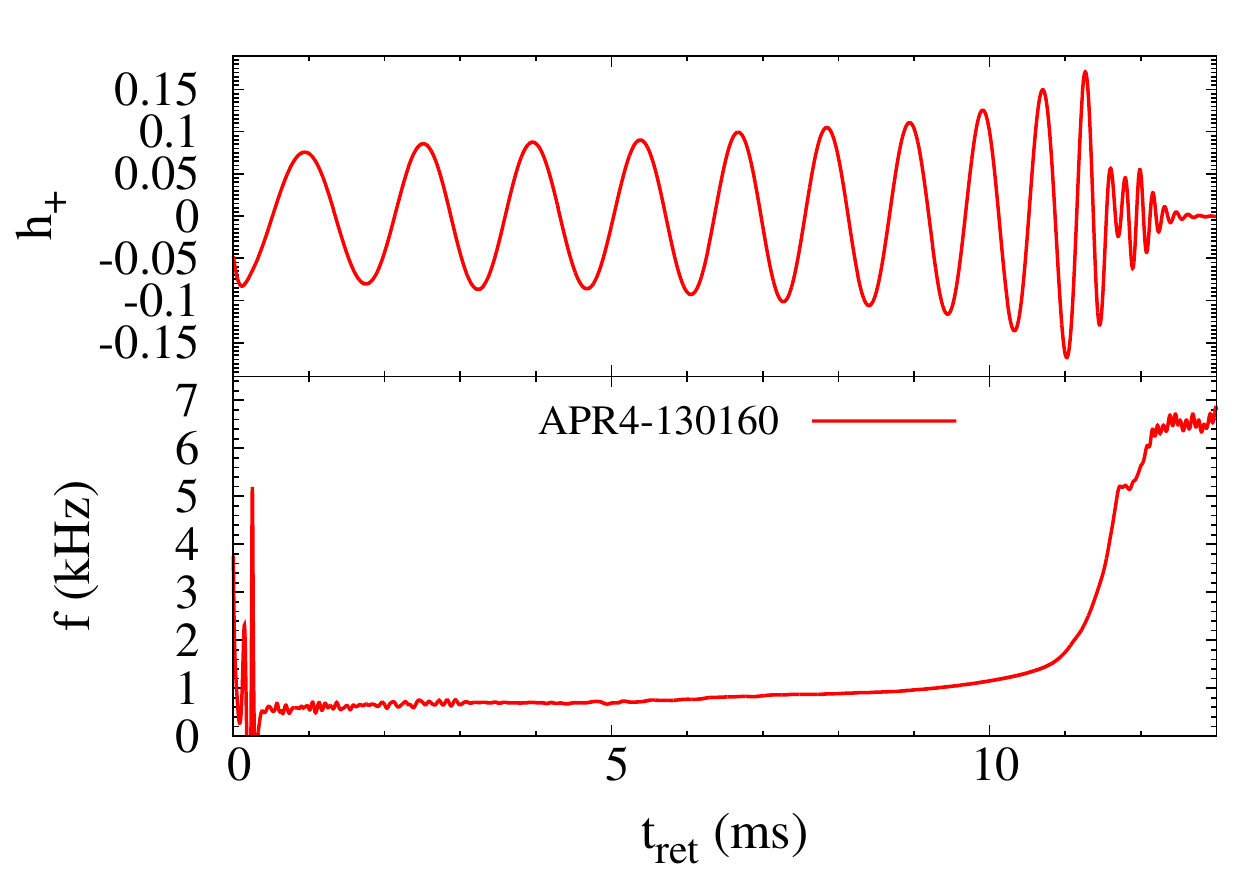}~~~~~~
\includegraphics[width=80mm,clip]{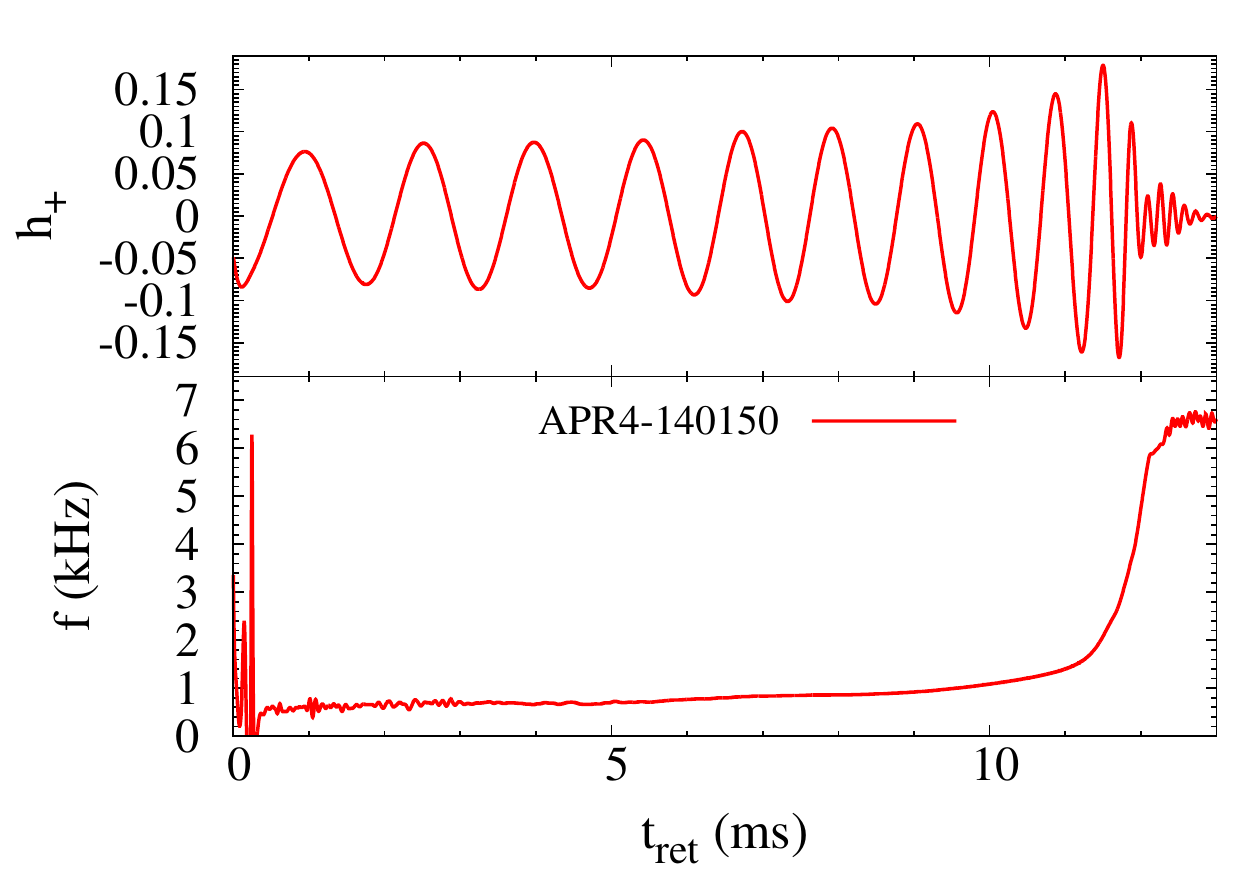}
\end{tabular}
\caption{The same as Fig.~\ref{figGW} but for 
models APR4-130160 (left) and APR4-140150 (right).  }
\label{fig16}
\end{figure*}

Figure~\ref{fig16} plots the gravitational waveforms for APR4-130160
and APR4-140150. For these models (also for APR4-145145), the
gravitational waveform is characterized by the inspiral waveform and
subsequent ringdown waveform. The frequency of gravitational waves
monotonically increases and eventually reaches the value of the
fundamental quasinormal mode of the formed black hole. For all three
cases, the frequency of gravitational waves associated with the
quasinormal mode is $6.55 \pm 0.05$~kHz (the mass and spin of the
formed black holes are 2.8 -- $2.83M_{\odot}$ and 0.77 -- 0.78,
respectively), which agrees with the frequency of the quasinormal mode
analytically derived~\cite{QNM}. Because HMNS is not formed, no
feature for the quasiperiodic oscillation associated with the HMNS
formation is found.

\section{Summary and discussion} \label{sec:summary}

\subsection{Summary}\label{sec5.1}

We reported our latest numerical-relativity studies for the material
ejected in the merger of binary neutron stars. We explored the
properties of the ejected material for a variety of EOSs, total
masses, and mass ratios of binary neutron stars, and found the
following facts. First, we summarize the results for the case that a
HMNS is formed:
\begin{itemize}
\item For the canonical total mass of the binary neutron stars $2.6$
-- $2.8M_{\odot}$, the total rest mass and kinetic energy of the
ejected material are approximately in the range $10^{-4}$ --
$10^{-2}M_{\odot}$ and $10^{49}$ -- $10^{51}$~ergs, respectively,
depending on the EOS, total mass, and mass ratio of binary neutron
stars.
%%%%%%%%%%%%%%%%%
\item The total rest mass and kinetic energy of the ejected material
depend strongly on the EOS. They are in general larger for binaries
composed of an EOS that yields compact (small-radius) neutron stars
(e.g., APR4).  They also depend on the efficiency of shock heating
(i.e., depend on $\Gamma_{\rm th}$): For many cases, smaller values of
$\Gamma_{\rm th}$ results in a larger ejected mass and kinetic
energy. 
%%%%%%%%%%%%%%%%%
\item The total rest mass and kinetic energy could depend also
strongly on the mass ratio of binary neutron stars. The dependence on
the mass ratio is in particular strong when the radius of neutron
stars is relatively large (i.e., for stiff EOSs such as H4 and
MS1). For many models in such EOSs, the ejected rest mass and kinetic
energy are larger for smaller mass ratios (for more asymmetric binary
neutron stars). By contrast, when the EOS is relatively soft (i.e.,
for APR4), the dependence of these quantities on the mass ratio is
weak.
%%%%%%%%%%%%%%%%%
\item The total rest mass and kinetic energy depend also on the total mass
of binary neutron stars. For many models in the present study, these
quantities are larger for the larger total mass irrespective of the
EOS. However, there are some exceptions for a class of EOS and for a
massive system.
%%%%%%%%%%%%%%%%%
\item The total rest mass ejected is in the wide range between $\sim
5\times 10^{-4}$ (H4) and $\sim 7 \times 10^{-3}M_{\odot}$ (APR4) for
equal-mass binaries with the total mass $m=2.7M_{\odot}$. For the
unequal-mass case with $q \approx 0.8$, it is in a rather narrow range
between $\sim 3\times 10^{-3}$ (MS1 and H4) and $\sim 8 \times
10^{-3}M_{\odot}$ (APR4) for $m=2.7M_{\odot}$.  This indicates that
the ejection of the material is induced by two different processes,
the torque exerted the HMNS and the shock heating.
%%%%%%%%%%%%%%%%%
\item The total kinetic energy is also in the wide range between
$\sim 10^{49}$~ergs (MS1 and H4) and $\sim 10^{51}$~ergs (APR4).  The
typical maximum velocity of the ejected material is 0.5 -- $0.8c$, and
the average velocity is 0.15 -- $0.25c$. For EOSs that yield a 
compact neutron star, the velocity of escaping material is larger. 
\end{itemize}

In our present study, a black hole is promptly formed for three
models; APR4-145145, APR4-140150, and APR4-130160.  For the case of
the prompt black-hole formation, a significant mass ejection occurs
only for the asymmetric binaries. For APR4-145145, the rest mass of
the ejected material is $\sim 10^{-4}M_{\odot}$. By contrast, it is
rather large for APR4-140150 and APR4-130160 as $6 \times
10^{-4}M_{\odot}$ and $2 \times 10^{-3}M_{\odot}$. Thus, a massive and
asymmetric binary can eject a large amount of the material even if a
black hole is promptly formed (even in the absence of a HMNS). For
this case, the average velocity of the ejected material is larger than
that for the case of the HMNS formation, and thus, a unique feature
may be seen in the observed electromagnetic signal (see next
section). However, we should note that the prompt black-hole formation
will occur only for a large total mass, because the latest discovery
of a high-mass neutron star PSR J1614-2230 with mass $1.97 \pm 0.04
M_{\odot}$~\cite{twosolar} indicates that the EOS should be rather
stiff and with such stiff EOSs, a HMNS is the canonical outcome for
the canonical-mass binary with $m=2.6$ -- $2.8M_{\odot}$.

The distribution of the matter around the remnant object depends
strongly on the merger process. For the case that a HMNS is formed, a
dense (physical) atmosphere (including the ejected material) is formed
around it (see Figs.~\ref{fig3A} -- \ref{fig3C}). The atmosphere is
distributed in a weakly anisotropic manner, and thus, the matter is
present even in the spin axis of the HMNS. Such dense atmosphere will
be present even after the HMNS collapses to a black hole. HMNS or
black hole subsequently formed will emit a huge amount of neutrinos
(e.g.,~\cite{SKSS}), and may drive a jet from the central region
through the fireball production via neutrino-antineutrino pair
annihilation.  To drive a SGRB for this case, however, the jet has to
penetrate the dense atmosphere and ejected material. Whether it is
possible or not is a question to be answered by the future
research. By contrast, for the case of the black hole formation, the
atmosphere is not very dense around the spin axis of the black hole
(see Fig.~\ref{fig14}). Thus, for this case, a SGRB would be driven,
if an energetic jet is launched as studied, e.g., in~\cite{AJM05}.

\subsection{Discussion}

We here briefly argue possible electromagnetic signals emitted by the
material ejected from the merger of binary neutron stars, referring to
the numerical results in the present work. As already mentioned, a
recent discovery of a high-mass neutron star PSR J1614-2230 suggests
that the maximum mass of spherical neutron stars should be larger than
$1.97 \pm 0.04 M_\odot$~\cite{twosolar}.  This indicates that a
long-lived HMNS would be a canonical outcome of the merger of binary
neutron stars, if the binaries were composed of neutron stars of
canonical mass of $1.3$ -- $1.4M_\odot$ with the total mass $\sim 2.6$
-- $2.8M_{\odot}$~\cite{hotoke}. The present numerical results
indicate that from the long-lived HMNS, a fraction of the
material could be ejected with large kinetic energy.

References~\cite{LP1998,Shri2005,Metzger2010,MB2012} discuss the
signals by the radioactive decay of $r$-process nuclei, which would be
produced from the neutron-rich material in the
outflow~\cite{RKLR2011,GBJ2011,Rosswog1999,Freiburghaus1999,Rosswog2000,Rosswog2012},
and subsequently decay and emit a
signal that may be observable by current and future-planned optical
telescopes such as PTF~\cite{PTF}, Pan-STARRs~\cite{PanStar}, and
LSST~\cite{LSST}. In this scenario, the typical duration of the peak
luminosity is of order a day or less as
\cite{LP1998}
\beqn
t_{\rm peak} \approx 0.1{\rm d}\biggl({\beta_0 \over 0.2}\biggr)^{-1/2}
\biggl({M_{*{\rm esc}} \over 10^{-3}}\biggr)^{1/2}, 
\eeqn
and the associated peak luminosity is
\beqn
L_{\rm peak} &\approx& 7 \times 10^{41}~{\rm ergs/s}
\biggl({f_{\rm eff} \over 3 \times 10^{-6}}\biggr) \nonumber \\
&&~~~~~~~~~~~~~
\times\biggl({\beta_0 \over 0.2}\biggr)^{1/2}
\biggl({M_{*{\rm esc}} \over 10^{-3}M_{\odot}}\biggr)^{1/2}
\eeqn
where $f_{\rm eff}$ denotes the conversion rate of the energy per
rest-mass energy in the ejected material through the radioactive decay
process, which is $\sim 3 \times 10^{-6}$ according to the results
of~\cite{Metzger2010}. $\beta_0c$ is the typical velocity of the
ejected material.  The result of~\cite{Metzger2010} suggests that if
the total ejected mass is $\agt 10^{-3}M_\odot$, the signal will be
detected by large optical surveys such as LSST for a typical distance
to sources $\sim 100$~Mpc.  Our numerical results indicate that
$\beta_0=0.15$ -- 0.25, and the total ejected mass is $\sim 10^{-3}$
-- $10^{-2}M_{\odot}$ for binaries composed of neutron stars with a
small radius $\sim 11$ -- 12.5~km, and $\sim 0.3
\times 10^{-3}$ -- $5 \times 10^{-3}M_{\odot}$ for binaries composed
of neutron stars with a larger radius $\sim 13.5$ -- 14.5~km for
plausible values of $\Gamma_{\rm th}=1.6$ -- 2.0.  Thus, if the EOS is
a rather ``soft'' one that yields a small-radius neutron star, an
observable optical signal due to the radioactive decay can be expected
with a duration of several hours.  Taking into account the short
duration of the signal, rapid follow-up searches and an efficient
coverage for the error circle of the direction of the
gravitational-wave events are required.  If the EOS is a ``stiff'' one
that yields a large-radius neutron star, the strength of the signal
will be weaker and the duration shorter, although it would be still
possible to detect the signal in particular for the merger of
unequal-mass (sufficiently asymmetric) neutron stars.

There is also another possible channel for the electromagnetic
emission. According to recent studies~\cite{Nakar2011,MB2012,PNR2012},
the ejected material, which is in the free expansion, will sweep up
the interstellar matter and form blast waves. During this process
turning on, the shocked material could generate magnetic fields and
accelerate particles that emit synchrotron radiation, for a
hypothetical amplification of the electromagnetic field and a
hypothetical electron injection.  The emission will peak when the
total swept-up mass approaches the ejected mass, because the blast
waves are decelerated and transit to the phase in which the motion of
the material is described by the (non-relativistic) Sedov-Taylor's
self-similar solution.  The predicted duration for the synchrotron
radiation depends on the total energy $E_0$ and speed of the ejected
material $\beta_0c$ as well as the number density of the interstellar
matter $n_0$. The duration to reach the peak luminosity is estimated
in~\cite{Nakar2011} as
\begin{equation}
\tau_{\rm radio} \sim 4~{\rm yrs}
\biggl({E_0 \over 10^{50}~{\rm ergs}}\biggr)^{1/3}
\biggl({n_0 \over 1~{\rm cm}^{-3}}\biggr)^{-1/3}
\biggl({\beta_0 \over 0.2}\biggr)^{-5/3}. \label{eq32}
\end{equation}
By the synchrotron radiation, a radio signal could be emitted as in
the late phase of supernovae and the afterglow of gamma-ray
bursts~\cite{Nakar2011}. Our numerical results indicate that the
typical velocity of the ejecta is $\beta_0=0.15$ -- 0.25 irrespective
of the EOS and masses of neutron stars in binaries.  However, $E_0$ is
in a wide range between $\sim 10^{49}$~ergs and $10^{51}$~ergs, depending
strongly on the EOS, mass ratio, and total mass of the binaries, and
its value is highly uncertain. Thus the predicted value of $\tau_{\rm
radio}$ is in a wide range $\sim 1$ -- 10~yrs, even for an optimistic
value of $n_0=1~{\rm cm}^{-3}$. For smaller values of $n_0$ which is
likely when the merger occurs outside the galactic plane, the value of
$\tau_{\rm radio}$ is much longer.

For the typical value of the ejecta velocity $\beta_0 \sim 0.2$, the
peak flux for the observed frequency is obtained at the deceleration
time described in Eq.~(\ref{eq32}). Specifically, the peak flux may be
obtained at the self-absorption frequency, $\sim 1$--2 hundreds MHz,
and the typical synchrotron frequency is sub-MHz. The peak flux for a
given observed radio-band frequency $\nu_\mathrm{obs}$ is 
\begin{align}
 F_\nu & \approx 90~\mu\mathrm{Jy} \left( \frac{E_0}{10^{50}~\mathrm{ergs}}
 \right) \left( \frac{n_0}{1~\mathrm{cm}^{-3}} \right)^{0.9} \left(
 \frac{\beta_0}{0.2} \right)^{2.8} \notag \\
 &~~~~~ \times \left( \frac{D}{200~\mathrm{Mpc}} \right)^{-2} \left(
 \frac{\nu_\mathrm{obs}}{1.4~\mathrm{GHz}} \right)^{-0.75},
\label{eq33}
\end{align}
where we assumed the power-law distribution of the electron's Lorentz
factor with the power 2.5. Equation~(\ref{eq33}) is applicable as long
as the observed frequency is higher than the typical synchrotron and
self-absorption frequency at the deceleration time, $\tau_{\rm
radio}$.  Equation (\ref{eq33}) indicates that for a hypothetical
event at a distance of 200~Mpc, $E_0 \sim 10^{50}$~ergs with
$n_0=1~{\rm cm}^{-3}$ is strong enough to be observed by
future-planned radio instruments (such as EVLA~\cite{EVLA},
ASKAP~\cite{ASKAP}, MeerKAT~\cite{MeerKAT}, and Apertif for which the
root-mean square value of the background noise for one hour
observation is smaller than 50~$\mu\mathrm{Jy}$ as shown
in~\cite{Nakar2011}).  Therefore, the mass-ejection mechanism could
supply a large amount of the kinetic energy which generates an
observable strong radio signal, if the EOS is rather soft (i.e., the
neutron-star radius is fairly small) or the binary is significantly
asymmetric.

In this scenario, the duration to reach the peak luminosity and the
strength of the radio signal depend strongly on the value of $n_0$.
In nature, the value of $n_0$ will depend strongly on the site where
the merger of binary neutron stars happens. If it is in a galactic
disk, $n_0$ would be typically $\sim 1~{\rm cm}^{-3}$, whereas if it
is outside a galaxy, the value is much smaller as $\sim 10^{-3}~{\rm
cm}^{-3}$. Equation (\ref{eq33}) shows that for a smaller value of
$n_0 \ll 1~{\rm cm}^{-3}$, $F_{\nu} \alt 1 \mu\mathrm{Jy}$ even for
$E_0=10^{51}~{\rm ergs}$. Our numerical simulation shows that the
maximum value of $E_0$ is at most $10^{51}$~ergs.  Therefore, for the
low value of $n_0 \sim 10^{-3}~{\rm cm}^{-3}$, this type of
electromagnetic signals may not be observable as a counterpart of the
gravitational-wave signal~\cite{MB2012}.

We here note the following point. We used the total kinetic energy and
average velocity as $E_0$ and $\beta_0$, when estimating the radio flux
estimated in Eq.~(\ref{eq33}). However, as we found in this paper, the
ejected material has a wide range of the velocity, and the amount of the
kinetic energy which the material of a given value of $\beta_0$ has
depends on the value of $\beta_0$. Thus, each material has a different
deceleration time and flux $F_{\nu}$, and therefore, the light curve
will have a complicated structure depending on the distribution of the
differential mass as a function of the ejecta velocity (see also
\cite{KIS2012}). In a subsequent paper, we plan to study the luminosity
curve in more detail following~\cite{PNR2012}.

As summarized in Sec.~\ref{sec5.1}, the properties of the ejected
material depend strongly on the EOS, mass ratio, and total mass of the
binary. This suggests that the observed electromagnetic signal depends
on them. The observation of gravitational waves in the inspiral phase
of binaries, which will be observed in the advanced gravitational-wave
detectors~\cite{LIGOVIRGO}, will carry the information of the mass
ratio and total mass. The observation of gravitational waves from the
final inspiral phase and HMNS could constrain the EOS of neutron
stars. Together with these information by the gravitational-wave
observation, the observation of the electromagnetic signals will be
used for clarifying the dynamics of the binary merger and ejected
material.  In addition, clarifying the spectroscopic properties of
electromagnetic emission associated with the decay of $r$-process
nuclei may be interesting. If there is a characteristic
emission/absorption feature in this emission, it will be helpful to
determine the cosmological redshift of the source event through a
spectroscopic observation. The redshift determined will be
subsequently used for determining the distance to the source (and
thus, Hubble constant) and the physical mass of binary systems through
the analysis of observed inspiral gravitational waves~\cite{schutz86}.
However, for the clarification, detailed theoretical studies for the
ejected material and electromagnetic radiation are necessary for a
variety of the EOS, mass ratio, and total mass.  These are new and
interesting tasks for the community of numerical relativity.

\acknowledgments

We are grateful to T. Piran for suggesting to explore the mass
ejection in detail, and K. Ioka, E. Nakar, and Y. Suwa for helpful
discussions.  This work was supported by Grant-in-Aid for Scientific
Research (21340051, 21684014, 23740160, 24244028, 24740163), by
Grant-in-Aid for Scientific Research on Innovative Area (20105004), and
HPCI Strategic Program of Japanese MEXT. The work of Hotokezaka is
supported by the Grant-in-Aid of JSPS.

\appendix

\section{Convergence}

The rest mass and kinetic energy of the ejected material with
different grid resolutions for selected models are listed in
Table~\ref{table:conv}.  This shows that for the unequal-mass models,
a convergence is well achieved (even for $m_1=1.3M_{\odot}$ and
$m_2=1.4M_{\odot}$), whereas for the equal-mass models, the
convergence is poor. In this case, the results with $N > 60$ could be
modified by a factor of $\sim 2$ from the results with $N=60$.  As
mentioned in Sec.~\ref{sec4.1}, the convergence is poor for the case
that a strong shock is formed at the merger and it plays a primary
role in the mass ejection. The possible reason for this is that (i)
the shocks are always computed by the first-order accuracy, and hence,
the accuracy is low and (ii) the ejected mass is a tiny part of the
entire system, and hence, a random error for the entire system
computed with a low accuracy significantly (and randomly) affects a
tiny amount of the ejected material. By contrast, when the 
tidal torque and hydrodynamical torque exerted by the HMNS play 
an important role in the mass ejection, the convergence is good. 

The averaged frequency of gravitational waves emitted by 
HMNSs is also listed for three grid resolutions. This shows that 
the frequency is obtained within the error of $\alt 0.1$~kHz, 
which is smaller than the physical dispersion of the 
frequency associated with the quasiradial oscillation of the HMNSs 
and the secular change of the density and velocity profiles of the 
HMNSs caused by the angular momentum transport. 

\begin{table*}[t]
\caption{Convergence for the rest mass and kinetic energy of 
the ejected material and average gravitational-wave frequency for
selected models. For each column, $(M_{*{\rm esc}}, E_{*{\rm esc}},
f_{\rm ave,5ms},f_{\rm ave,10ms})$ in units of $10^{-3}M_{\odot}$
$10^{50}$~ergs, and kHz are listed. In this table, the values for the
ejected material are shown in 2 significant digits. }
%%%%%%%%%%%%%%%%
{\begin{tabular}{c|cccc} \hline
Model & $N=40$ & $N=48$ & $N=50$ & $N=60$
\\ \hline \hline
APR4-130160 & $(2.3, 1.8,~\mbox{---}~,~\mbox{---}~)$ 
& $(2.5, 2.0,~\mbox{---}~,~\mbox{---}~)$ & ---  
& $(2.0, 1.5,~\mbox{---}~,~\mbox{---}~)$  \\
APR4-120150 & $(8.0, 5.4, 3.29, 3.30)$ & $(8.4, 5.7, 3.30, 3.28)$ 
& ---  & $(8.0, 5.2, 3.41, 3.35)$  \\
APR4-130140 & $(8.4, 5.7, 3.33, 3.34)$ & $(7.8, 5.0, 3.32, 3.29)$ 
& ---  & $(8.0, 4.8, 3.30, 3.27)$  \\
APR4-135135 & $(11,~ 7.0, 3.40, 3.40)$ & --- & 
$(6.6, 3.6, 3.34, 3.34)$  & $(6.5, 3.2, 3.31, 3.31)$  \\
ALF2-120150 & $(4.5, 2.5, 2.65, 2.68)$ & --- & 
$(4.8, 2.3, 2.75, 2.78)$  & $(5.4, 2.9, 2.70, 2.71)$  \\
ALF2-130140 & $(1.7, 0.7, 2.72, 2.72)$ & --- 
& $(1.7, 0.9, 2.71, 2.75)$  & $(1.6, 0.8, 2.73, 2.75)$  \\
ALF2-135135 & $(1.0, 0.5, 2.77, 2.82)$ & --- & 
$(1.5, 0.9, 2.79, 2.82)$  & $(2.8, 1.5, 2.75, 2.76)$  \\
H4-120150   & $(3.5, 1.6, 2.27, 2.27)$ & --- 
& $(3.8, 1.8, 2.28, 2.28)$  & $(3.5, 1.8, 2.30, 2.31)$  \\
H4-135135   & $(0.3, 0.1, 2.47, 2.51)$ & --- 
& $(0.3, 0.1, 2.48, 2.52)$  & $(0.5, 0.2, 2.44, 2.48)$ \\
MS1-120150  & $(3.4, 1.4, 2.08, 2.08)$ & --- 
& $(3.1, 1.4, 2.10, 2.09)$  & $(3.4, 1.5, 2.08, 2.09)$  \\
%%%%MS1-130140  & $(0.6, 0.2)$ & --- & (1.2,0.5) & $(0.6, 0.2)$  \\
MS1-135135  & $(0.6, 0.2, 2.08, 2.07)$ & --- & $(0.8, 0.3, 2.00, 1.97)$ 
& ($1.6, 0.6, 1.98, 1.95)$  \\
\hline
\hline
\end{tabular}
}
\label{table:conv}
\end{table*}

\end{document}